\documentclass[twocolumn,tighten]{aastex61}

\received{\today}
\revised{???, 2017}
\accepted{???, 2017}

\submitjournal{ApJ}

\shorttitle{Lopsidedness of Satellite Galaxy Systems in $\Lambda$CDM simulations}
\shortauthors{Pawlowski, Ibata \& Bullock}

\begin{document}

\defcitealias{2016ApJ...830..121L}{L16}

\title{The Lopsidedness of Satellite Galaxy Systems in $\Lambda$CDM simulations}

\correspondingauthor{Marcel S. Pawlowski}
\email{marcel.pawlowski@uci.edu}

\author{Marcel S. Pawlowski}
\altaffiliation{Hubble Fellow}
\affil{Department of Physics and Astronomy, University of California, Irvine, CA 92697, USA}

\author{Rodrigo A. Ibata}
\affil{Universit\'e de Strasbourg, CNRS, Observatoire astronomique de Strasbourg, UMR 7550, F-67000 Strasbourg, France}

\author{James S. Bullock}
\affil{Department of Physics and Astronomy, University of California, Irvine, CA 92697, USA}

\begin{abstract}
The spatial distribution of satellite galaxies around pairs of galaxies in the Sloan Digital Sky Survey (SDSS) have been found to bulge significantly towards the respective partner. Highly anisotropic, planar distributions of satellite galaxies are in conflict with expectations derived from cosmological simulations. Does the lopsided distribution of satellite systems around host galaxy pairs constitute a similar challenge to the standard model of cosmology? We investigate whether such satellite distributions are present around stacked pairs of hosts extracted from the $\Lambda$CDM simulations Millennium-I, Millennium-II, ELVIS, and Illustris-1. By utilizing this set of simulations covering different volumes, resolutions, and physics, we implicitly test whether a lopsided signal exists for different ranges of satellite galaxy masses, and whether the inclusion of hydrodynamical effects produces significantly different results. All simulations display a lopsidedness similar to the observed situation. The signal is highly significant for simulations containing a sufficient number of hosts and resolved satellite galaxies (up to $5\,\sigma$ for Millennium-II). We find a projected signal that is up to twice as strong as that reported for the SDSS systems for certain opening angles ($\sim16\%$\ more satellites in the direction between the pair than expected for uniform distributions). Considering that the SDSS signal is a lower limit owing to likely back- and foreground contamination, the $\Lambda$CDM simulations appear to be consistent with this particular empirical property of galaxy pairs.
\end{abstract}

\keywords{galaxies: dwarf --- galaxies: halos --- galaxies: statistics --- galaxies: structure }

\section{Introduction} \label{sec:intro}

The study of anisotropy in the distributions of satellite galaxies is gaining increasing attention. One major driver is the debate about planes of satellite galaxies.
First evidence for an anisotropic distribution of satellite galaxies around the Milky Way was reported for the by \citet{1976RGOB..182..241K} and \citet{1976MNRAS.174..695L}. This early indication has since then been corrobated \citep{2012MNRAS.423.1109P} and similar structures have been found around the Andromeda galaxy \citep{2013Natur.493...62I} and in the Local Group \citep{2013MNRAS.435.1928P}. There also is mounting evidence for a prevalence of similar structures beyond the Local Group: \citet{2013AJ....146..126C} reported that the dwarf Spheroidal (dSph) satellite galaxies around M81 lie in a flattened distribution, \citet{2015ApJ...802L..25T} identified two parallel satellite galaxy planes around Centaurus\,A, and \citet{2014Natur.511..563I} performed a statistical analysis of the kinemativs of satellite galaxy pairs which appear consistent with as prevalence of planes of co-orbiting satellite galaxies. The existence \citep{2015MNRAS.453.3839P, 2016A&A...595A.119M, 2015ApJ...805...67I}, origin \citep{2013MNRAS.431.3543H, 2015ApJ...799L..13C, 2015MNRAS.452.1052L, 2016ApJ...818...11S}, and compatibility with cosmological expectations \citep{2005A&A...431..517K, 2013MNRAS.429.1502W, 2014MNRAS.442.2362P, 2015MNRAS.452.3838C} of these structures are heavily discussed.  Most recently, \citet{2017ApJ...843...62M} and \citet{2017arXiv170200497M} have claimed that the satellite plane identified around the Milky Way may be a mere artifact caused by selection effects. \citet{2017arXiv170206143P} strongly critisized this claim for lacking a statistical basis and for neglecting the effects of observational biases and uncertainties (that work to wash out tight correlations). They also showed that the cosmological simulation of a single host in \citet{2017arXiv170200497M} does not show a kinematic correlation among satellites that is as strong as that observed among the Milky Way satellite galaxies.

Planar distributions, however, are not the only signature of anisotropy that has been investigated. There are numerous studies focusing on the satellite distributions relative to their host \citep{1969ArA.....5..305H, 2005ApJ...628L.101B, 2008ApJ...675..146F, 2006MNRAS.369.1293Y}, or in relation to the surrounding large-scale structure \citep{2011MNRAS.411.1525L, 2015MNRAS.452.1052L, 2015MNRAS.450.2727T}. 

The satellite galaxy system of M31 is highly lopsided, as about 80\% of its satellites lie on the side facing the Milky Way \citep{2013ApJ...766..120C}. Motivated by this finding, \citet[][hereafter L16]{2016ApJ...830..121L} have searched for a similar anisotropic signal in satellite systems around a larger sample of more distant galaxy pairs, chosen have $r$-band magnitudes similar to the Local Group giants ($-23.5 \le M_r \le -21.5$). They specifically analyzed stacked, photometrically-identified satellite candidates seen in projection around pairs selected from the Sloan Digital Sky Survey \citep[SDSS, ][]{2000AJ....120.1579Y}. Comparing the number of satellites within a given opening angle around the direction to their host's partner galaxy, they demonstrate that the satellite systems show a significant bulging in this direction. The $\approx 20^\circ$\ towards the second primary contain about 8\% more satellites than expected for a uniform distribution. Since \citetalias{2016ApJ...830..121L} count all galaxies that are close to their hosts in projection, with no consideration of satellite photometric or spectroscopic redshift, their sample is likely contaminated by foreground and background galaxies.  Therefore, the intrinsic signal is most likely higher. 

In order to test whether this observed signal is in conflict with expectations based on the standard model of cosmology ($\Lambda$CDM), we search for a similar signal of lopsided satellite distributions around galaxy pairs selected from cosmological simulations.

\section{Simulation Data} \label{sec:simdata}

\begin{deluxetable*}{l|cccccccc}
\tablecaption{Simulation samples and results. \label{tab:table}}
\tablehead{
\colhead{Simulation} & 
\colhead{$N_\mathrm{host}$} & 
\colhead{$N_\mathrm{pairs}$} & 
\colhead{$N_\mathrm{sat}$} & 
\colhead{$f_\mathrm{lopsided}$} & 
\colhead{$p_\mathrm{KS}$} & 
\colhead{$<\sigma>$} & 
\colhead{$\sigma_\mathrm{max}$} & 
\colhead{$\mathrm{angle~of}~\sigma_\mathrm{max}$}
}
\colnumbers
\startdata
3D analysis ($r_\mathrm{max} = 350\,\mathrm{kpc}$) &  &  &  &  &  &  &  & $\cos(\theta)$ \\
Millennium I & 402,669 & 7,573 & 59,809 & 0.5027 & $4.3 \times 10^{-31}$ & 8.7 & 13.7 & 0.8 \\
Millennium I (no orphans) &            &          & 7,091 & 0.5167 & $3.0 \times 10^{-6}$ & 3.6 & 5.1 & 0.6 \\
Millennium II & 5,519 & 147 & 43,146 & 0.5000 & $5.6 \times 10^{-8}$ & 3.9 & 6.7 & 0.7 \\
Millennium II (no orphans) &         &          & 14,386 & 0.5047 & $2.1 \times 10^{-2}$ & 1.9 & 3.0 & 0.6 \\
ELVIS & 24 & 12 & 15,793 & 0.4981 & $4.9 \times 10^{-3}$ & 2.1 & 4.0 & 0.7 \\
Illustris & 2,551 & 99 & 2,461 & 0.5091 & $4.2 \times 10^{-1}$ & 1.0 & 2.0 & 0.8 \\
\hline
2D analysis ($r_\mathrm{max} = 250\,\mathrm{kpc}$) &  &  &  &  &  &  &  & $\theta~\mathrm{in}~^\circ$ \\
Millennium I & 402,669 & 6,387 & 55,074 & 0.5091 & $1.2 \times 10^{-15}$ & 5.9 & 8.5 & 36.0 \\
Millennium I (no orphans) &         &          & 9,319 & 0.5215 & $7.4 \times 10^{-6}$ & 3.2 & 4.0 & 63.0 \\
Millennium II & 5,519 & 123 & 39,901 & 0.5036 & $1.9 \times 10^{-8}$ & 4.4 & 6.5 & 27.0 \\
Millennium II (no orphans) &          &          & 15,259 & 0.5118 & $3.1 \times 10^{-4}$ & 3.3 & 5.3 & 18.0 \\
Illustris & 2551 & 83 & 2304 & 0.5243 & $2.0 \times 10^{-2}$ & 0.9 & 1.6 & 72.0 \\
\hline
SDSS (observed) &          & 12,210 & 46,043 &          & $1.04 \times 10^{-5}$ & 4.4 & $>5$ & $\sim20$ \\
\enddata
\tablecomments{The first lines for the Millennium simulations include all satellite galaxies, the second ones only those with types 0 and 1. 
$N_\mathrm{host}$\ is the total number of host galaxies in the selected $r$-band magnitude range, $N_\mathrm{pairs}$\ is the number of sufficiently isolated pairs identifed among those, which in turn host a total of $N_\mathrm{sat}$\ satellite galaxies. $f_\mathrm{lopsided}$\ is the average overall lopsidedness between the hemisphere pointing to the partner galaxy and the opposite hemisphere. $p_\mathrm{KS}$\ is the KS-test $p$-value of the angular distribution of satellite galaxies being drawn from a uniform one. $<\sigma>$\ is the average significance of the cumulative angular distribution relative to the partner galaxy, compared to a uniform distribution, while $\sigma_\mathrm{max}$\ is the peak significance value, reached at an opening angle of $\mathrm{angle~of}~\sigma_\mathrm{max}$.
}
\end{deluxetable*}

\begin{figure*}
\plottwo{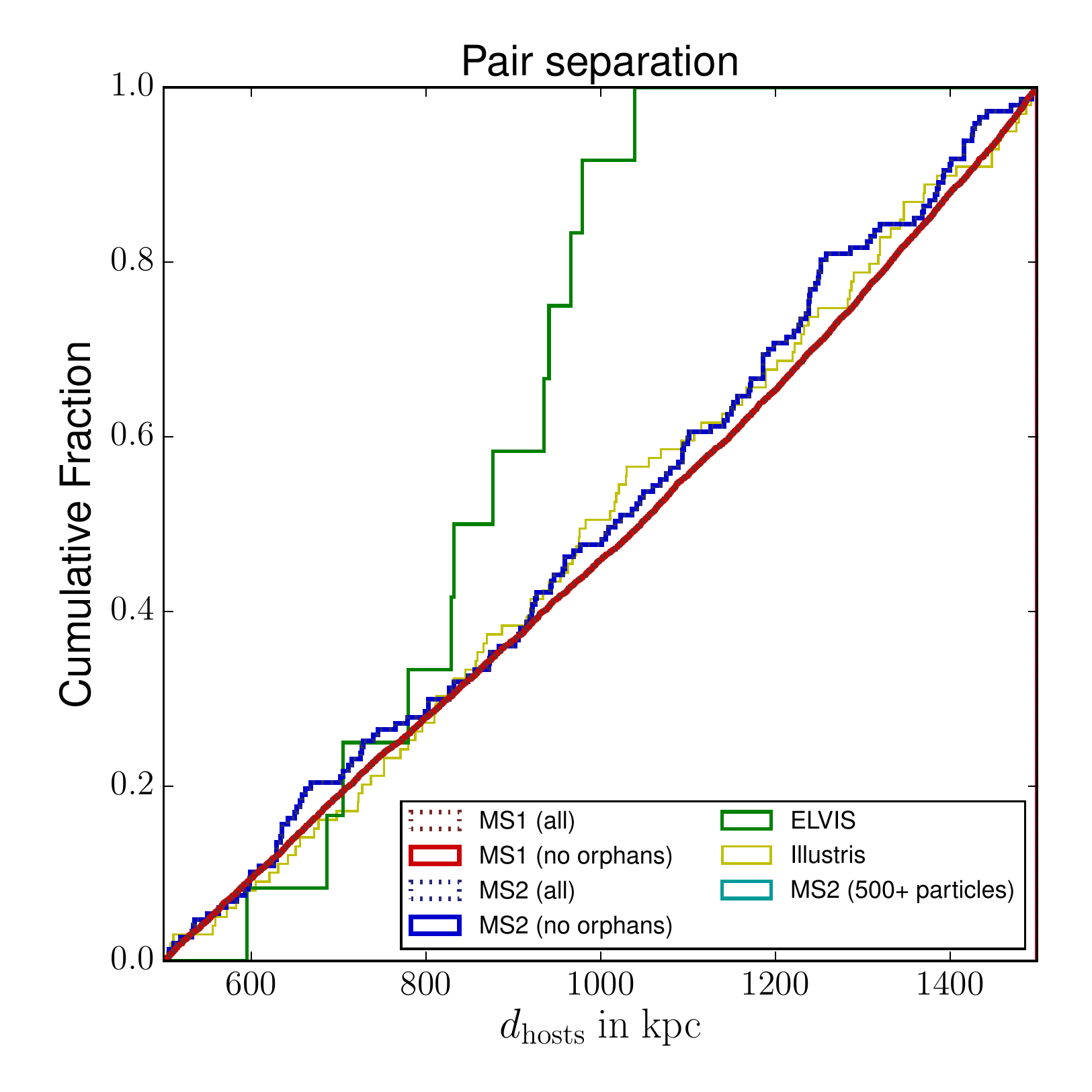}{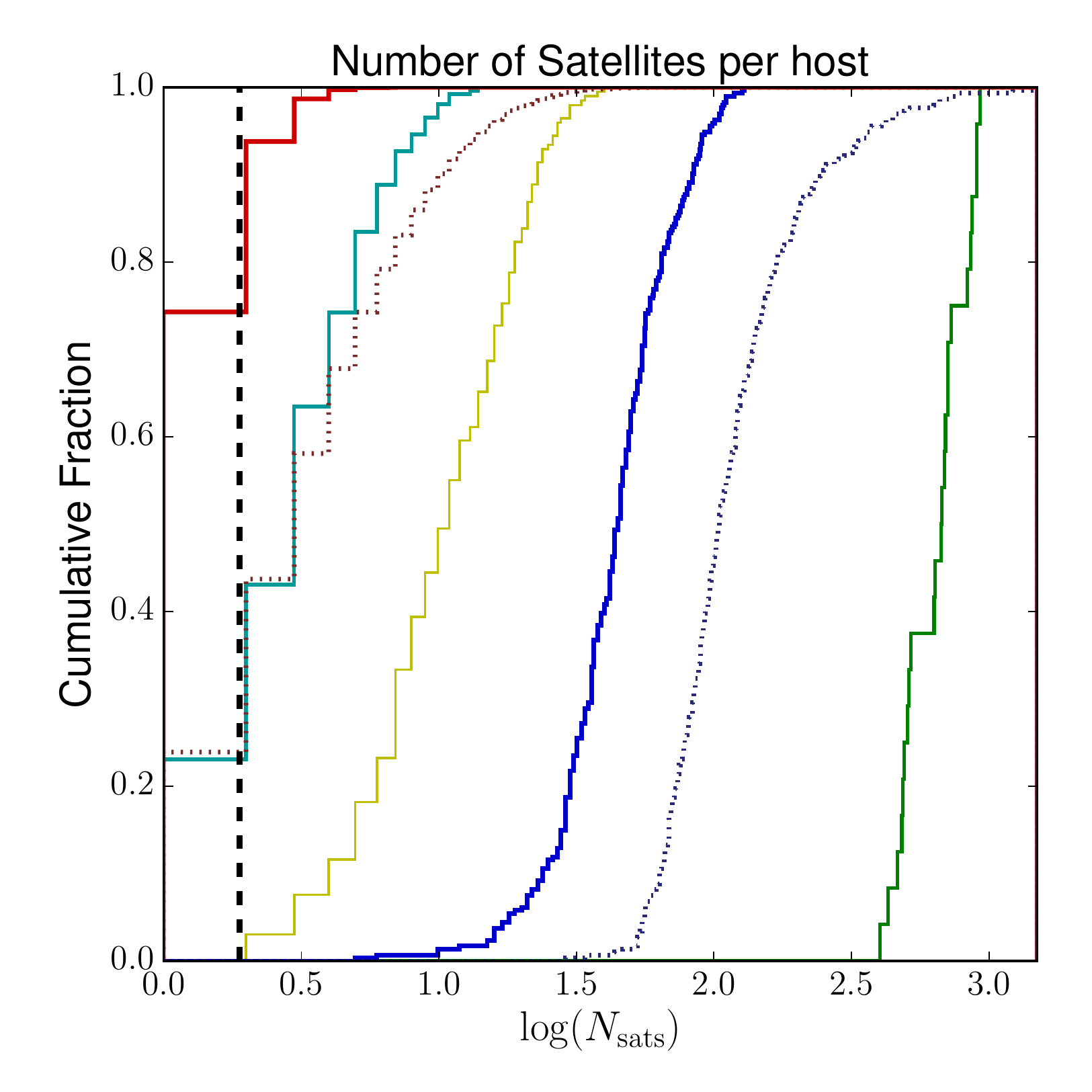}
\caption{Cumulative distributions of the separations (left) and number of satellites (right) for the host galaxy pairs selected from the simulations. The ELVIS simulations were selected to have separations close to that of the Milky Way and M31. The black dashed vertical line corresponds to the average number of satellites per host in the observed sample.
\label{fig:pairprop}}
\end{figure*}

The study of \citetalias{2016ApJ...830..121L} is based on a fiducial sample of 12,210 galaxy pairs with 46,043 potential satellite galaxies selected from the SDSS. An optimal comparison requires that the considered cosmological simulations contain similar numbers of host galaxy pairs and resolved satellite galaxies. 

We chose the Millennium-I \citep{2005Natur.435..629S} and Millennium-II \citep{2009MNRAS.398.1150B} simulations for our main comparison, using the redshift zero galaxy catalogues made publicly available in the Millennium Database \citep{2006astro.ph..8019L}. These catalogues were created by populating the dissipationless, dark-matter-only (DMO) simulations with galaxies via semi-analytic galaxy formation models \citep{2013MNRAS.428.1351G}, after the simulations were scaled to WMAP7 cosmological parameters \citep{2011ApJS..192...14J}.

Millennium-I covers a volume of $522\,h^{-1}\,\mathrm{Mpc}$\ side length, with a mass resolution of $1.06 \times 10^9\,\mathrm{M}_{\sun}$\ per particle. It contains a large number of host galaxies and pairs, and has been used previously to study galaxy pairs \citep[e.g.][]{2008MNRAS.384.1459L}. 
Millennium-II covers a considerably smaller volume ($104\,h^{-1}\,\mathrm{Mpc}$\ side length) at increased resolution (particle mass $8.5 \times 10^6\,\mathrm{M}_{\sun}$). It thus resolves more satellites per host. By comparing the two simulations we can test whether the lopsidedness of satellites depends on their mass.
We expand this further by also analysing the simulations in the Exploring the Local Volume in Simulations (ELVIS) suite, which have even higher resolution (particle mass $1.9 \times 10^5\,\mathrm{M}_{\sun}$). These DMO zoom-simulations focus on 12 pairs of main halos selected to resemble the Local Group in masses, separations, and relative velocities, and are also based on WMAP7 cosmology.
Finally, we include the hydrodynamic Illustris-1 simulation \citep{2014Natur.509..177V, 2015A&C....13...12N} to test whether baryonic effects such as gas dynamics and feedback processes affect the results. Specifically, Illustris models gas cooling, star formation, stellar evolution (including supernova explosions and metal enrichment of the surrounding gas), stellar wind feedback (with velocity scaled to the local dark matter density), as well as feedback from super-massive black holes. Illustris-1 offers a large enough volume ($75\,h^{-1}\,\mathrm{Mpc}$\ side length) to identify several pairs of host galaxies in the desired magnitude range, while also resolving a number of satellites around them (dark matter particle mass $6.3 \times 10^6\,\mathrm{M}_{\sun}$).

We select isolated host galaxy pairs according to the following criteria motivated by those in \citetalias{2016ApJ...830..121L}. All potential hosts with an $r$-band magnitude in the range ($-23.5 \leq M_r \leq -21.5$) are considered. We identify potential pairs by selecting those potential hosts which have exactly one other potential host within a 3D distance of ($0.5 \leq d_{\mathrm{hosts}} \leq 1.5\,\mathrm{Mpc}$). The potential pairs are furthermore required to be isolated by rejecting those which have a third galaxy with $M_r$\ brighter than 1 magnitude fainter than the less-luminous partner within 1.5\,Mpc of the pair's midpoint. Finally, only isolated pairs with a magnitude difference $\leq1$\ are retained. This is our sample of paired primaries. The ELVIS project was designed to simulated pairs of dark matter halos that mimic the Milky Way / M31 system in mass and separation.  All 12 pairs in the suite are used. 

As satellites, we identify all non-primary galaxies in the Millennium-I, Millenium-II, Illustris-1, and all dark matter (sub-)halos in the ELVIS halo catalogues, if they are within a three-dimensional distance of 350\,kpc from the closest host (for the 3D analysis in Sect. \ref{sec:3Danalysis}), or within a projected distance of 250\,kpc from their host but within the $\pm 1.5\,\mathrm{Mpc}$\ box around the primary pair (in the 2D analysis in Sect. \ref{sec:2Danalysis}). This includes galaxies that lie beyond the virial radius of their host and would thus not qualify as a satellite in the strictest sense. However, we aim to mimic an observational approach which does not have direct access to properties of the host's dark matter halo. All satellite galaxies are selected irrespective of their mass or luminosity, which is necessary to obtain sufficiently large samples of satellites that allow statistically meaningful conclusions, and effectively tests whether an anisotropic signal is present for satellites of different masses.

The numbers of total potential hosts ($N_{\mathrm{hosts}}$), of isolated primary pairs ($N_{\mathrm{pairs}}$), and of satellites ($N_{\mathrm{sat}}$) for the simulations are compiled in Table \ref{tab:table}. Figure \ref{fig:pairprop} summarizes the pair separations, as well as the distribution of the number of satellites per host which highlights the substantially different resolutions.

Note that the \citetalias{2016ApJ...830..121L} sample from the SDSS have on average ~1.9 satellite candidates per host, as indicated by the vertical dashed line in the right panel of Figure \ref{fig:pairprop}.  The intrinsic count in the data is likely smaller than this since many of these potential satellites are likely foreground or background galaxies. The intrinsic ratio of satellite galaxies per host in the data is therefore best matched to the resolution of the  Millennium-I catalogs. The Illustris, Millennium-II, and ELVIS simulations, on the other hand, are resolving satellite galaxies that are much less massive than can be seen in the SDSS data.  As we discuss below, lopsidedness appears to be only mildly sensitive to the mass ratio of the satellite to host, with more massive satellites showing some preference for more lopsidedness.

In the Millennium Simulations, galaxies are assigned one of three types representing the relation to their dark matter halo: 
\begin{enumerate}
\item Type 0 are central galaxies occupying the main halo of a Friend-of-Friends group;
\item Type 1 are galaxies in subhalos associated with a Friend-of-Friends group;
\item Type 2 (``orphan'') galaxies lost their subhalos to tidal disruption after accretion on a more massive halo. Their positions are assigned to the most bound particle of their subhalo before disruption.
\end{enumerate}
To select galaxies as closely as they would be selected in an observational study, we do not differentiate between type 0 and type 1 for the host galaxies. Thus, a host can be the satellite of its partner if the other selection criteria, in particular the similarity in luminosity and their mutual distance, are fulfilled\footnote{If only type 0 hosts are used, the sample size of primary pairs is reduced by about a factor of 2, which substantially affects the statistical significance of the found satellite excess in the direction of the partner galaxy. Furthermore, while the satellite excess is in general still present, its strength can be reduced by about a factor of two compared to the larger sample including hosts of type 1.}.

Since the orphan satellite positions are tied to single dark matter particles we consider their positions unreliable and therefore focus on type 0 and 1 satellites, but discuss results including type 2 galaxies for completeness.

\section{3D analysis} \label{sec:3Danalysis}

\begin{figure*}
\plottwo{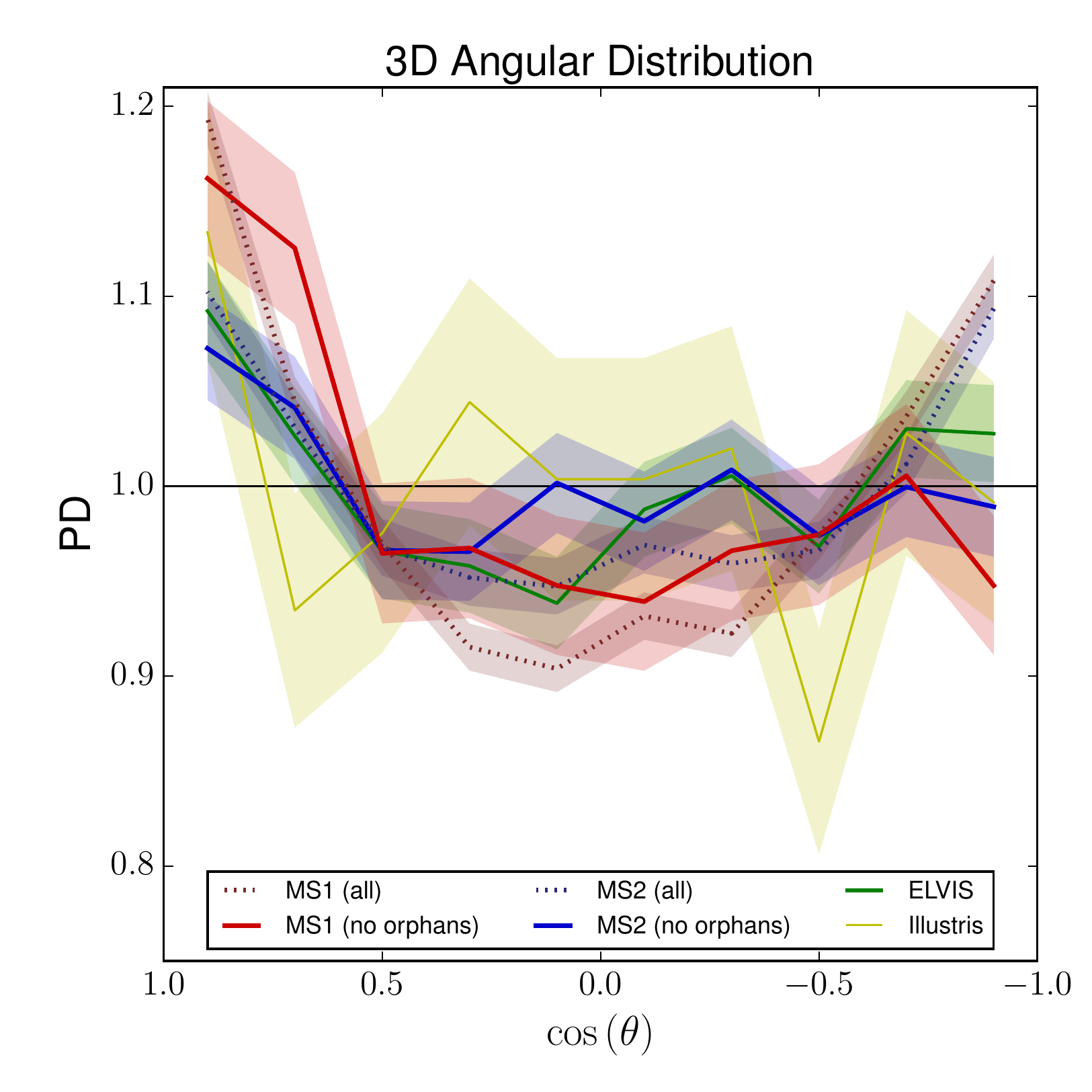}{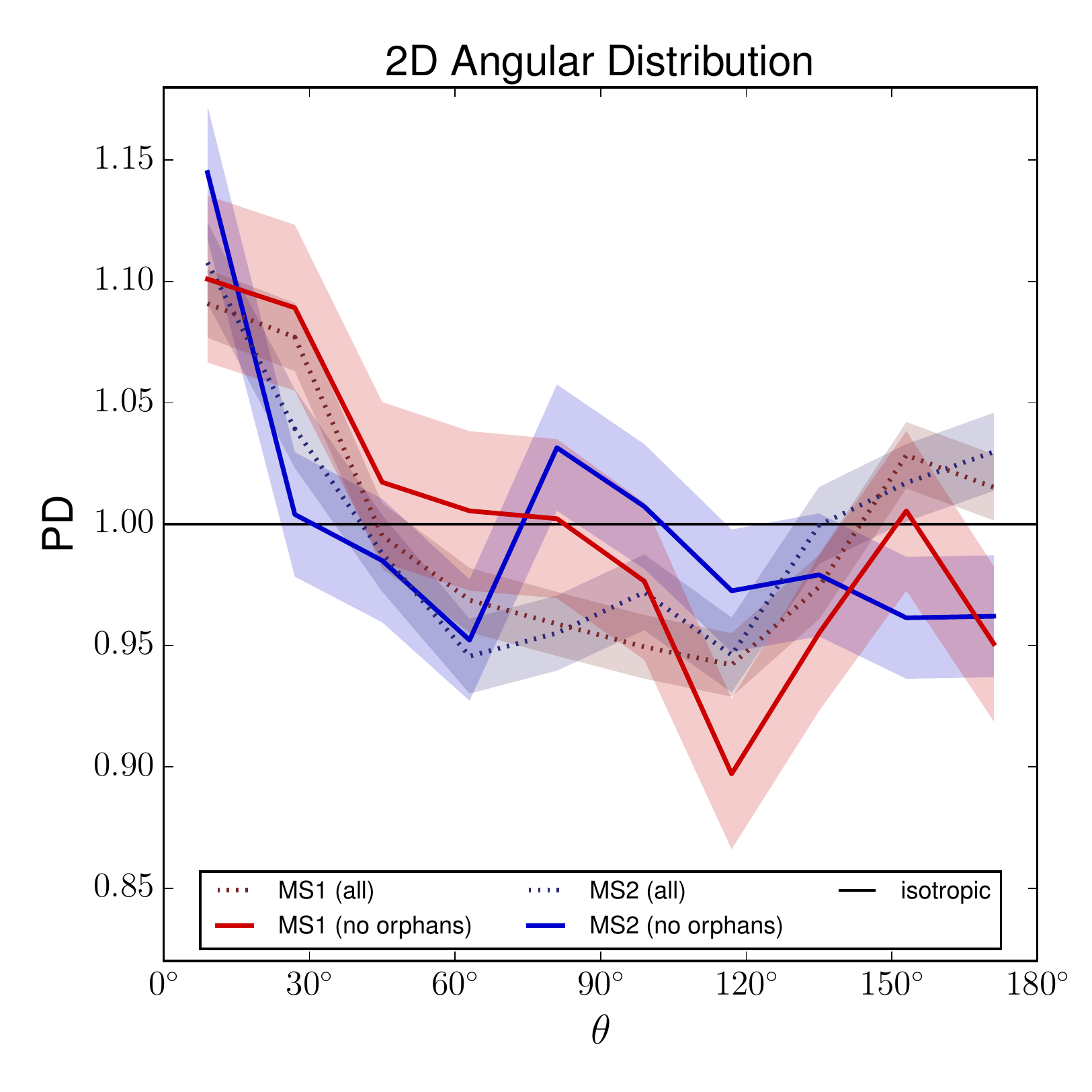}
\caption{Probability distributions of the angles between the satellite positions relative to their hosts and the lines connecting the two primaries. The \textit{left panel} uses the full three-dimensional positions, the \textit{right panel} measures the angle in projection. The shaded regions indicate the $\pm1\sigma$\ uncertainty, the black lines correspond to uniform distributions. All simulations show their largest excess in the bin closest to the direction facing the partner ($\cos(\theta) = 1$\ and $\theta = 0^\circ$). The results for the different simulations are largely consistent with each other (the shaded regions overlap), with the exception of MS1 and MS2 if unresolved ``orphan'' satellites are included (dotted lines), in which case there is a second excess of satellites in the direction facing away from the other primary.
\label{fig:angdistr}}
\end{figure*}

\begin{figure*}
\plottwo{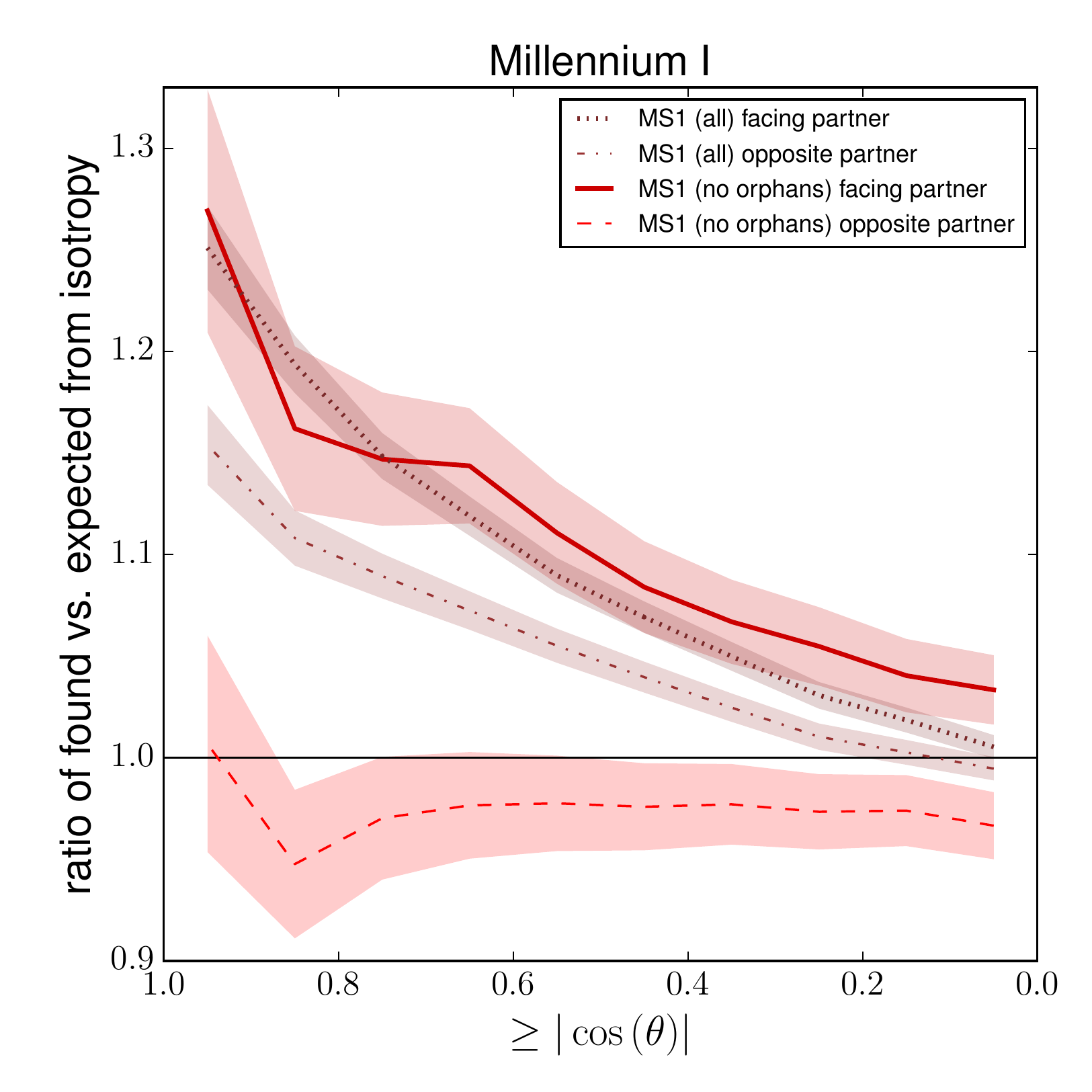}{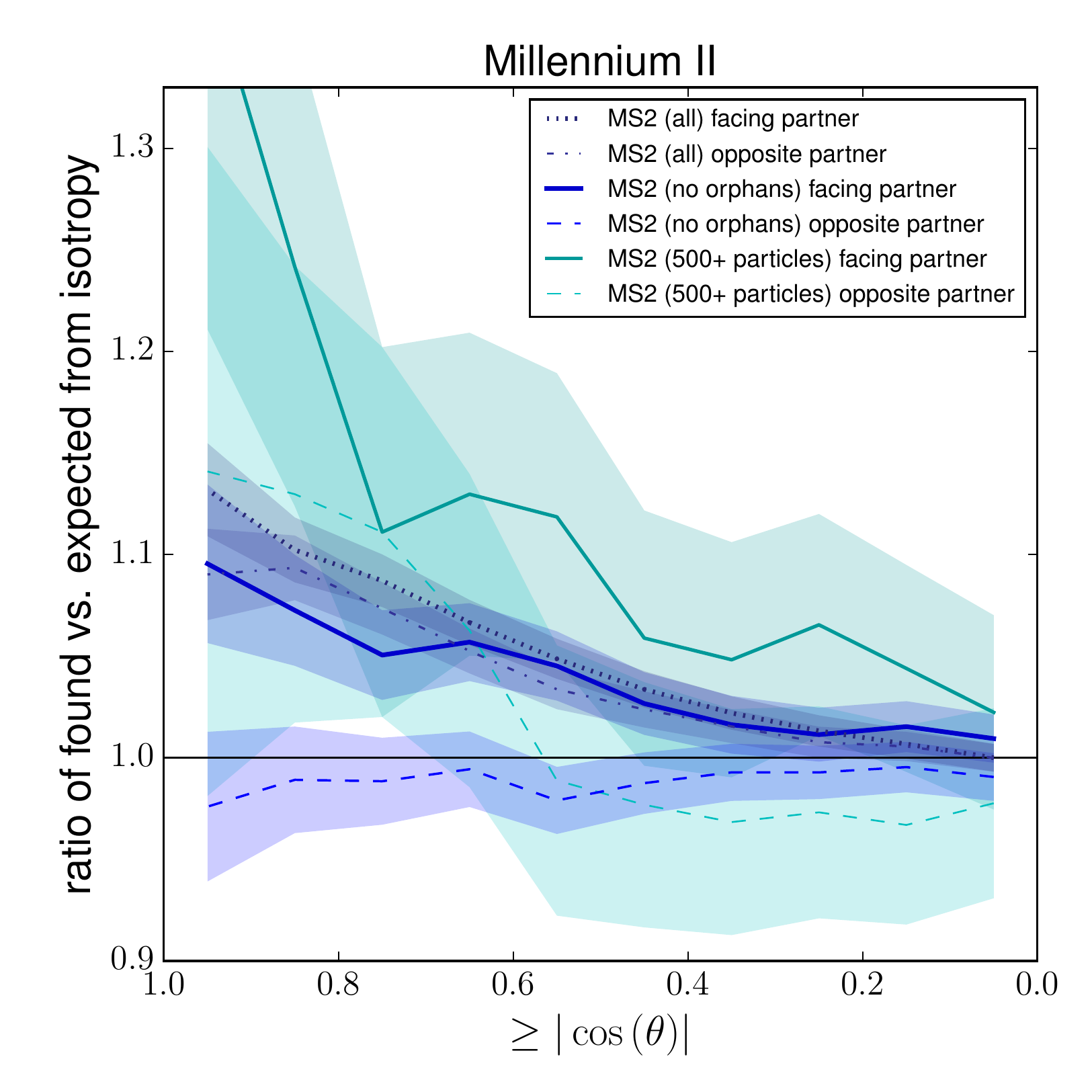}
\plottwo{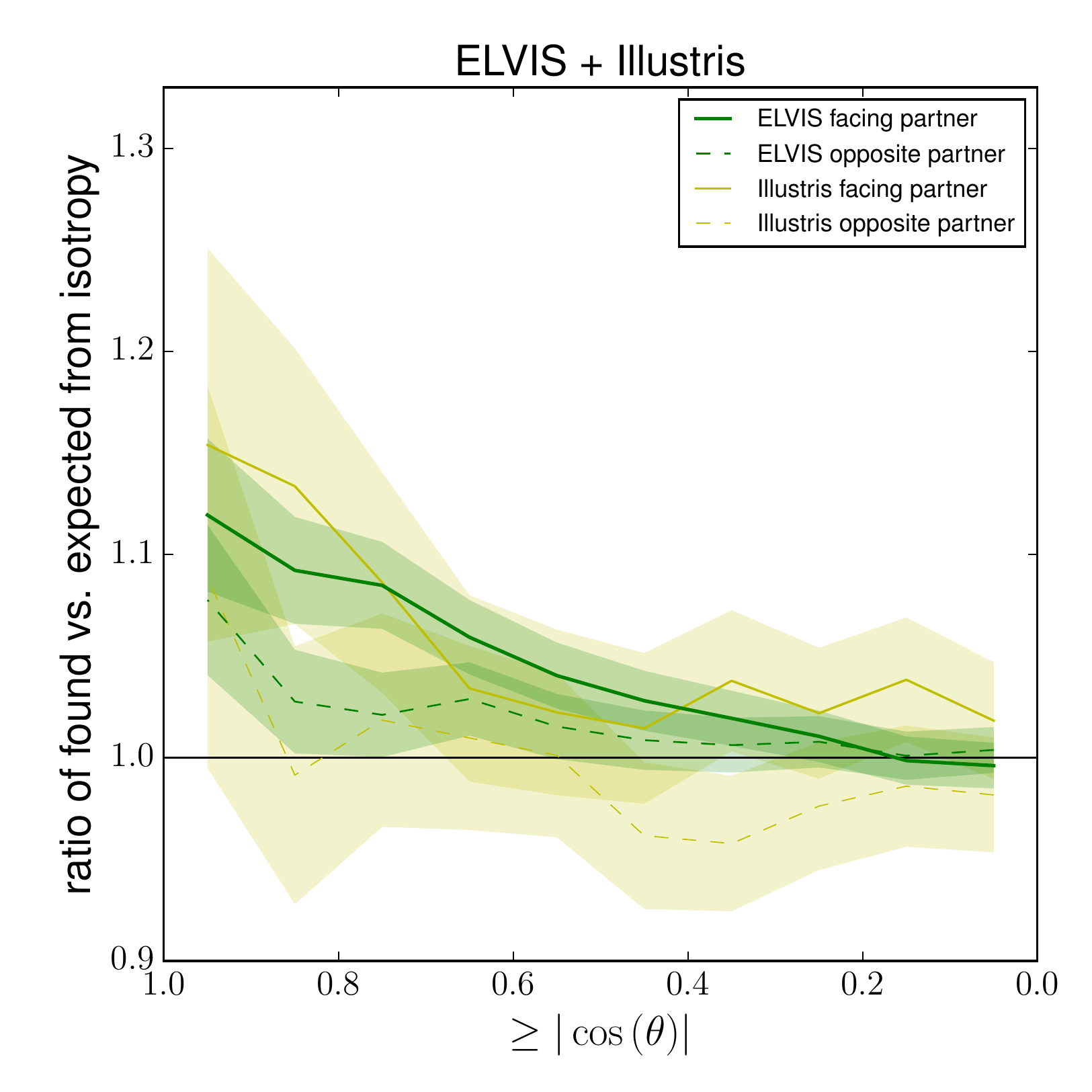}{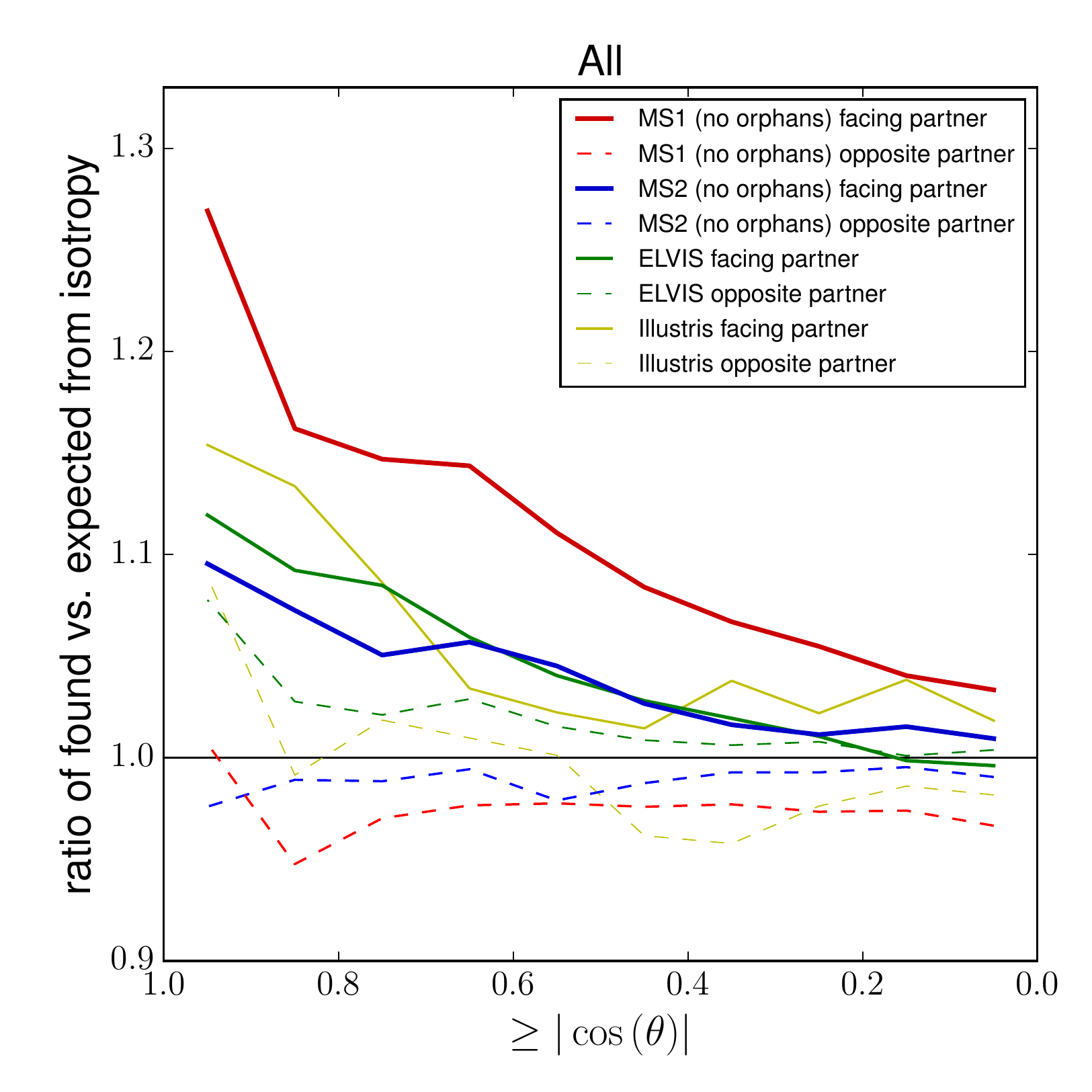}
\caption{The abundance of satellite galaxies within cones with opening angle $\cos(\theta)$\ towards the other primary (solid lines) and away from it (dashed lines), relative to that expected for a uniform distribution (black line). The shaded areas give the $\pm1\sigma$\ spread expected for random distributions of the same number. Panels (a) to (c) show the results for the Millennium-I, Millennium-II, as well as ELVIS and Illustris-1 simulations. Panel (d) combines them for easier comparison. 
 \label{fig:3D}}
\end{figure*}

Even though \citetalias{2016ApJ...830..121L} only worked with projected positions, we first investigate the distribution of satellites in three dimensions. This constitutes a more direct comparison between the different simulations and avoids effects based on the choice of projection axes, which could particularly affect simulations having only a small samples of pairs. For better comparability with the observational findings we analyse projected satellite systems in Sect. \ref{sec:2Danalysis}. 

For a given pair of host galaxies, all satellites within a maximum radius of $r_\mathrm{max} = 350\,\mathrm{kpc}$\ or $r_\mathrm{max} = 0.5\,d_{\mathrm{hosts}}$\ (whichever is smaller) are selected\footnote{This is larger than the fiducial radius of 250\,kpc used by \citetalias{2016ApJ...830..121L}. However, they worked in projection and thus include satellites at larger 3D distances from the host. Selecting simulated satellites from a smaller volume results in too small samples to provide sufficient statistics. We follow their choice when we analyse the projected distribution in Sect. \ref{sec:2Danalysis}.}, and the angle $\theta$\ between their position with respect to the closest host and the line connecting the two hosts is measured. The distributions for the different simulations are shown in the left panel of Fig. \ref{fig:angdistr}. The number count per bin, normalized to that expected for a uniform distribution, is plotted against $\cos(\theta)$. The direction towards the partner galaxy corresponds to $\cos(\theta) = 1.0$, the opposite direction is $\cos(\theta) = -1.0$. The shaded regions indicate the $\pm1\sigma$\ scatter from poisson noise.
 
All simulations show an excess of satellites around $\cos(\theta) = 1.0$, with a maximum excess of 7 to 20\% in the bin closest to the other primary. We perform Kolmogorov-Smirnov (KS) tests to determine whether the distributions differ significantly from a uniform distribution. The resulting probabilities $p_{\mathrm{KS}}$\ of drawing the observed distributions given an isotropic distribution as the null hypothesis are compiled in column (6) of Table \ref{tab:table}. All simulations are inconsistent with this hypothesis at a $>95\%$ confidence level, except for Illustris. However, two-sample KS-tests comparing Illustris with each of the other simulations cannot reject the hypothesis that they are drawn from the same distribution. The Illustris distribution is thus also consistent with the other simulations, indicating that it lacks in informative power due to the substantially lower total number of satellites.

The \textit{overall} lopsidedness of the satellite systems can be expressed as the fraction of satellites within the hemisphere facing the other primary, $f_{\mathrm{lopsided}}$. In most simulations (except ELVIS) there is a weak ($\sim 1 \%$) preference for satellites to lie on the side facing the partner. The degree of this average lopsidedness is considerably lower than that found for the M31 satellite system, where $f_{\mathrm{lopsided}}^{\mathrm{M31}} \approx 80\%$ \citep{2013ApJ...766..120C}.

Fig. \ref{fig:3D} shows the abundance of satellites in the two hemispheres. Plotted as solid lines are the numbers of satellites, normalized to those expected for a uniform distribution, within cones that originate from the host galaxy and face towards the other primary with opening angles of $\theta$. The dashed lines correspond to cones facing in the opposite direction, away from the other primary. The shaded regions indicate the $1\sigma$\ Poisson scatter. Since the relative scatter is plotted, the width of the shaded region decreases for increasing $\theta$\ as the number of satellites included in the cones increases.

All simulations show an overabundance of satellites towards the other primary, amounting to between 10 and 27\%. The strongest overabundance is consistently found for the narrowest opening angle ($\theta \approx 25^\circ$). The most significant overabundance (or ``peak significance'': the highest significance compared to an isotropic distribution among all considered opening angles), is found for angles of $45^\circ$\ (see Table \ref{tab:table} for average\footnote{Following \citetalias{2016ApJ...830..121L} we also compile the significance averaged over all opening angles, $<\sigma>$, even though the different bins are not independent since subsequently larger opening angles contain all satellites included in the smaller ones.} and peak significances). The distributions facing the other primary are all inconsistent with a uniform distribution, while those facing away from this direction follow the black line of a uniform distribution.  

The Millennium-I simulation shows the strongest lopsidedness in its satellite distributions and the most significant deviation from uniform ($5.1\sigma$\ at $\cos(\theta) = 0.6$), the other three simulations are consistent with each other within the scatter. This points towards a stronger alignment for more massive satellites. We test this hypothesis by selecting the most massive satellites in the Millennium-II simulation (cyan lines in the top right panel of Figure \ref{fig:3D}). Halos in Millennium-I are required to consist of at least 20 particles. Accounting for the difference in resolution this corresponds to selecting satellites with at least 2500 particles in Millennium-II. Since this limit provides too few objects, we reduce the required number to 500 particles (20\% of the Millennium-I limit), which results in a similar number of satellites per host as in the Millennium-I case including orphan galaxies (see right hand panel in Figure \ref{fig:pairprop}). Despite this difference, the selected most-massive satellites in Millennium-II do indeed show a considerably stronger signal, consistent with that found for Millennium-I. However, due to the substantially reduced sample size this finding is not statistically significant. We thus tentatively conclude that there are indications for a mass-dependency of the lopsided signal, and will further investigate this possibility in Sect. \ref{sec:subsamples}.

If unresolved orphan galaxies are included for the two Millennium simulations, the overall strength of the lopsidedness in the direction of the other primary does not change, but an overabundance is now also found in the direction facing away from the other primary, in contrast to the observed situation reported by \citetalias{2016ApJ...830..121L}. This indicates that the two types of objects do not follow the same distributions. It will be interesting to investigate the origin of this marked difference in more detail, but such an investigation is beyond the scope of the current work.

\section{2D Analysis}\label{sec:2Danalysis}

\begin{figure*}
\plottwo{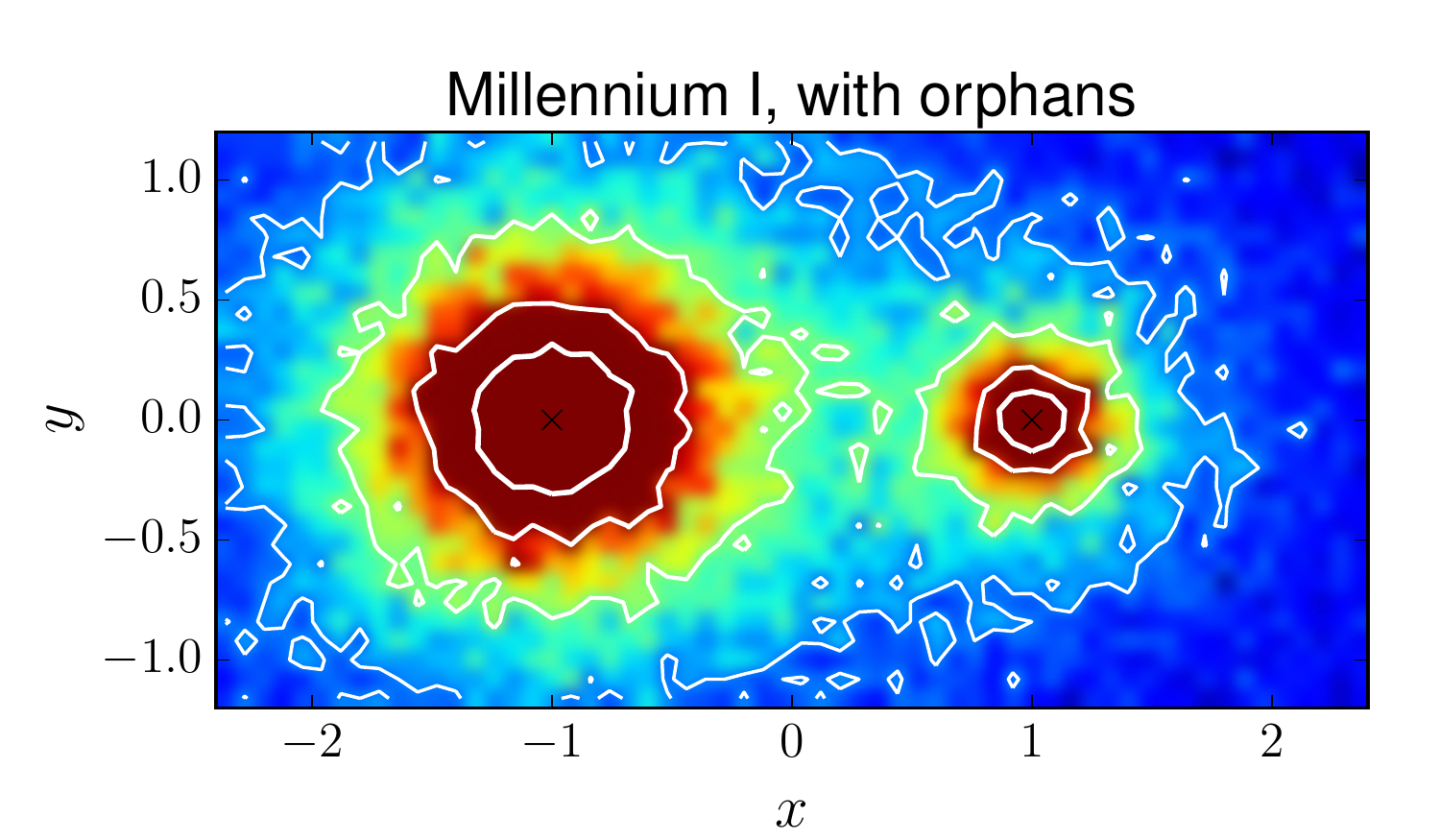}{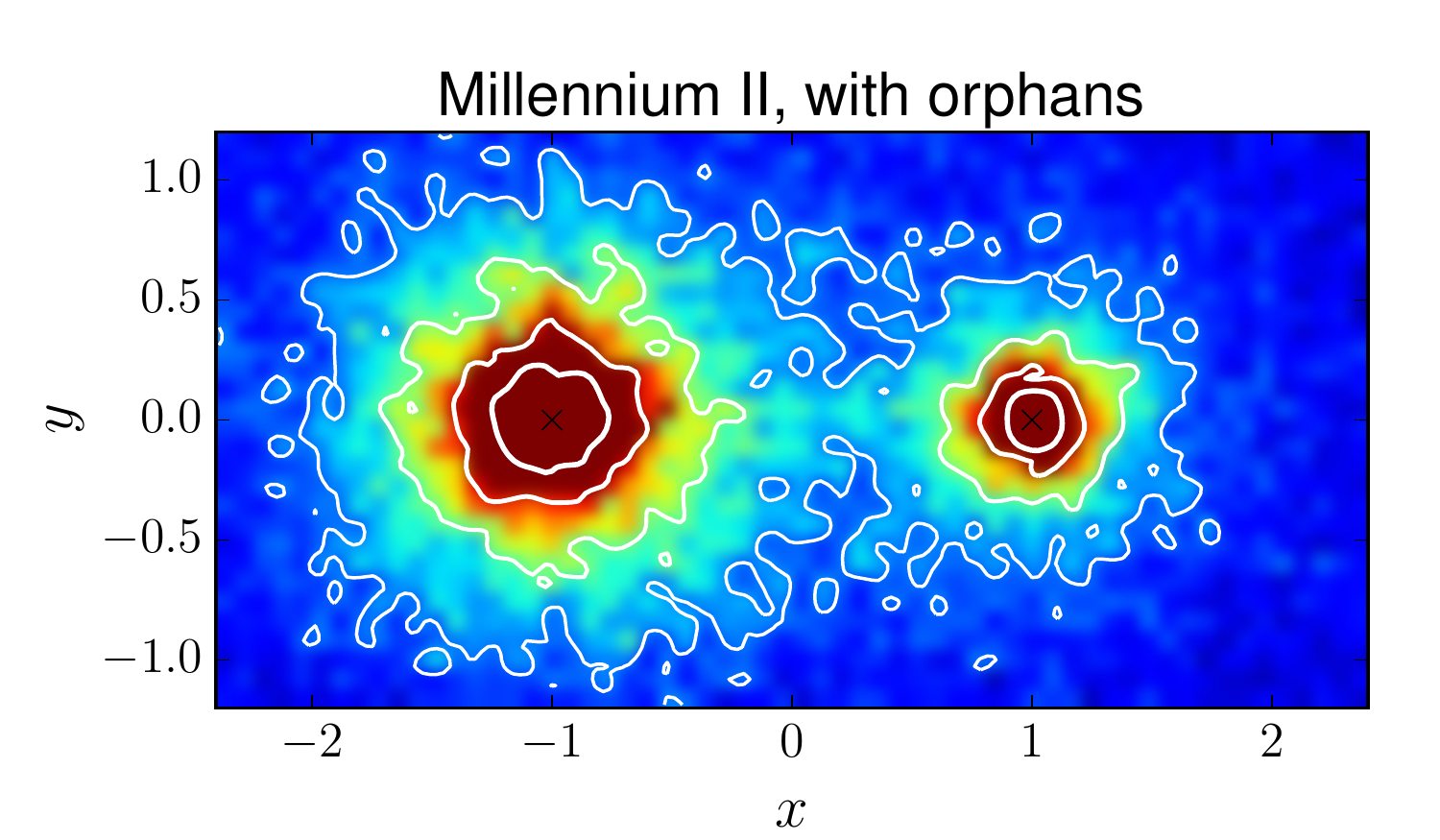}
\plottwo{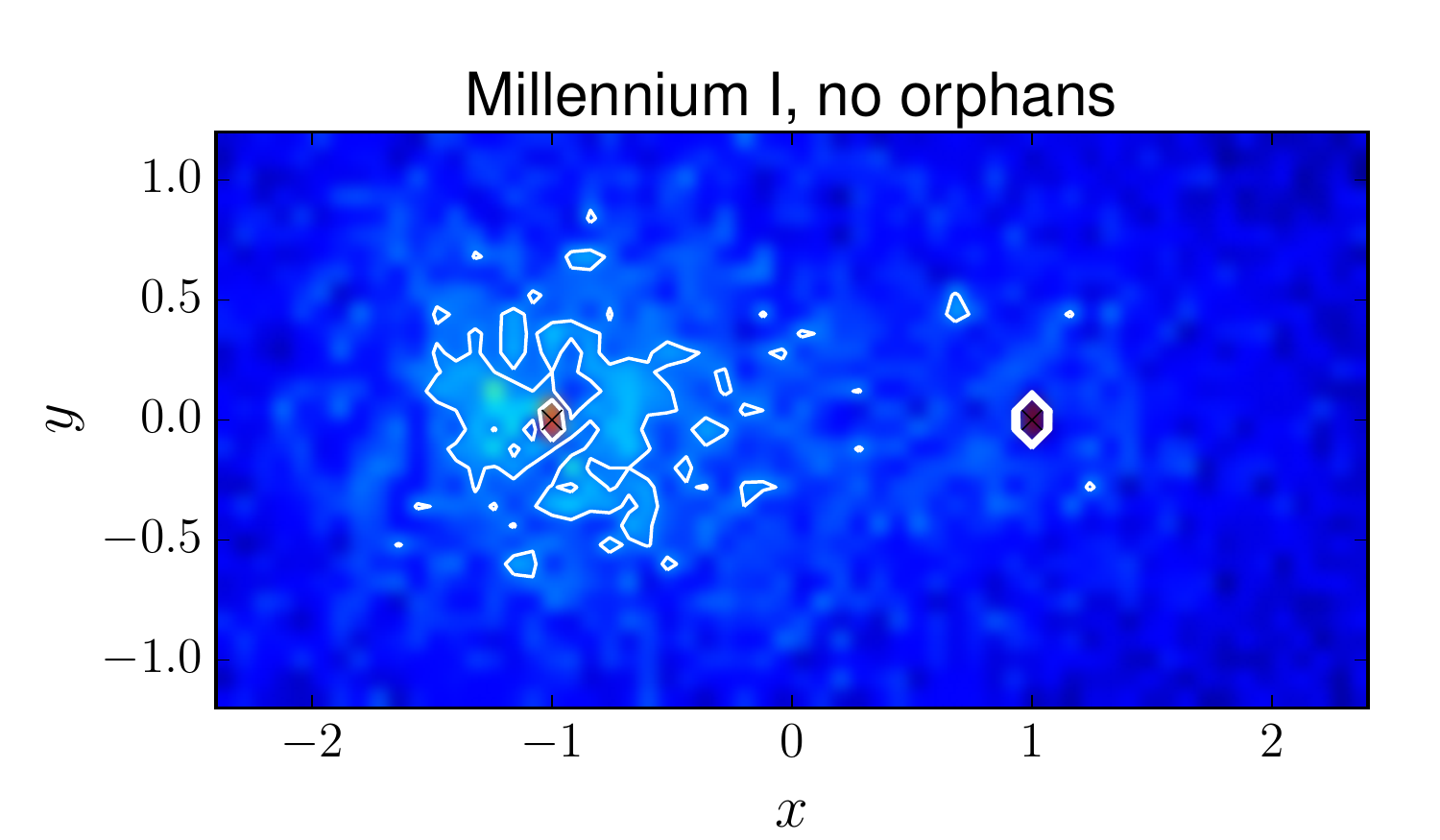}{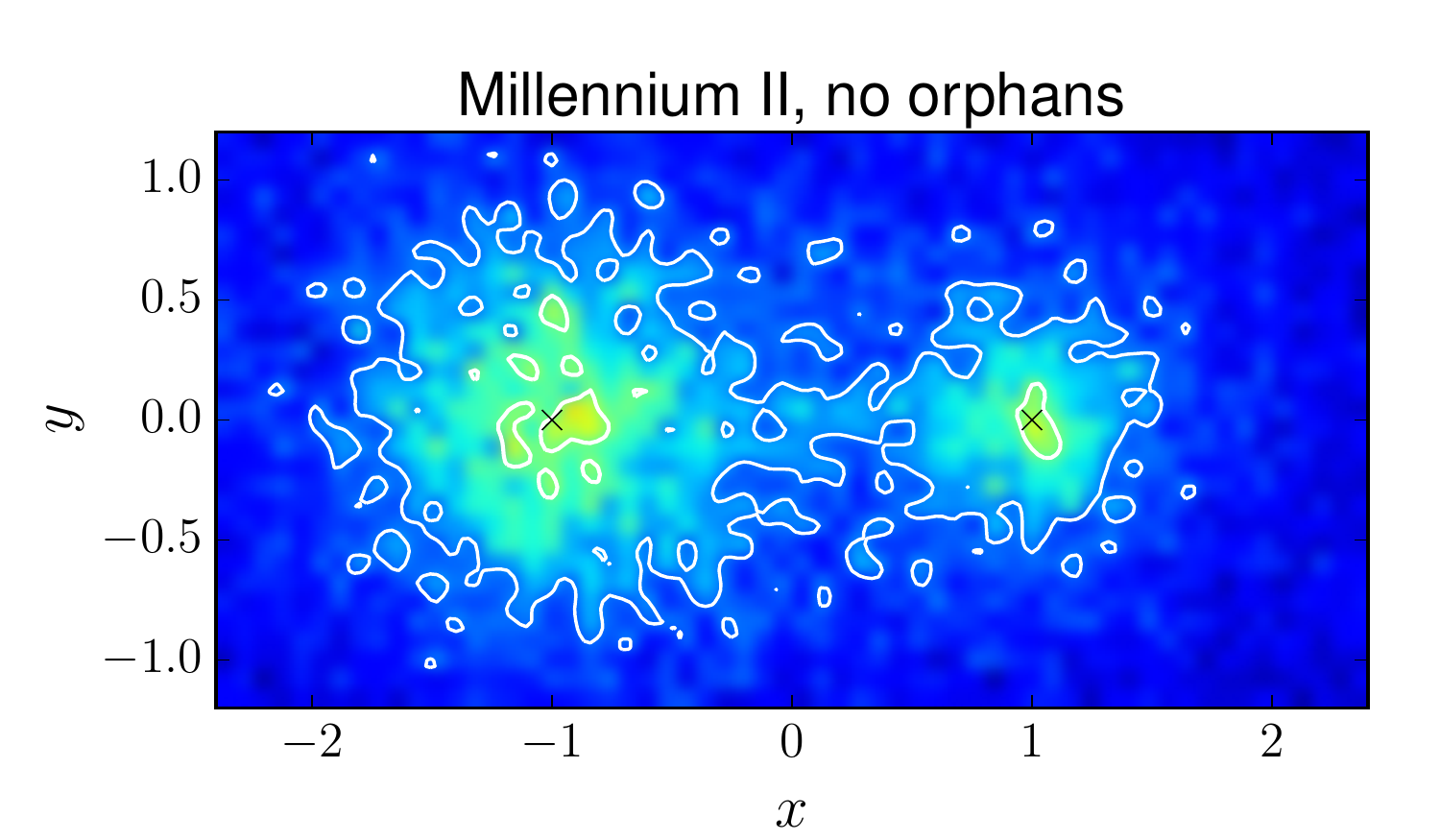}
\caption{Stacked density distributions of satellite galaxies in the projected samples. The simulated distributions are rotated and scaled such that the more luminous primary is at x = -1, while the less luminous partner is at x = 1 (black crosses). Compare to fig. 5 in \citetalias{2016ApJ...830..121L}.  The same arbitrary density scale is used in all four panels to illustrate the substantial lack of non-orphan galaxies, in particular in the Millennium-I simulation.
\label{fig:2Ddensitydistr}}
\end{figure*}

\begin{figure*}
\plottwo{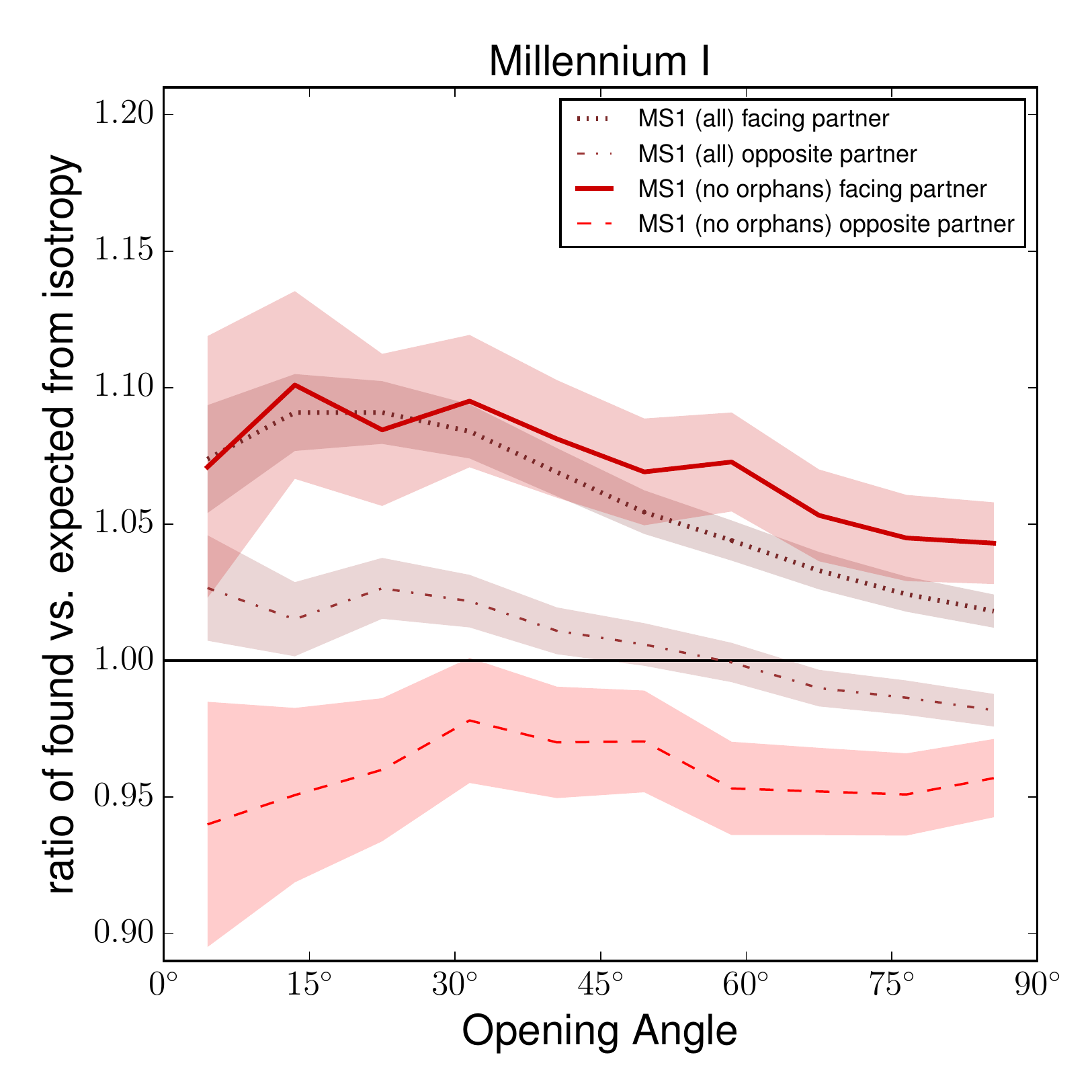}{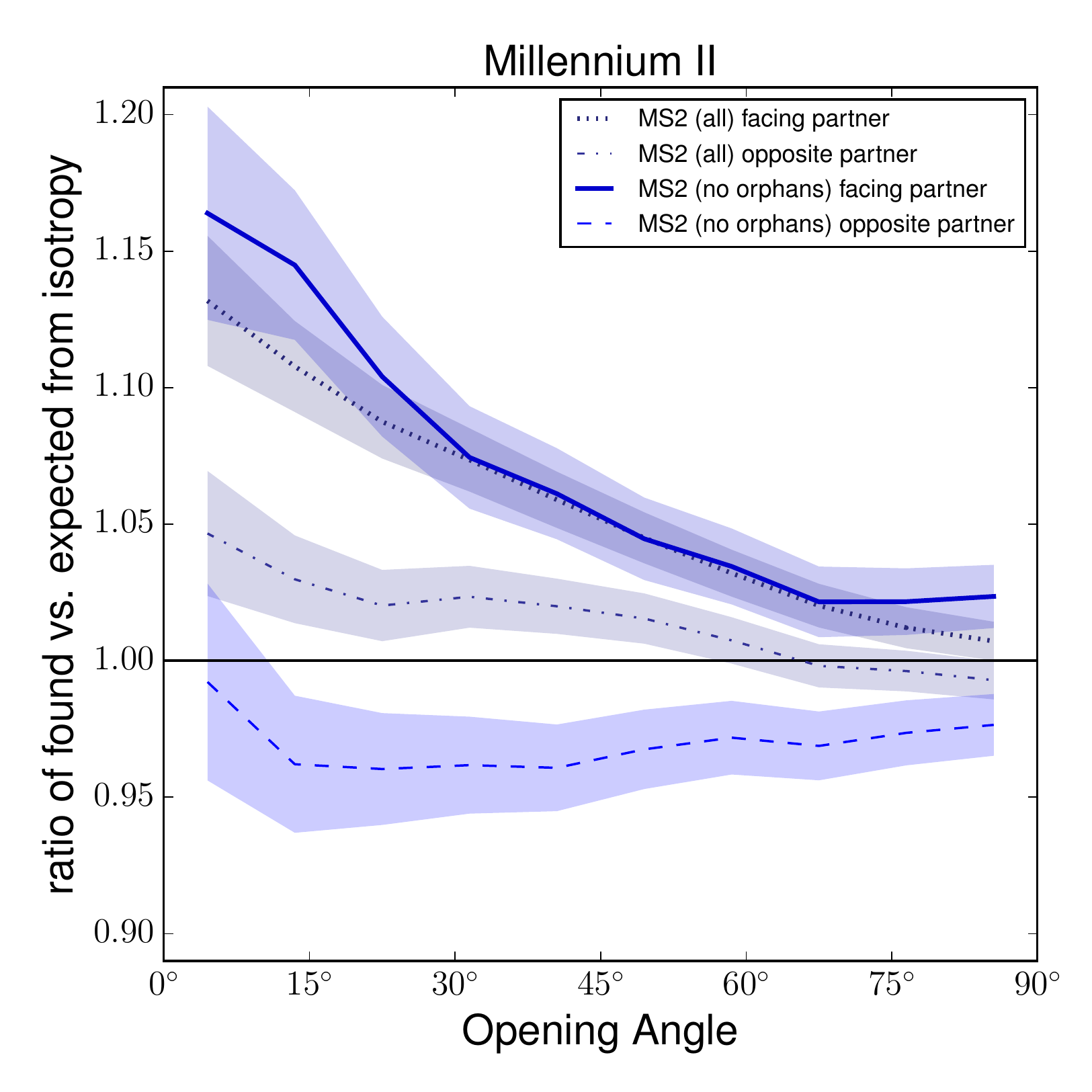}
\caption{Same as Fig. \ref{fig:3D}, but the opening angle $\theta$\ is measured in projection on a plane.  \label{fig:2D}}
\end{figure*}

\begin{figure}
\plotone{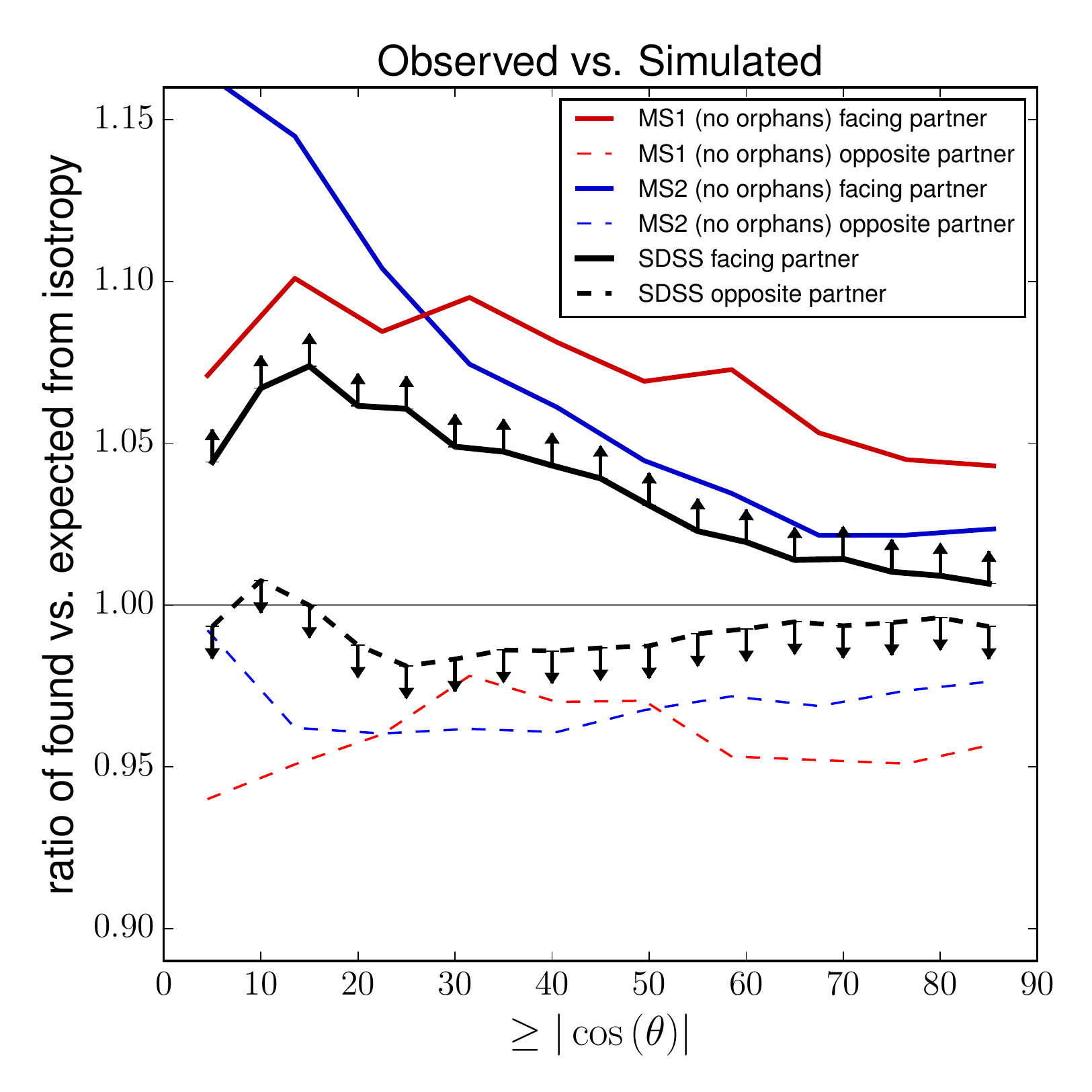}
\caption{
Comparison of projected 2D lopsidedness in observations and simulations. The signal reported by \citetalias{2016ApJ...830..121L} around observed SDSS galaxy pairs is shown in black, while those found in this work for simulated galaxies in the Millennium-I and Millennium-II simulations are plotted in red and blue, respectively. Note that the Millennium-I satellites resemble the observed sample more closely, while the Millennium-II simulations include satellites of much lower luminosity. Orphan galaxies are excluded in both cases. Because the SDSS satellites candidates are identified photometrically, the observed sample is expected to contain (evenly distributed) foreground and background contamination. The observed signal should therefore be considered as a lower/upper limit (facing towards/away from the other primary) on the true lopsided signal, as illustrated by the black arrows. 
 \label{fig:obsvstheory}}
\end{figure}

While finding a lopsidedness of simulated satellite galaxies in three dimensions is encouraging, it does not allow immediate conclusions on the compatibility with the observed situation. The SDSS observations do not provide distance measurements for the satellites that would allow one to reconstruct the full distribution of the satellite systems. \citetalias{2016ApJ...830..121L} therefore consider all possible satellites along the line-of-sight to a host. For a fair comparison, we restrict the analysis to projected 2D positions in the following. We only consider the Millennium simulations because the other simulation sets have too small sample sizes to be statistically meaningful.

For each isolated pair, we determine the projected separation $d_\mathrm{sep}$\ between the two primaries in the $x$-$y$, $y$-$z$, and $x$-$z$\ planes of the simulation coordinates, in that order. If the separation is found to be within $0.5 \leq d_\mathrm{sep} \leq 1.0\,\mathrm{Mpc}$\ (corresponding to the allowed projected separation for the SDSS primaries in \citetalias{2016ApJ...830..121L}), this projection is chosen for the further analysis. If none of the three projections results in a separation in this range, the pair is rejected. The numbers of accepted pairs are listed in column 3 of Table \ref{tab:table}. About 16\% of the previously considered pairs are excluded from this part of the analysis.

For a visual representation, Fig. \ref{fig:2Ddensitydistr} plots the density distribution of simulated galaxies around the stacked primary pairs, for both Millennium simulations. The pairs are rotated to align with the x-axis and scaled to a mutual separation of 2. An excess of galaxies in the direction between the hosts is apparent. In the cases including orphans, the distributions of satellites also appear to be slightly flattened in the y direction, which is consistent with the second excess on the side facing away from the partner primary.

For each host, all galaxies within a projected radius of $r_\mathrm{max} = 250\,\mathrm{kpc}$\ in the chosen plane, and within $\pm 1.5\,\mathrm{Mpc}$\ along a ``line-of-sight'' perpendicular to the plane (along the $z$, $x$, or $y$ axis) around the midpoint of the pair are included in the analysis. This therefore includes not only satellite galaxies, but also galaxies in the nearby fore- and background around the pairs. For simplicity, we call the objects possible satellites.

For these possible satellites, the angle in the chosen plane of their projected position relative to their host and the projection of the line connecting the two hosts is measured. The right panel of Fig. \ref{fig:angdistr} shows the resulting distribution. As in 3D, there is a clear overabundance of possible satellites towards the other primary, amounting to between 10 and 15\%. Similarly, if ``orphan'' galaxies are included the distributions again show a second over-abundance in the opposite direction, which is not present if only resolved galaxies are considered. This demonstrates that the anisotropic structure found in the 3D distribution can be recovered in a 2D analysis, strengthening the validity of the observational approach by \citetalias{2016ApJ...830..121L}. Despite the difference in resolution and covered volume between the two Millennium simulations, and the resulting difference in probed satellite masses, the angular distributions of possible satellites are largely consistent with each other within the $1\sigma$\ scatter bands. Two-sample KS-tests comparing the corresponding Millennium-I and Millennium-II distributions (either including or excluding orphans) cannot rule out that they were drawn from the same distribution. KS-tests do, however, show that the possibility that the distributions are drawn from a uniform distribution is highly unlikely ($p_{\mathrm{KS}} < 0.00031$, see column 6 in Table \ref{tab:table}). For all simulations, the overall measure of lopsidedness $f_{\mathrm{lopsided}}$\ of the satellite systems is slightly increased compared to the 3D analysis. 

Figure \ref{fig:2D} shows the abundance of possible satellites within opening angles, in analogy to Fig. \ref{fig:3D} for the 3D analysis and figure 4 in \citetalias{2016ApJ...830..121L}. Excluding orphan galaxies, the average significance of the deviation from uniform distributions are 3.2 and $3.3 \sigma$, respectively, while the bins with the most significant deviations reach 4.0 and $5.3 \sigma$. These numbers are comparable to those found by \citetalias{2016ApJ...830..121L}, despite the 3 to 5 times lower number of satellites in the simulations. This can be understood by comparing the strength of the signal: The maximum signals are 10\% for Millennium-I and 16\% for Millennium-II, compared to 8\% for the SDSS galaxies (see the comparison in Figure \ref{fig:obsvstheory}). We thus expect that the observations contain up to a factor of two more fore- and background galaxies beyond the nearby fore- and background included in our analysis. A factor of two of uniformly distributed background would reduce our signal by 50\%, and also halve the deviation of $f_\mathrm{lopsided}$\ from 0.5, bringing both into agreement with the SDSS signal. The analysis in \citetalias{2016ApJ...830..121L} of subsets of possible satellites selected to be close to their host galaxy pair in photometric redshift (see their figure 8) also indicates an agreement between simulations and observations. The most restrictive subset -- satellites with photometric redshift $\leq0.5\sigma$ off their host's spectroscopic redshift -- increases the observed signal to $\approx14\%$. This is in full agreement with the signal strengths found for the Millennium simulations.

The difference between the distributions with and without orphan galaxies is somewhat reduced in the 2D compared to the 3D analysis. This is to be expected if the lopsided signal extends beyond the immediate vicinity of the host galaxy because disrupted orphan galaxies are predominantly found closer to their host than non-disrupted galaxies, they are thus more (radially) concentrated on the hosts. This interpretation is supported by the slightly larger fraction of non-orphan galaxies among $N_{\mathrm{sat}}$\ in the 2D analysis compared to the 3D case.

\section{Subsamples}\label{sec:subsamples}

\begin{figure*}
\gridline{\fig{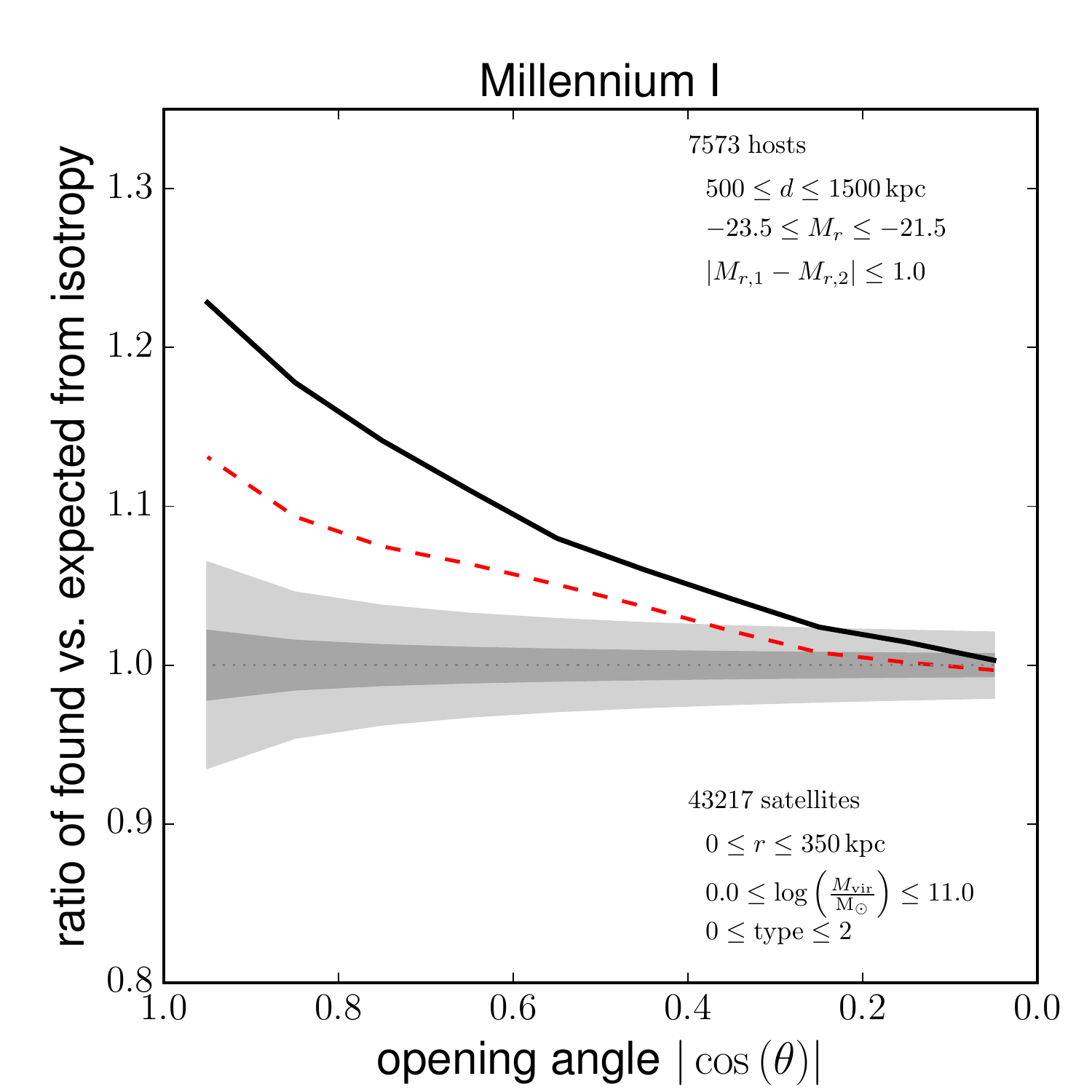}{0.25\textwidth}{(a)}
          \fig{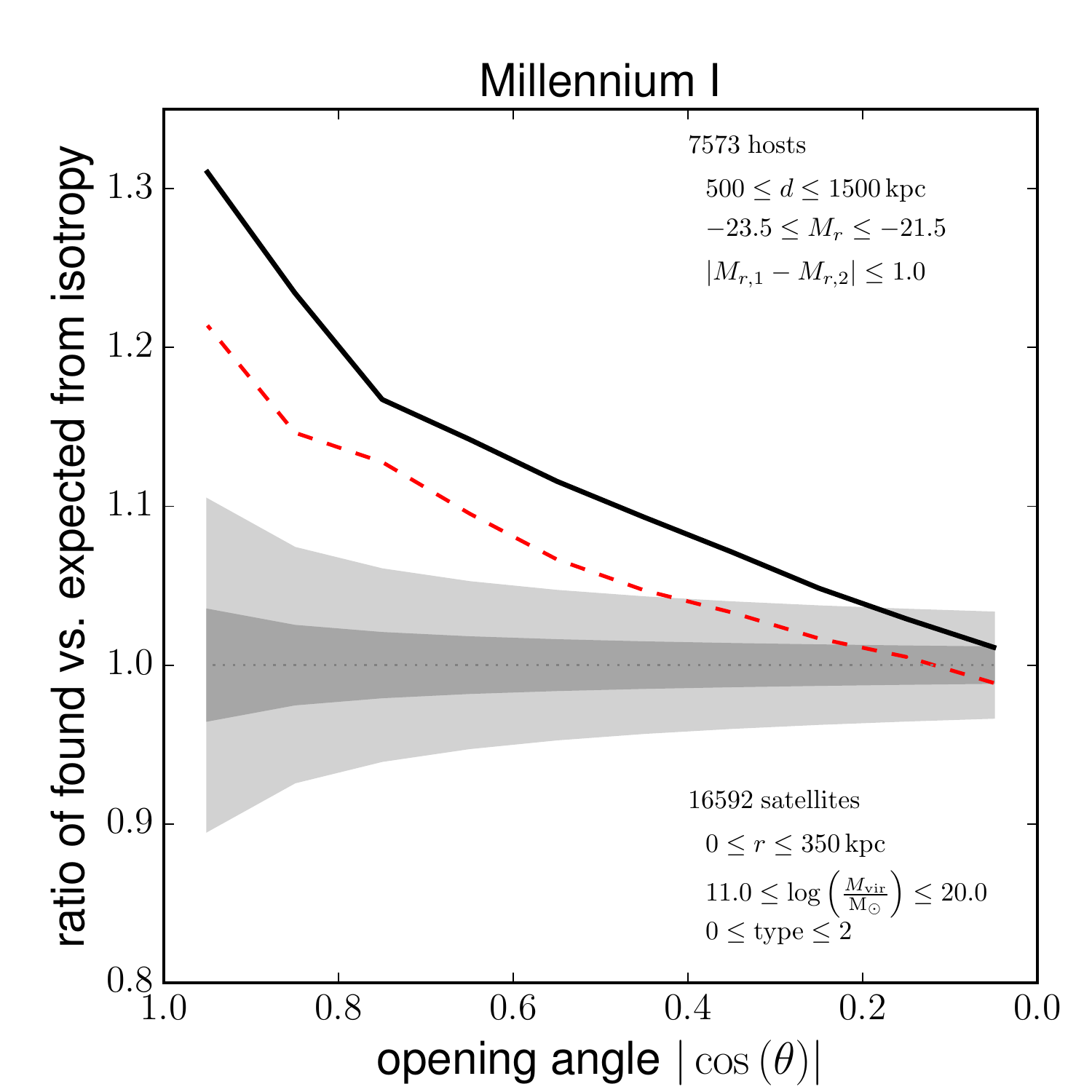}{0.25\textwidth}{(b)}
          \fig{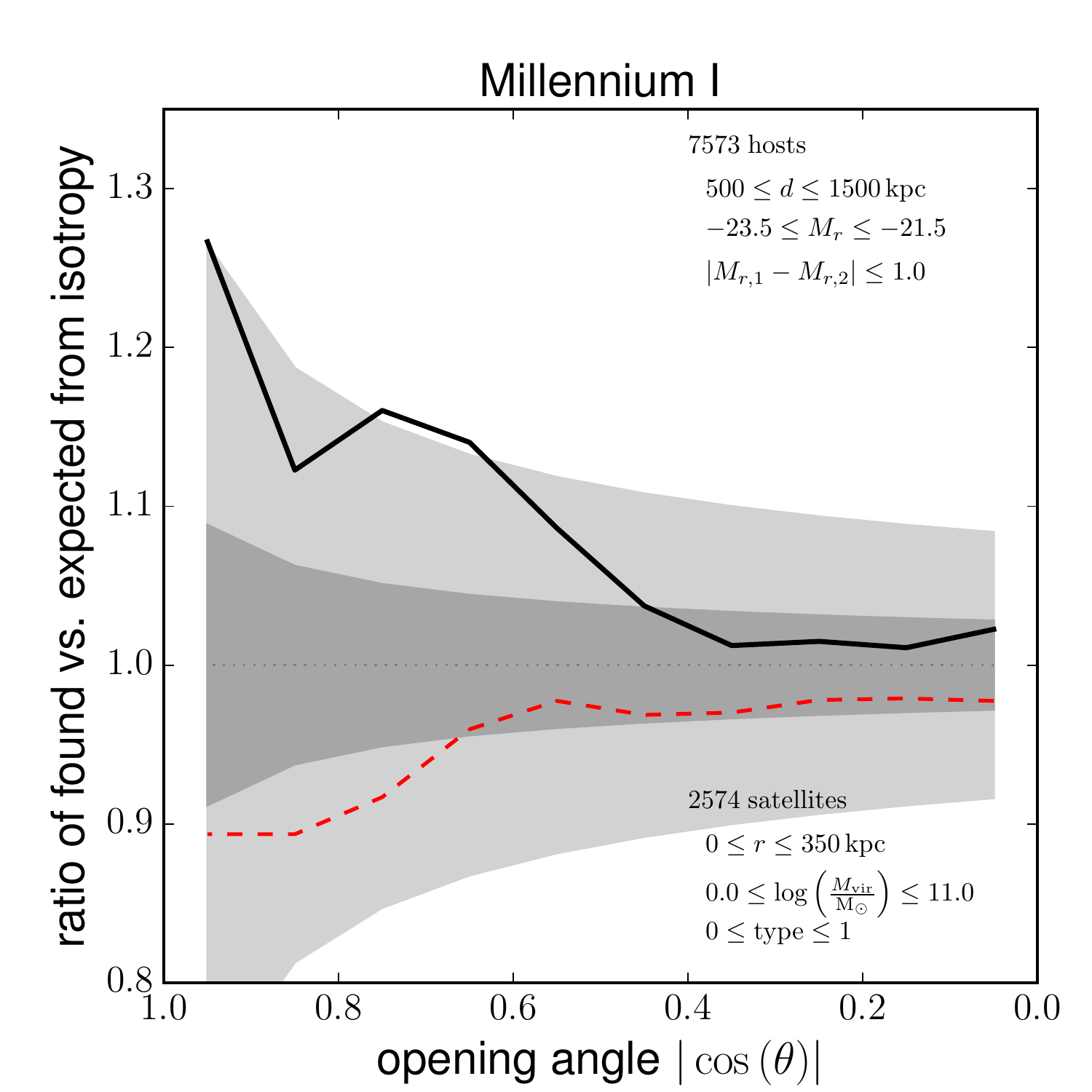}{0.25\textwidth}{(c)}
          \fig{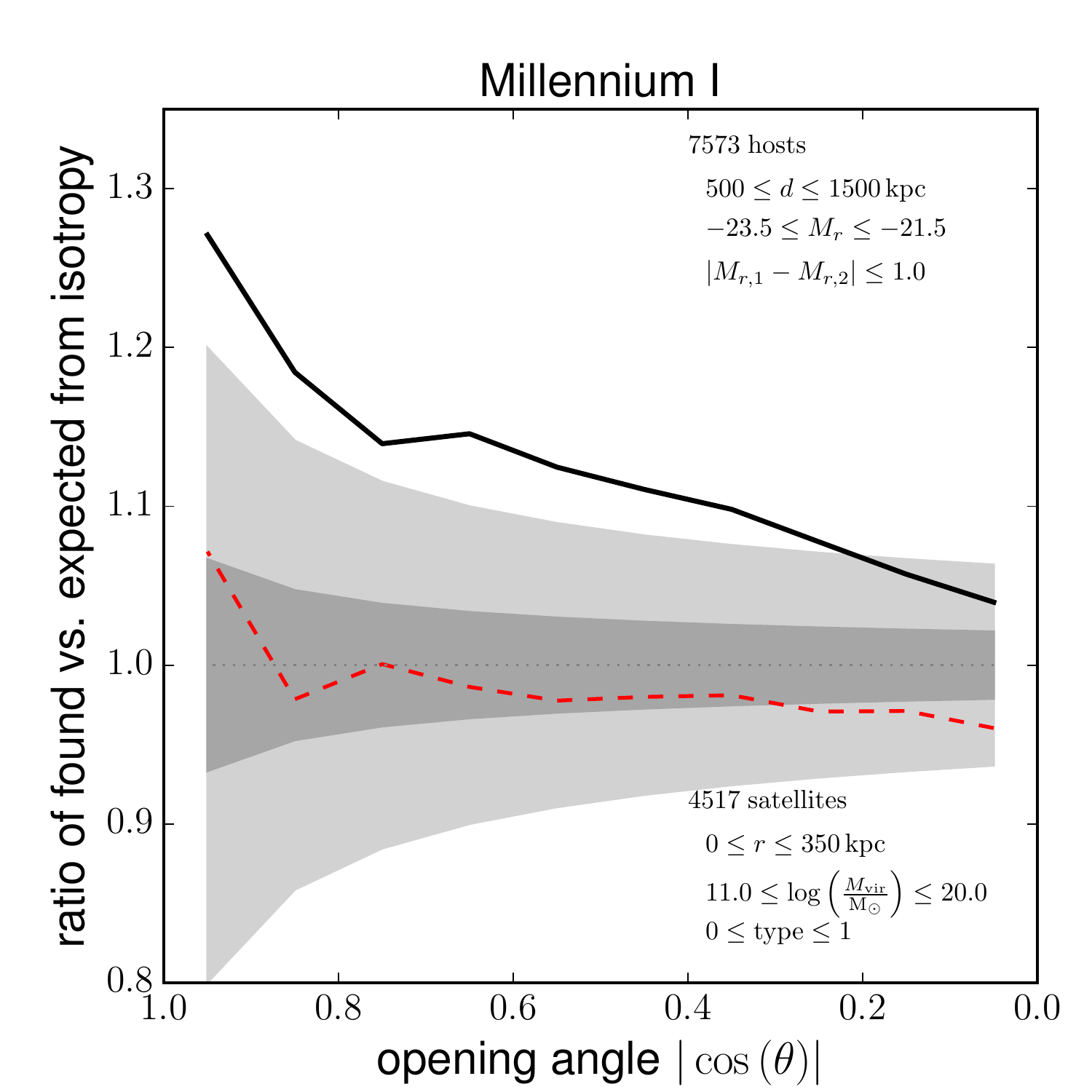}{0.25\textwidth}{(d)}
          }
\gridline{\fig{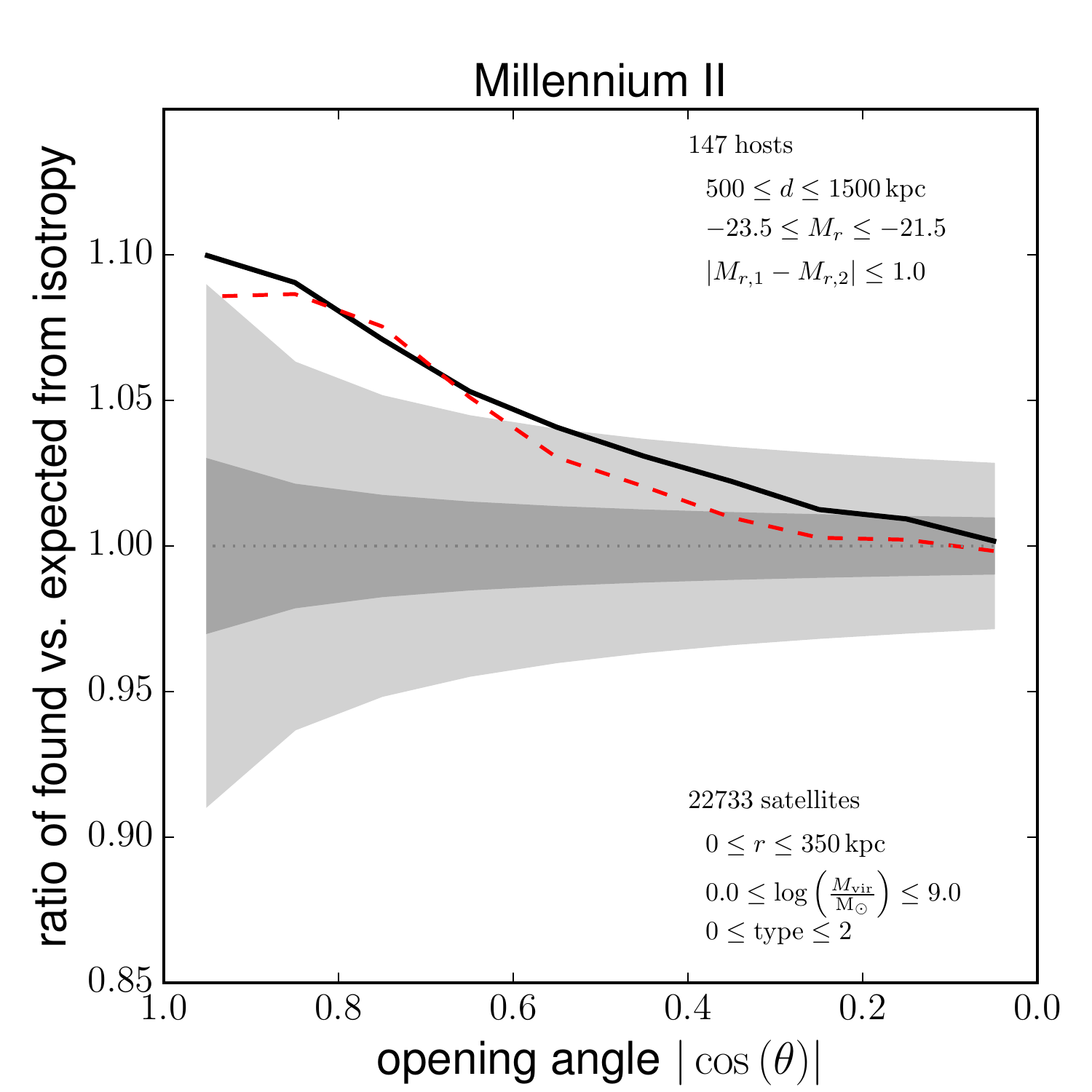}{0.25\textwidth}{(e)}
          \fig{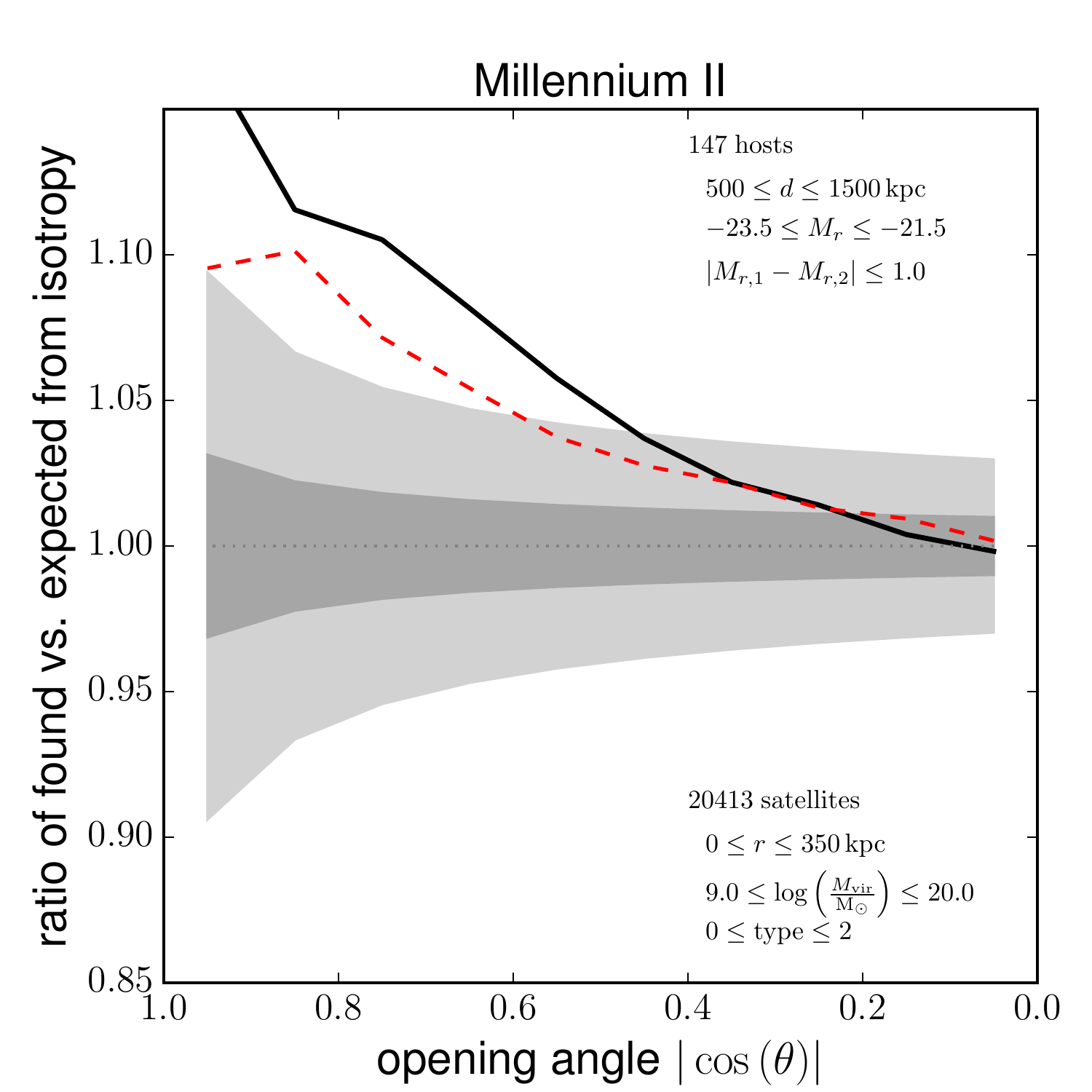}{0.25\textwidth}{(f)}
          \fig{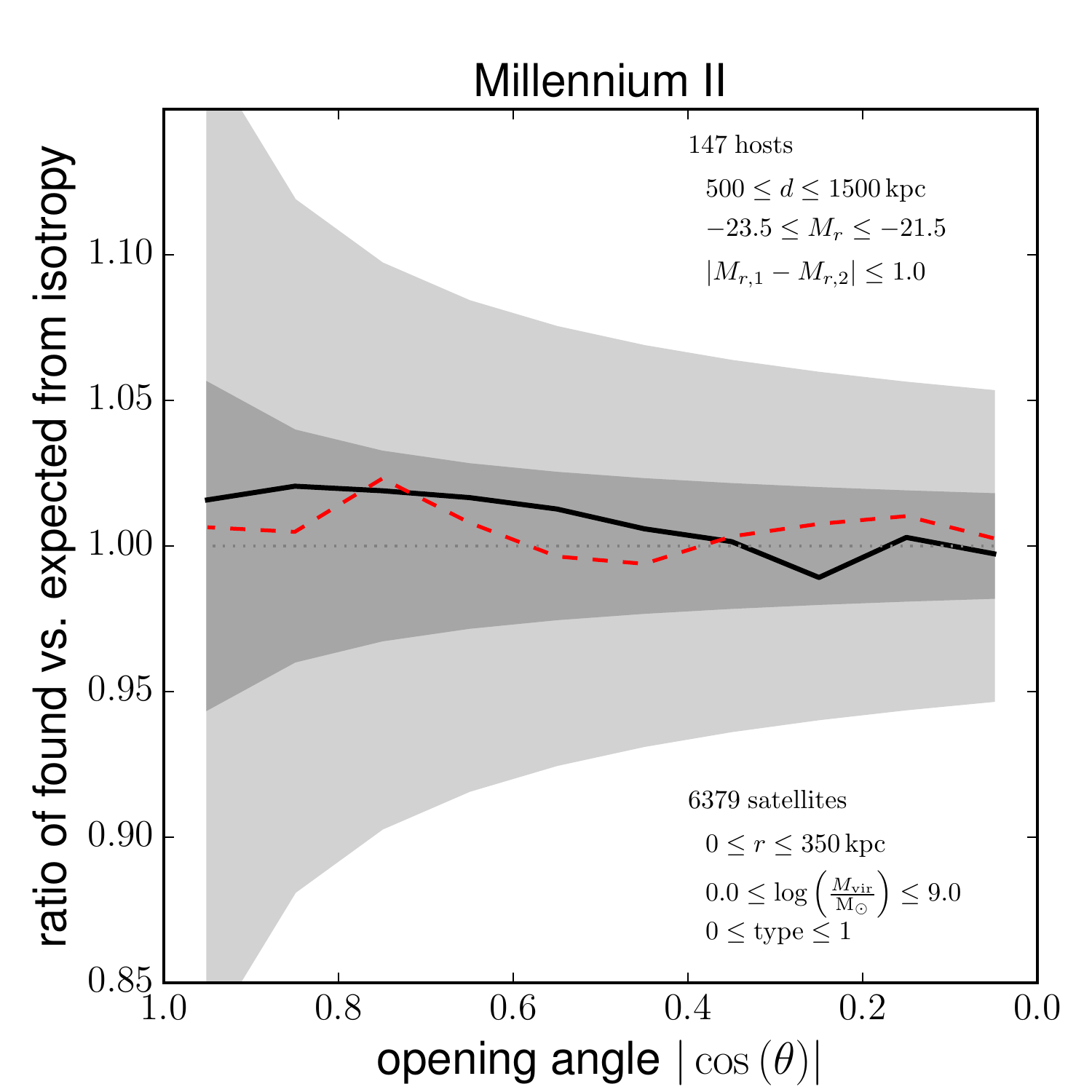}{0.25\textwidth}{(g)}
          \fig{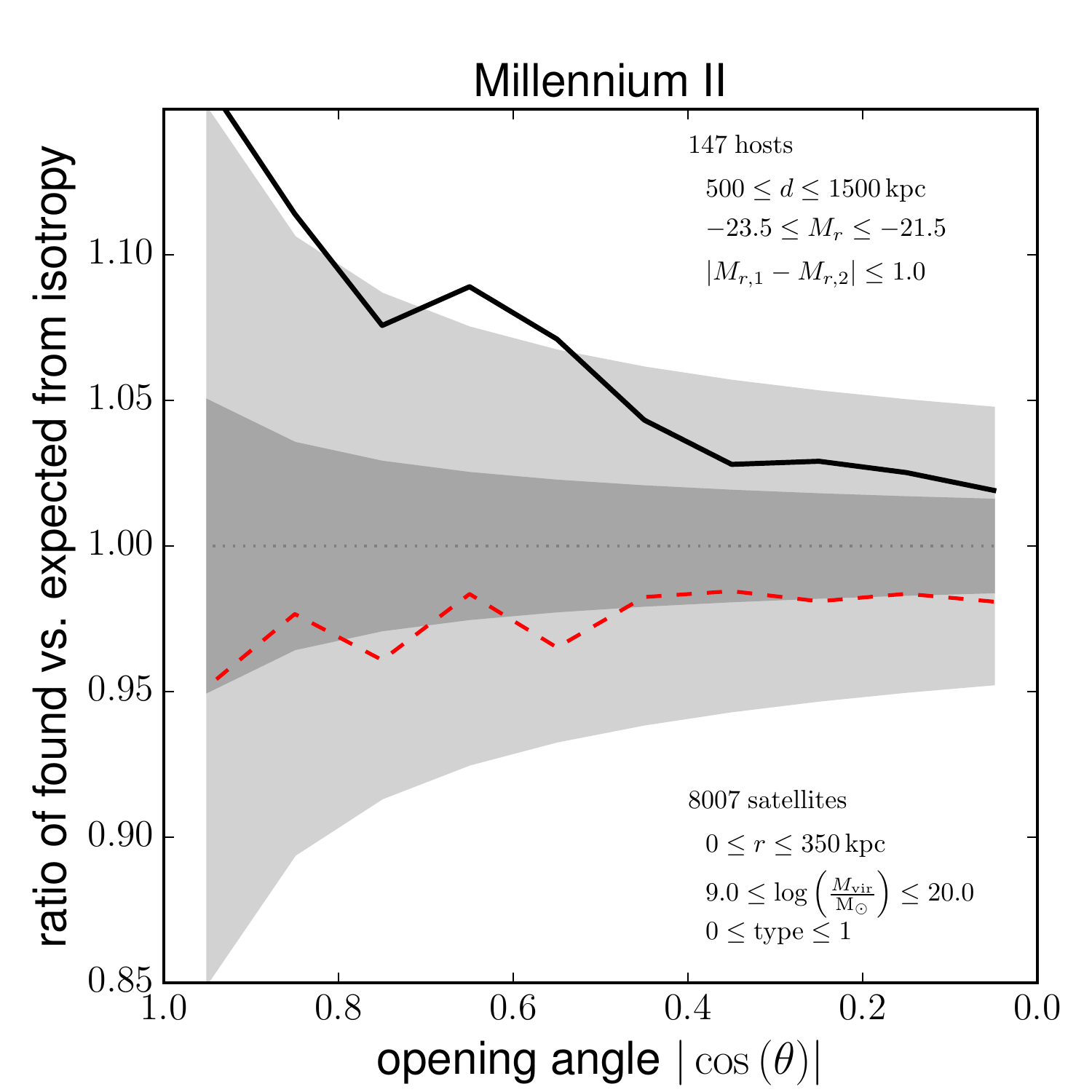}{0.25\textwidth}{(h)}
          }
\gridline{\fig{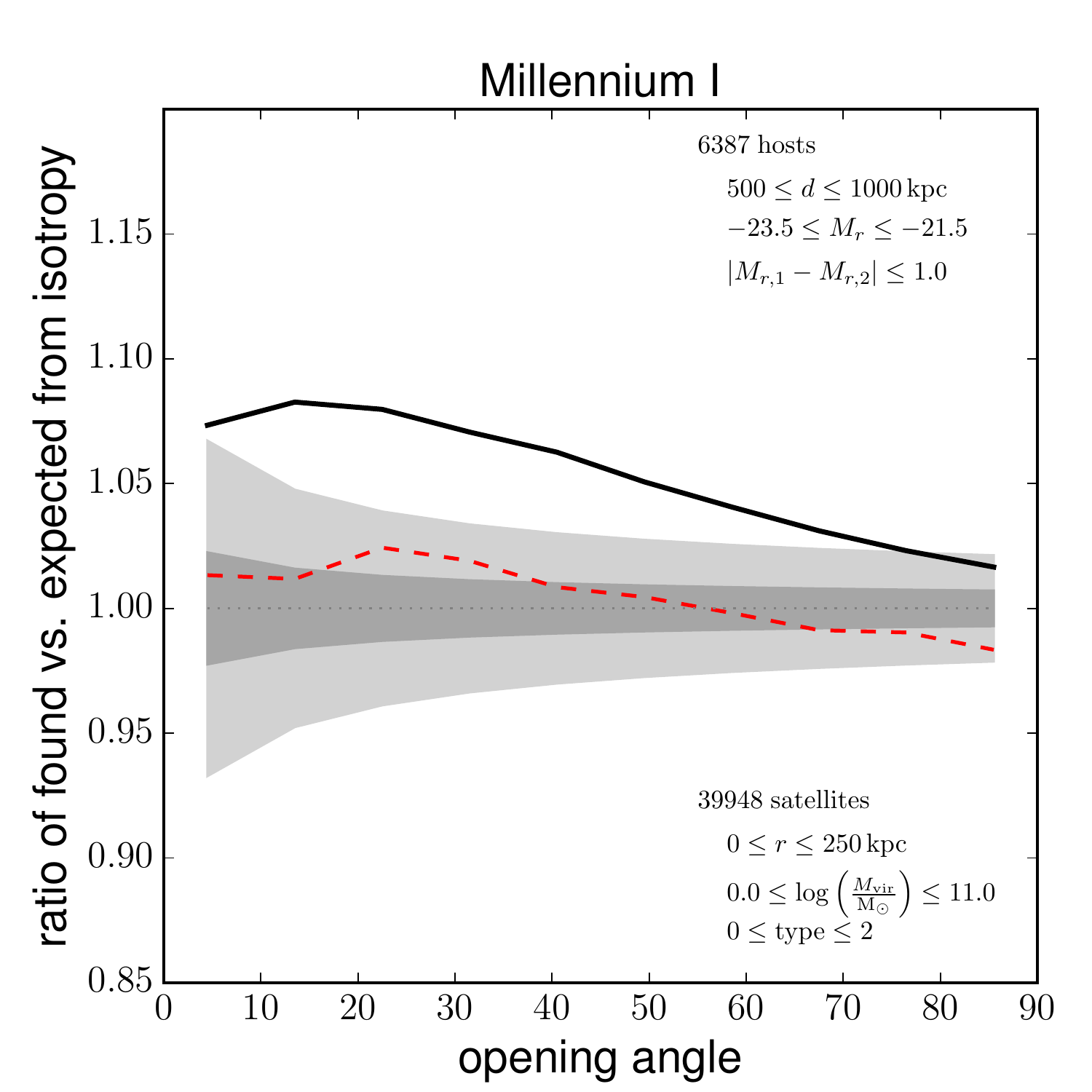}{0.25\textwidth}{(i)}
          \fig{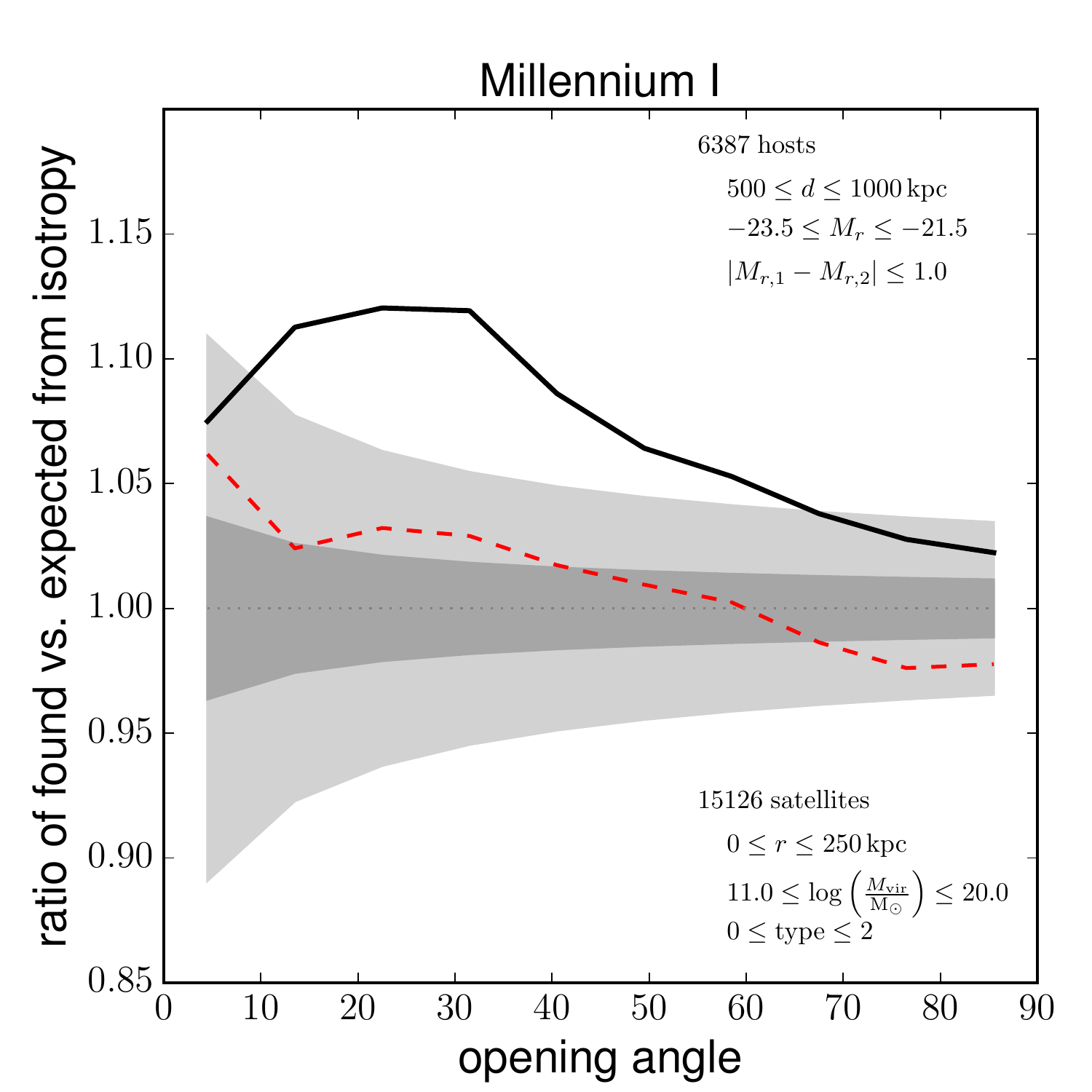}{0.25\textwidth}{(j)}
          \fig{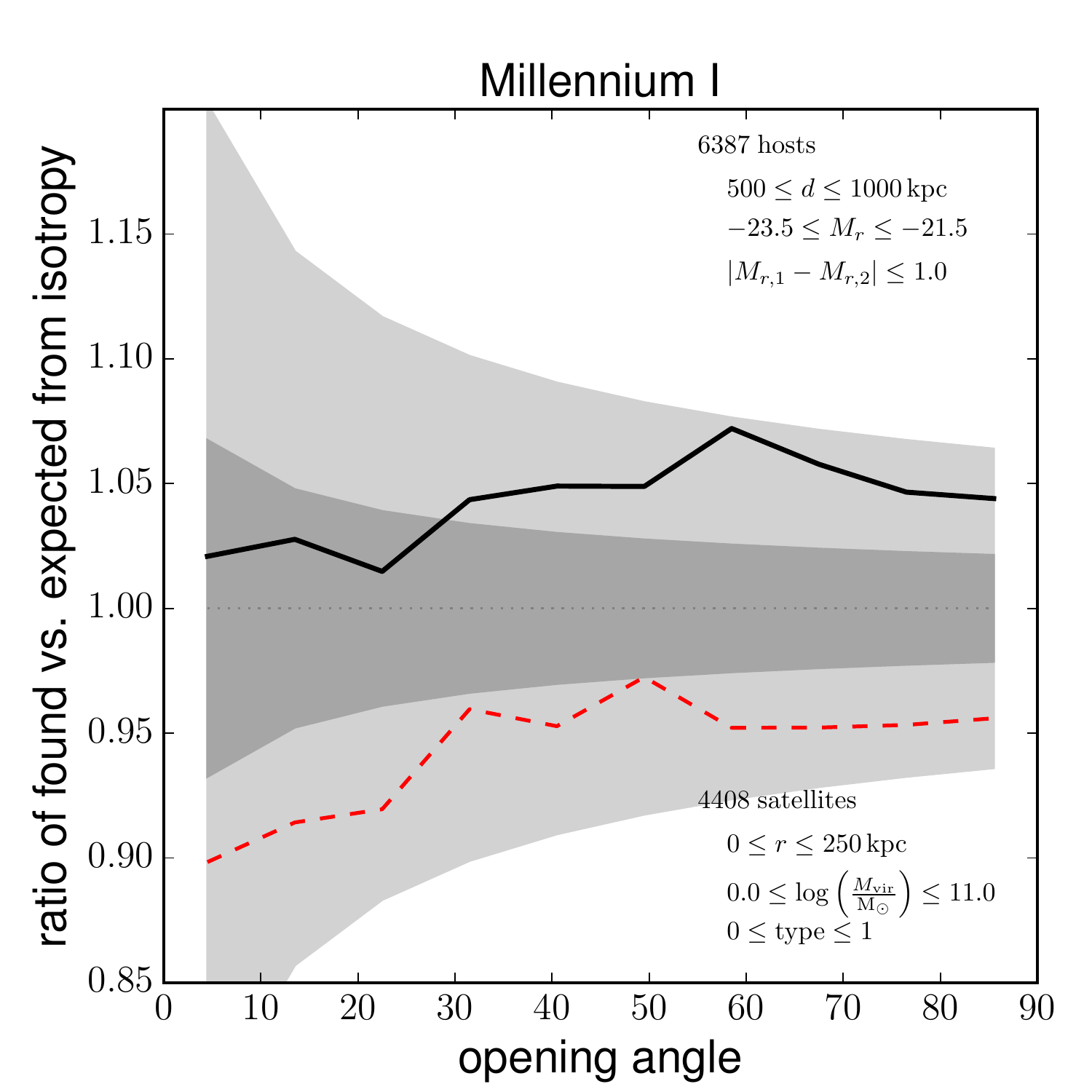}{0.25\textwidth}{(k)}
          \fig{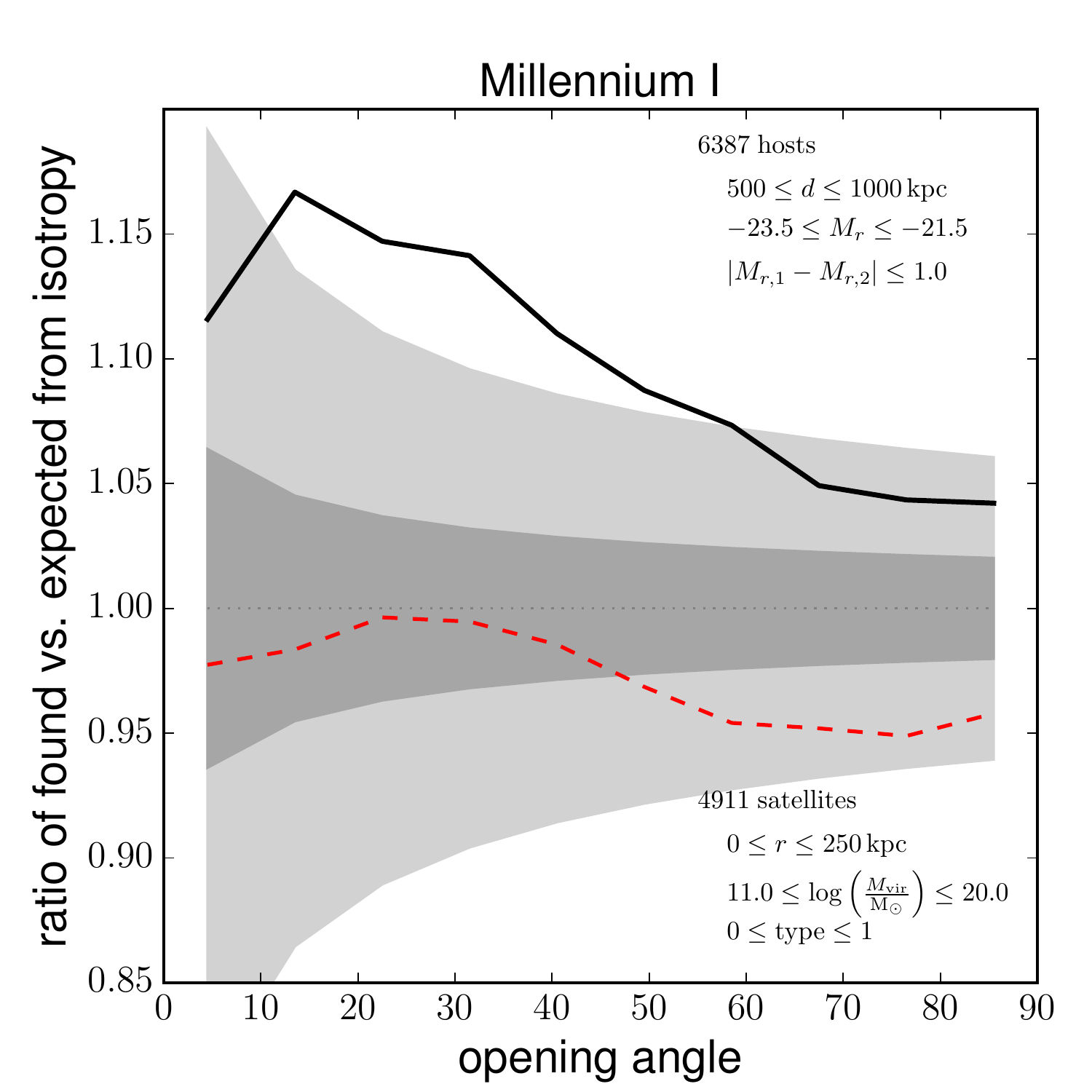}{0.25\textwidth}{(l)}
          }
\gridline{\fig{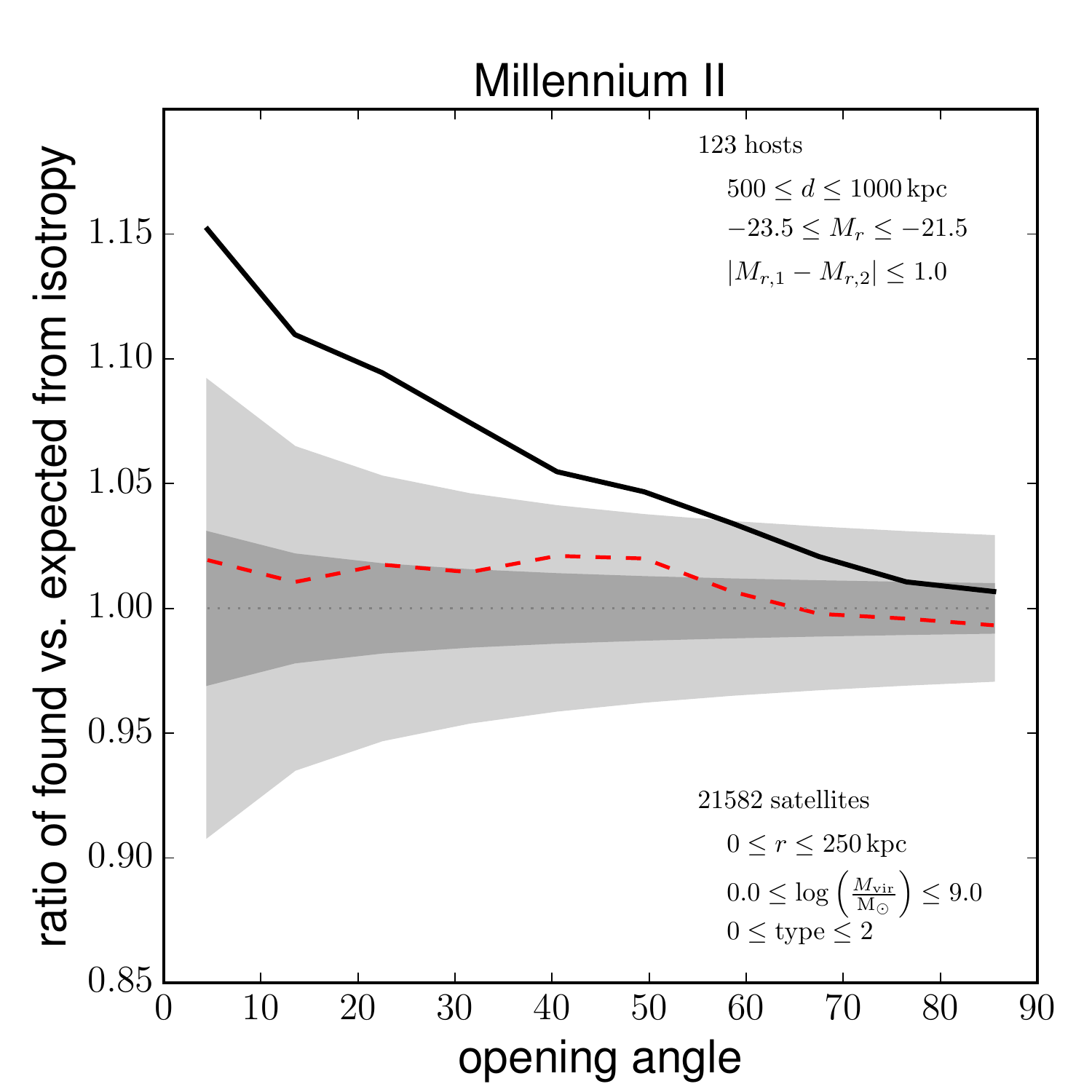}{0.25\textwidth}{(m)}
          \fig{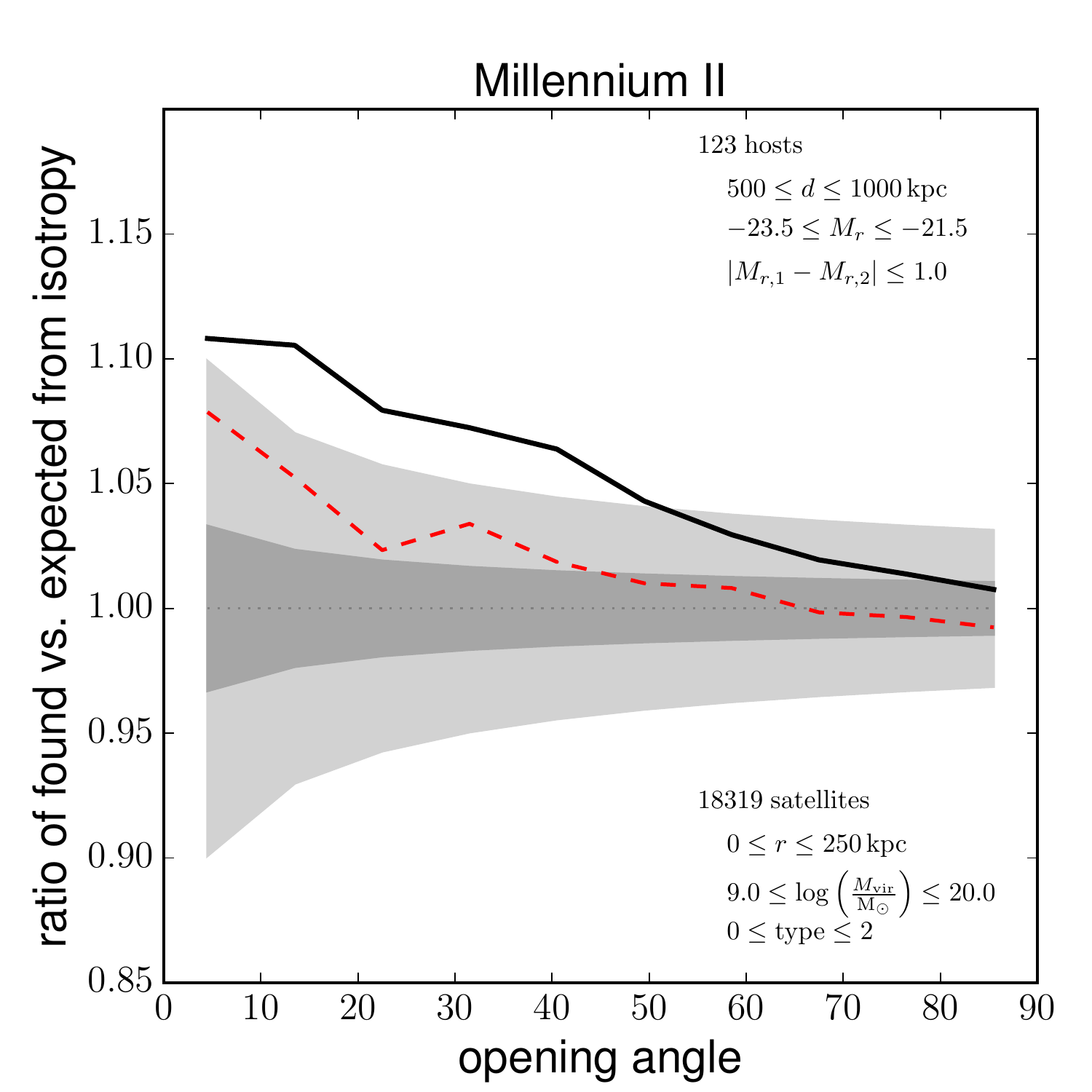}{0.25\textwidth}{(n)}
          \fig{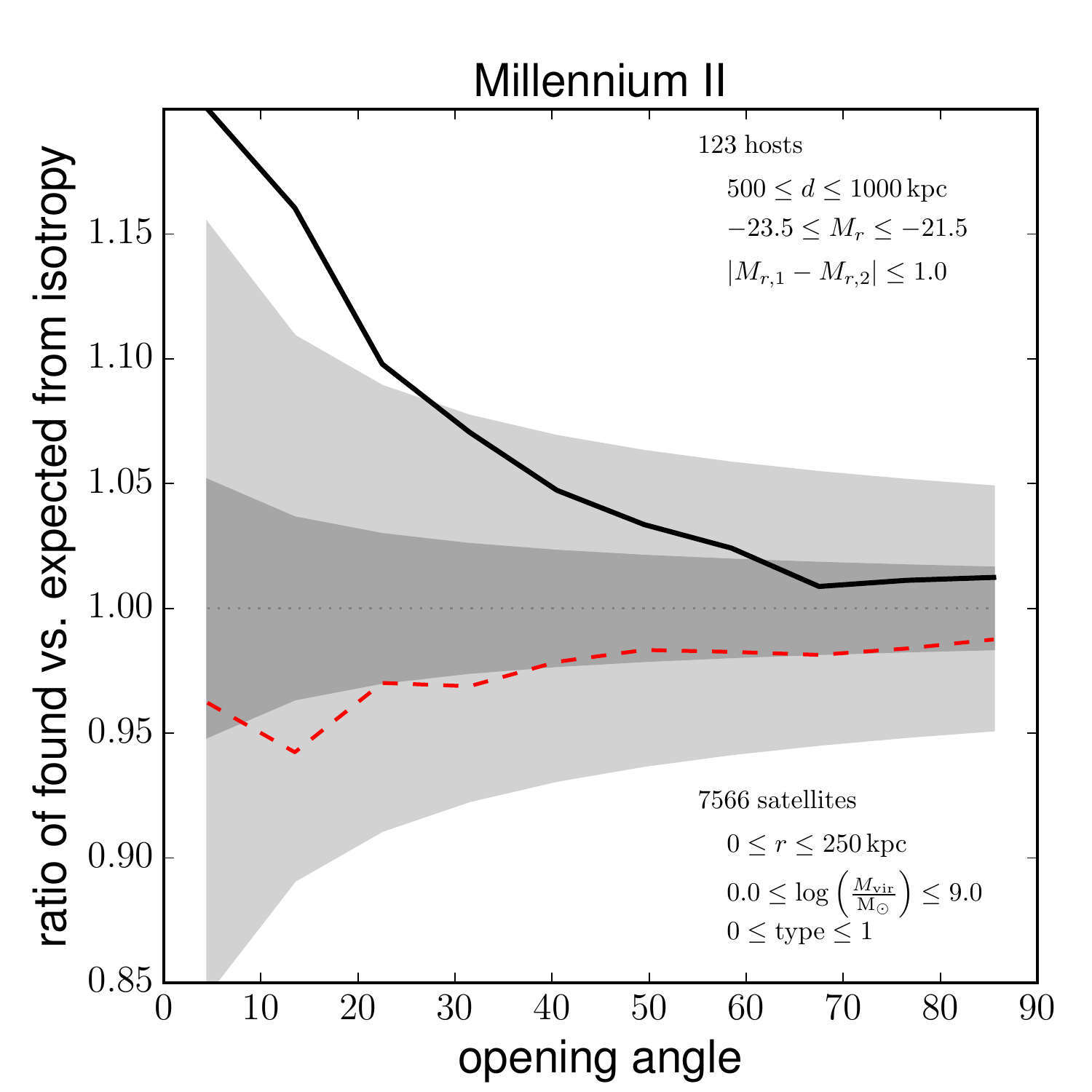}{0.25\textwidth}{(o)}
          \fig{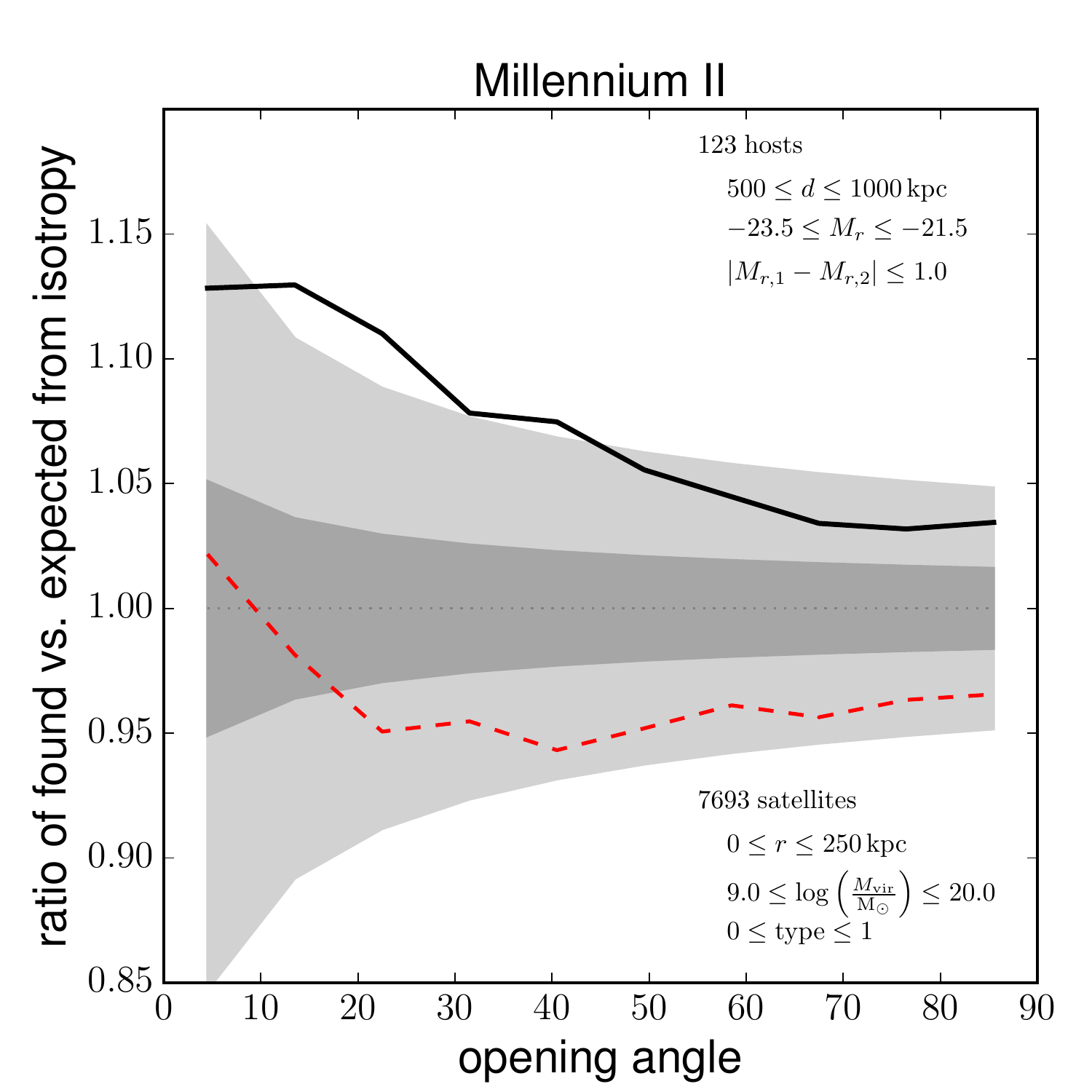}{0.25\textwidth}{(p)}
          }
\caption{
Overabundance of satellites towards (black solid lines) and away from (red dashed lines) the other primary. Similar to Figs. \ref{fig:3D} and \ref{fig:2D}, but for satellite sub-samples seperated by virial mass. The expected 1 and $3\,\sigma$\ scatter around an isotropic distribution is illustrated by the dark and light gray areas, respectively. The first two rows show the 3D distributions for MS1 and MS2, the last two rows the respective 2D distributions. The first two colums include orphan galaxies (type = 2), in the last two they are excluded.
\label{fig:masssubsamples}}
\end{figure*}

\begin{figure*}
\gridline{\fig{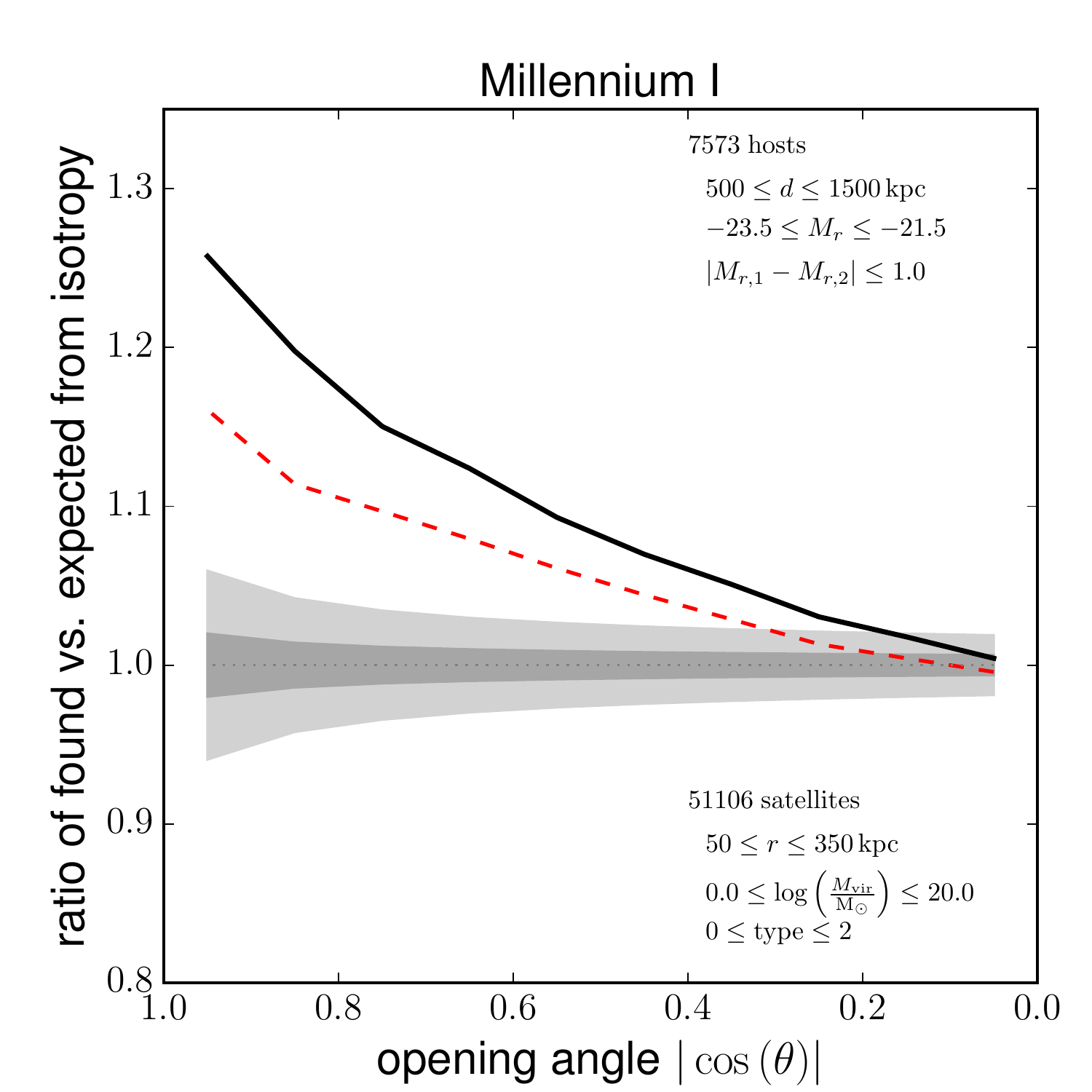}{0.25\textwidth}{(a)}
          \fig{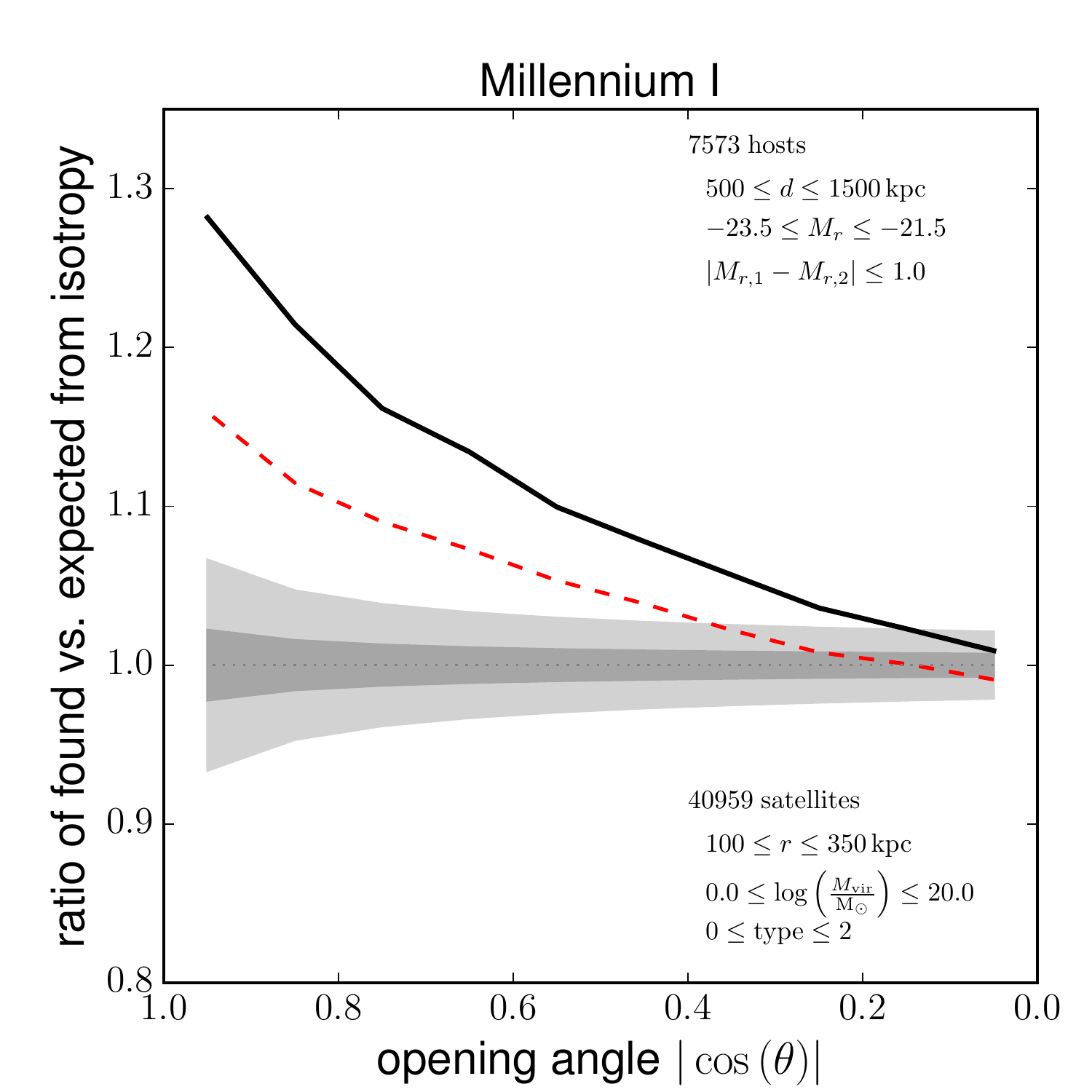}{0.25\textwidth}{(b)}
          \fig{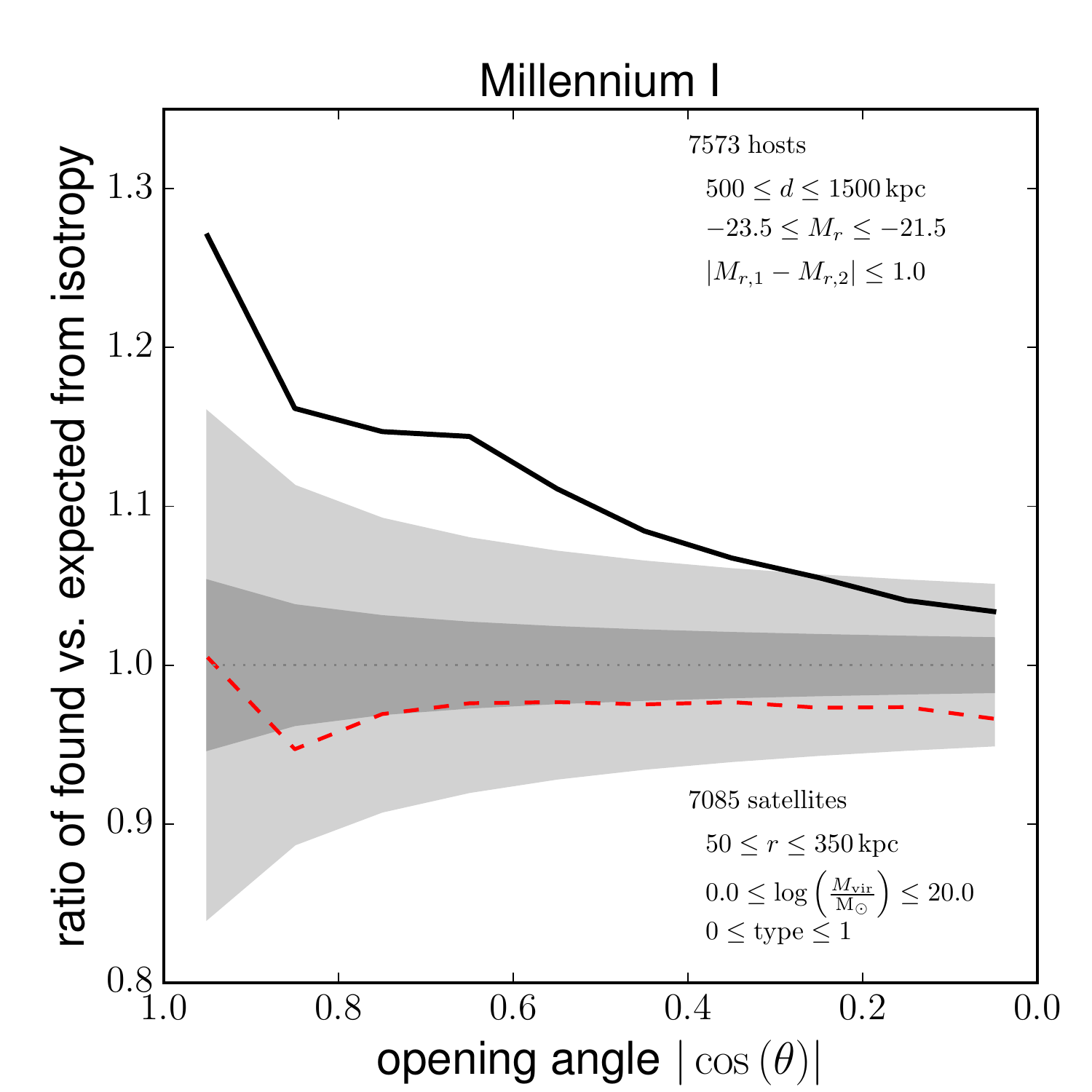}{0.25\textwidth}{(c)}
          \fig{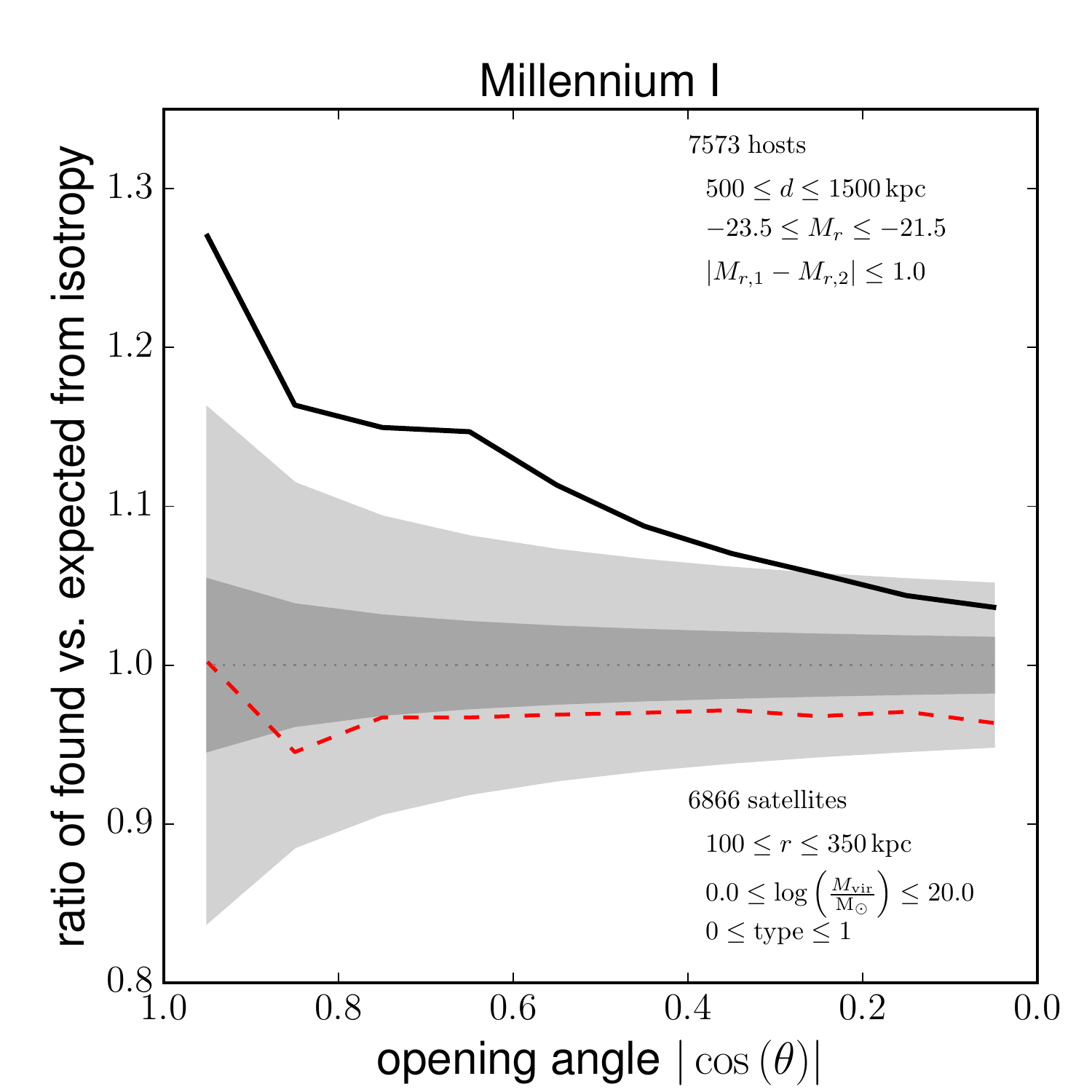}{0.25\textwidth}{(d)}
          }
\gridline{\fig{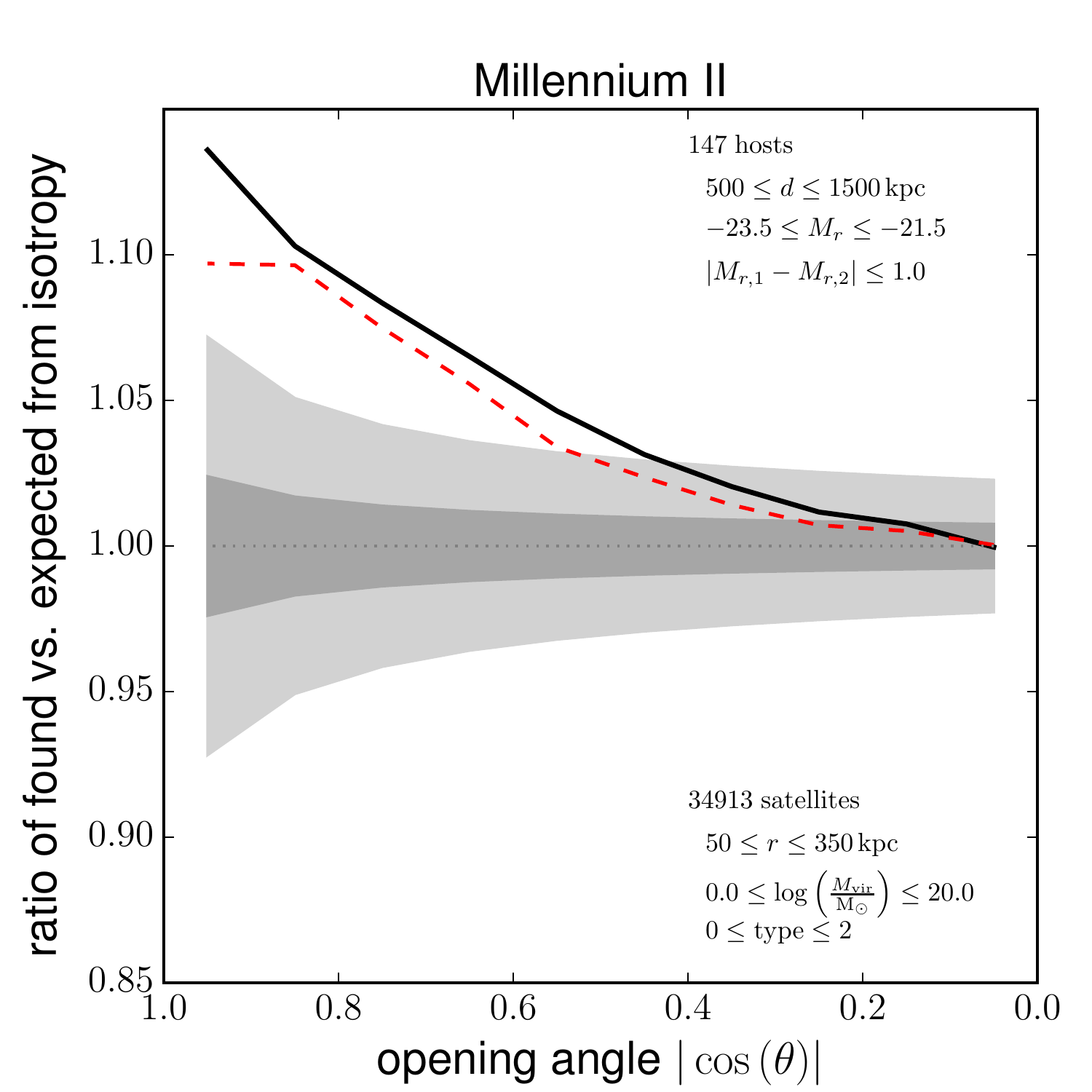}{0.25\textwidth}{(e)}
          \fig{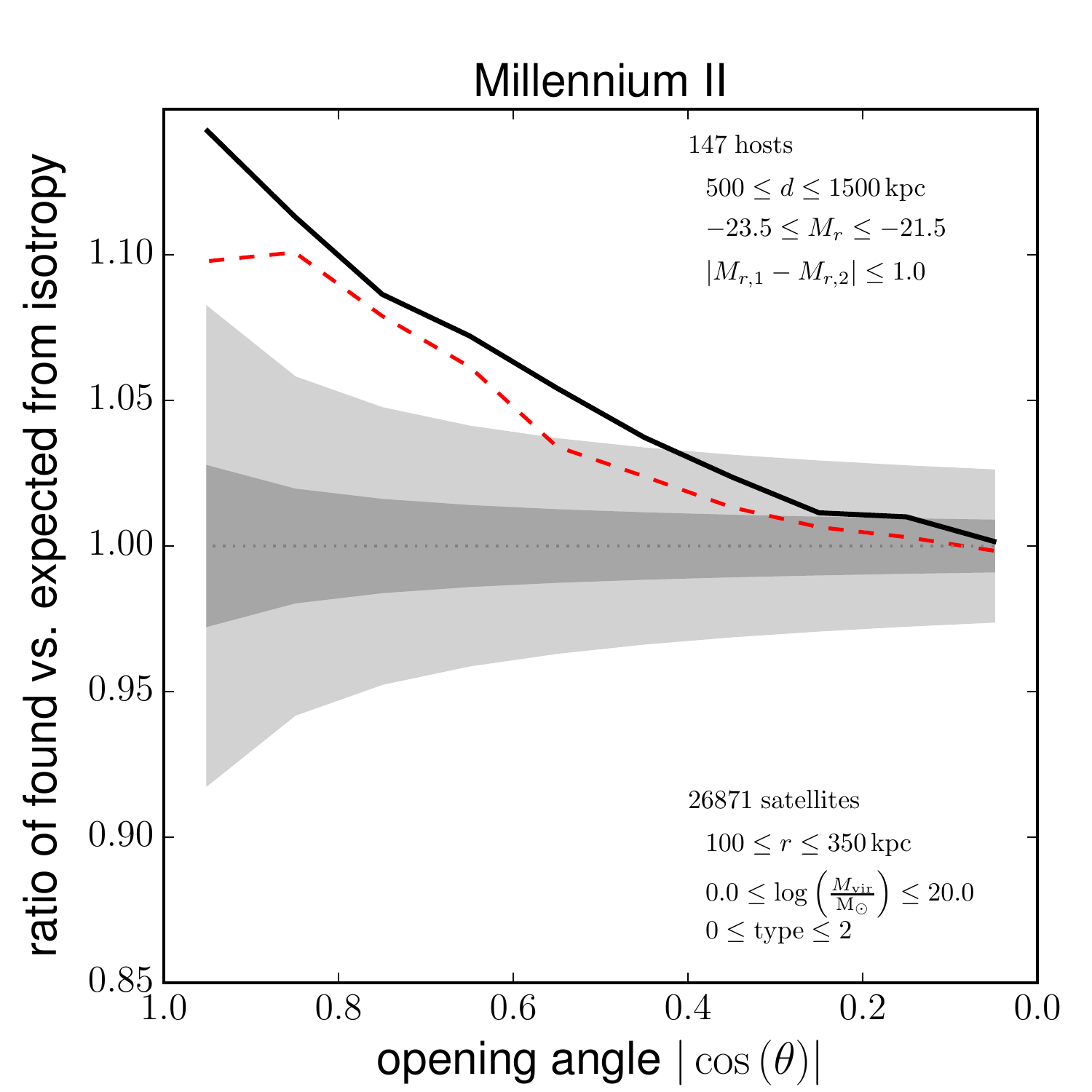}{0.25\textwidth}{(f)}
          \fig{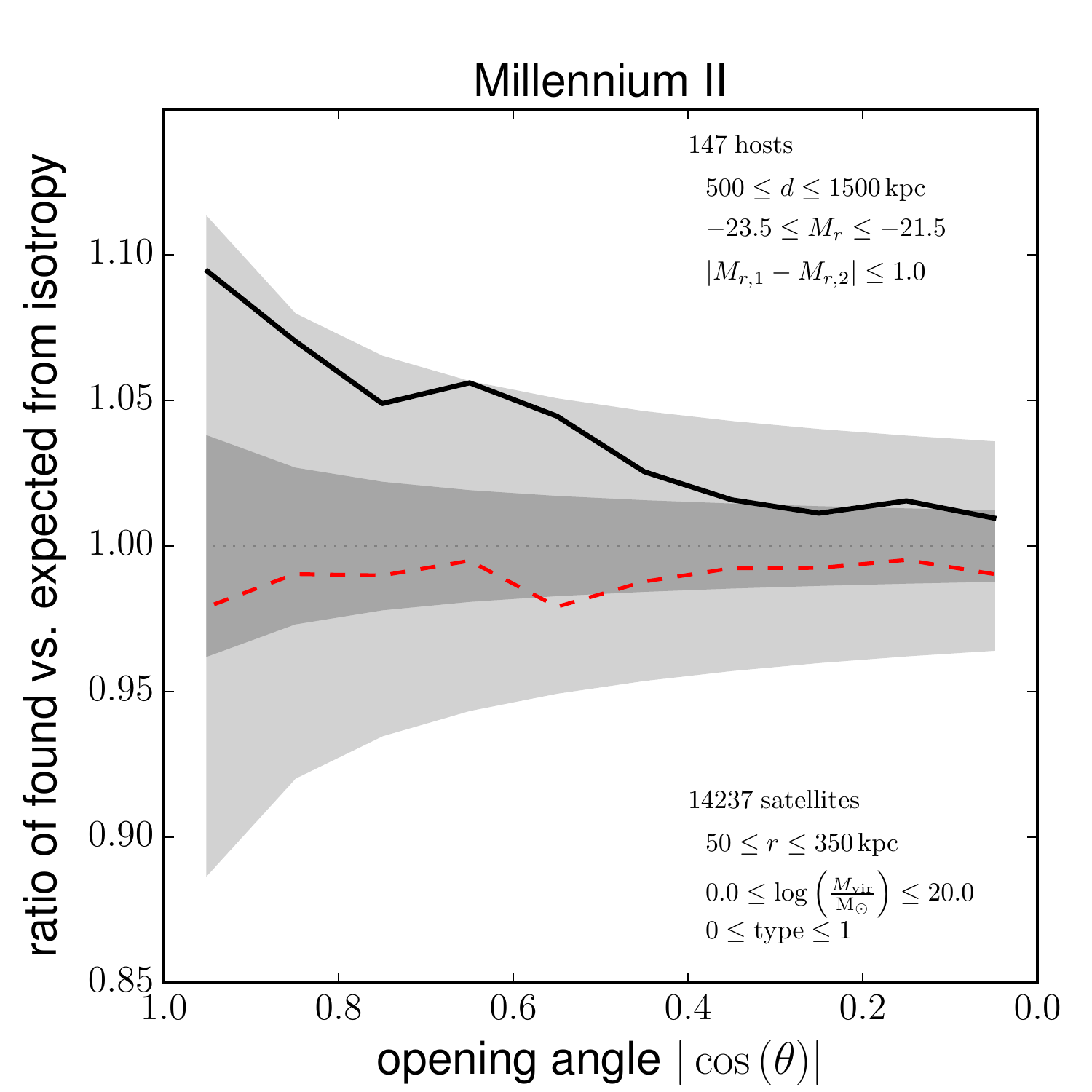}{0.25\textwidth}{(g)}
          \fig{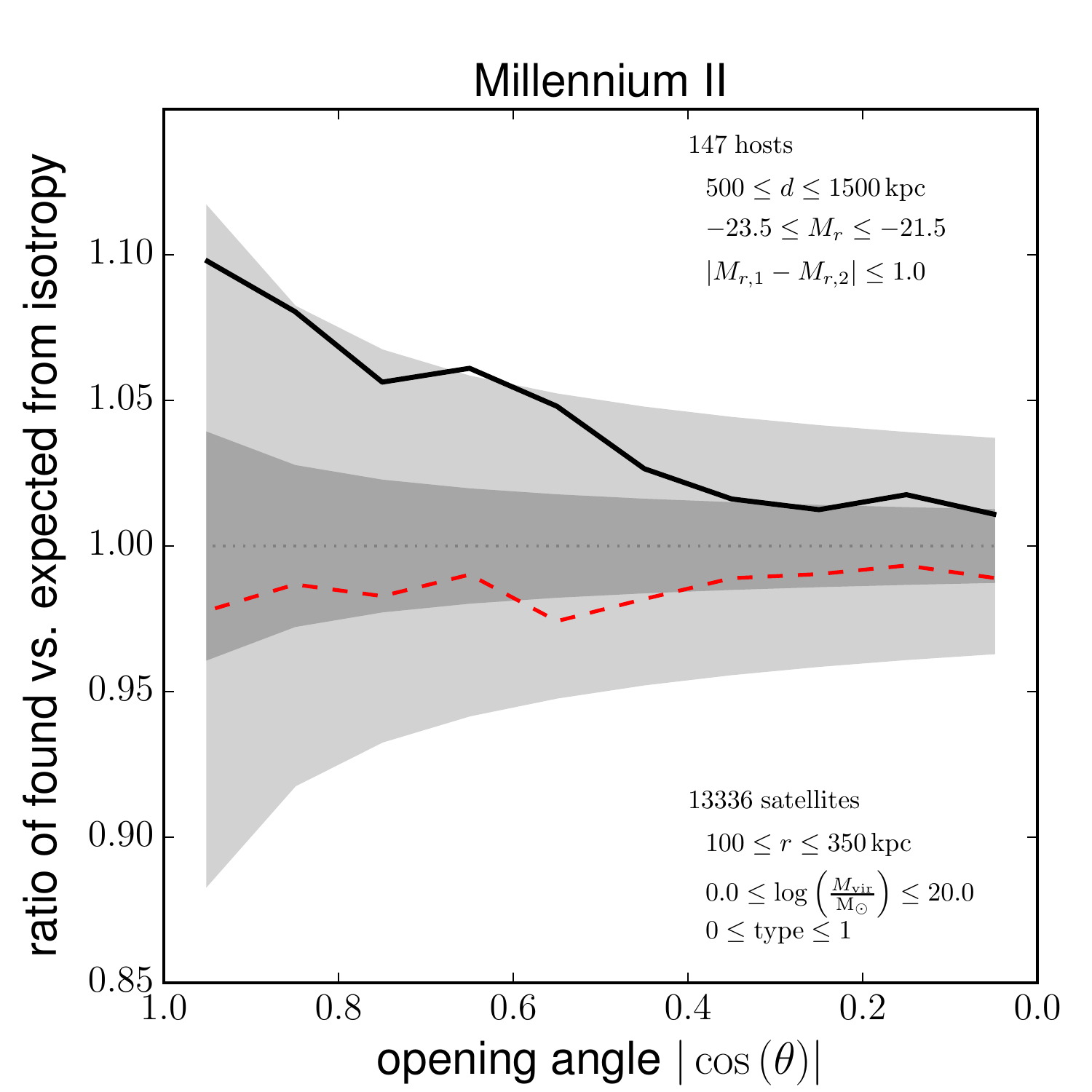}{0.25\textwidth}{(h)}
          }
\gridline{\fig{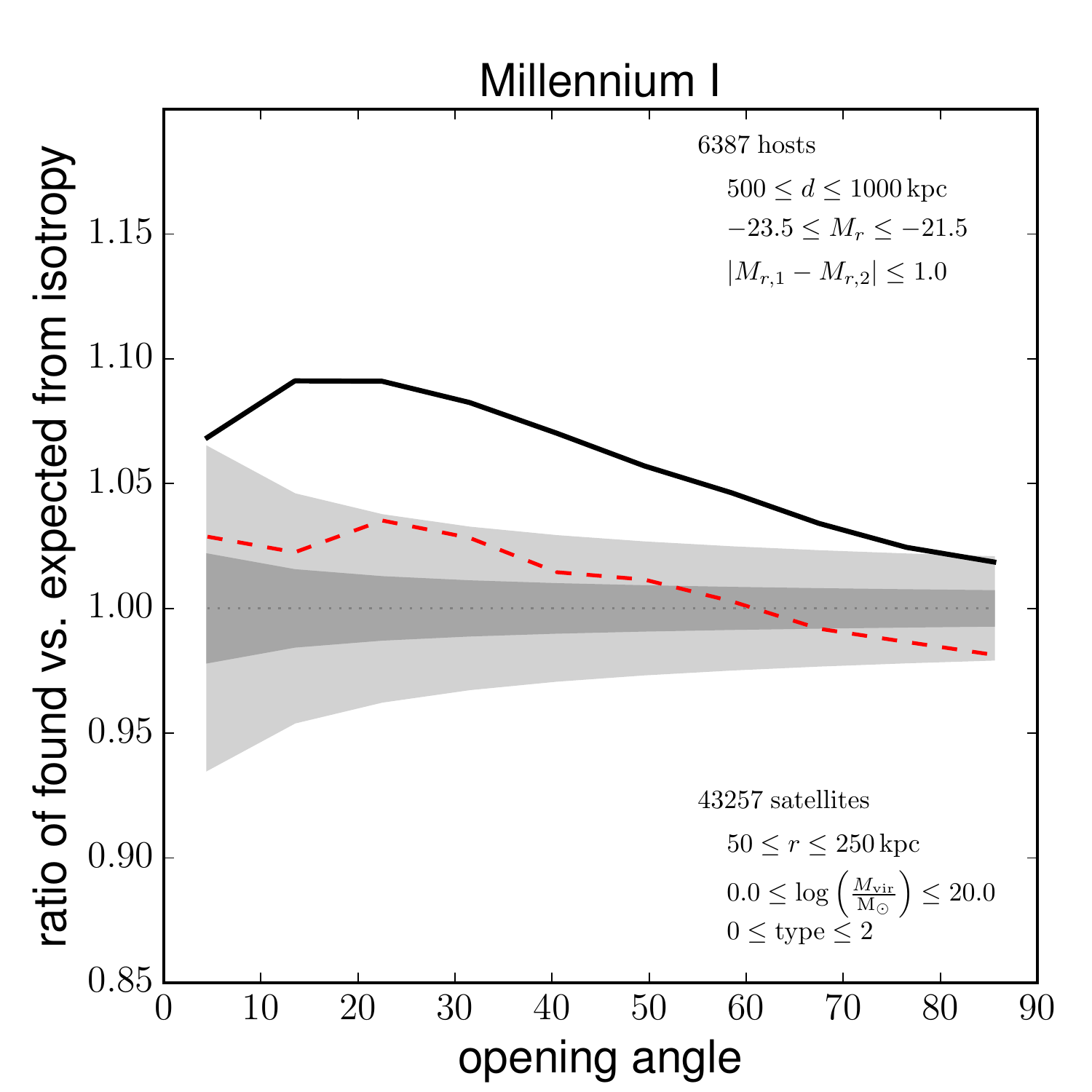}{0.25\textwidth}{(i)}
          \fig{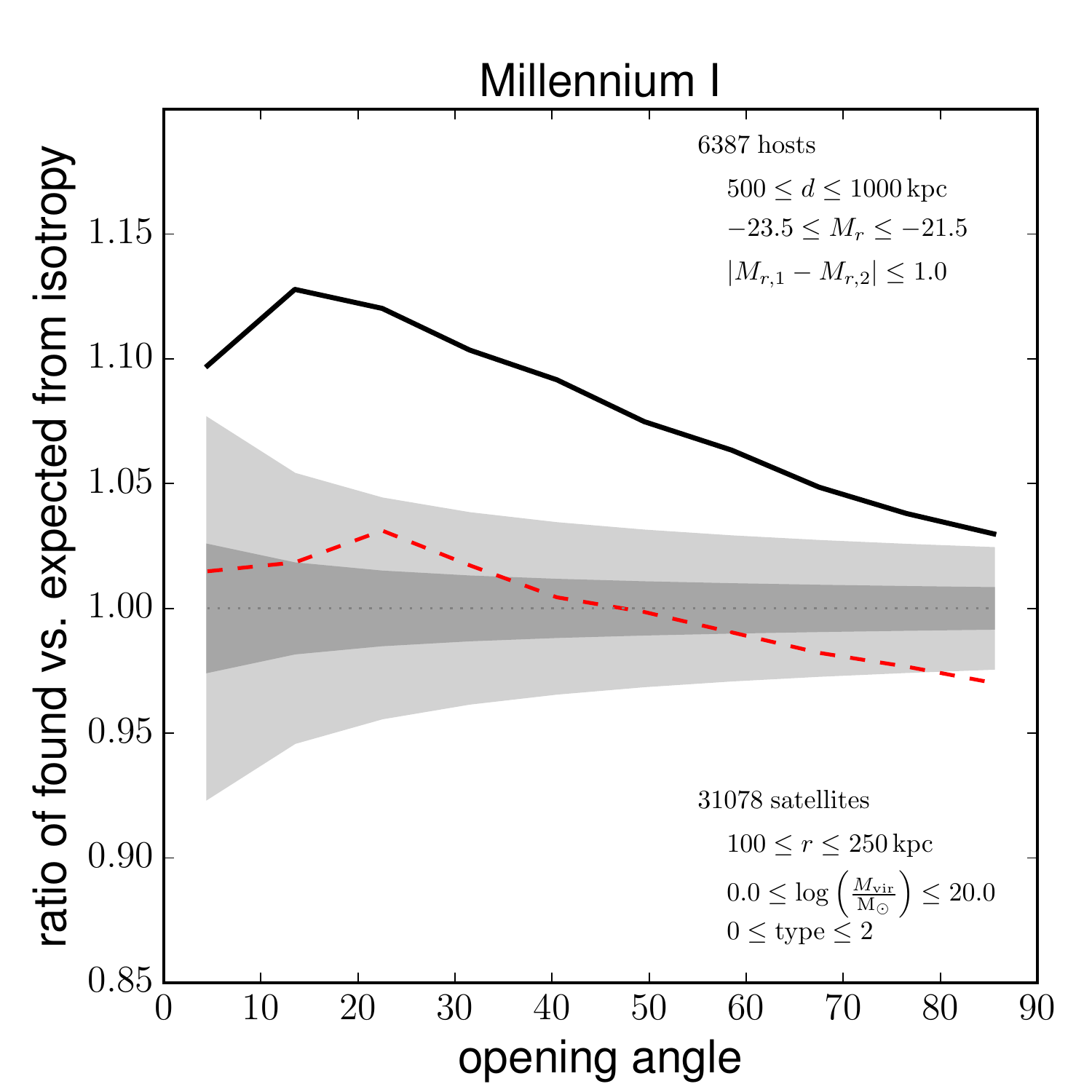}{0.25\textwidth}{(j)}
          \fig{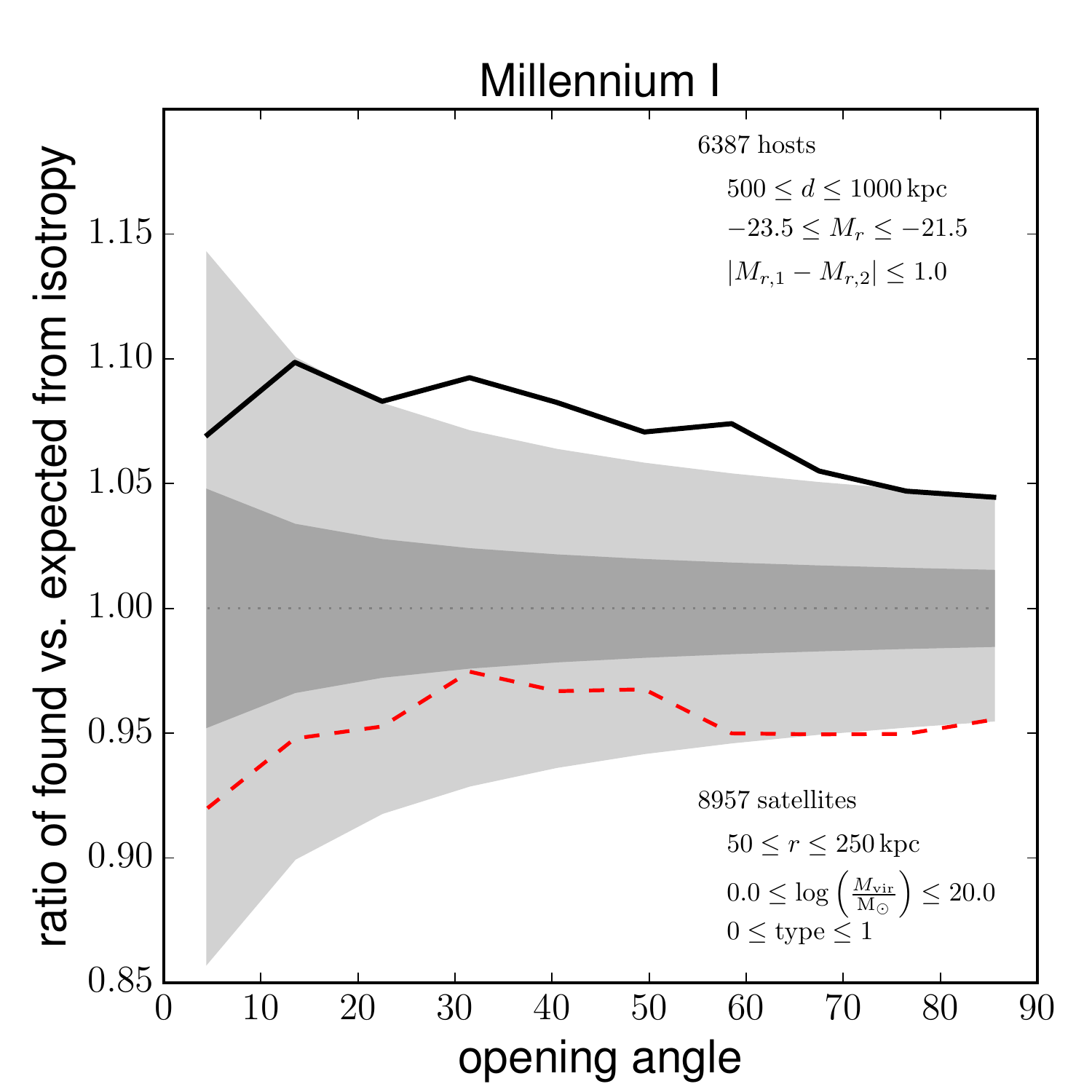}{0.25\textwidth}{(k)}
          \fig{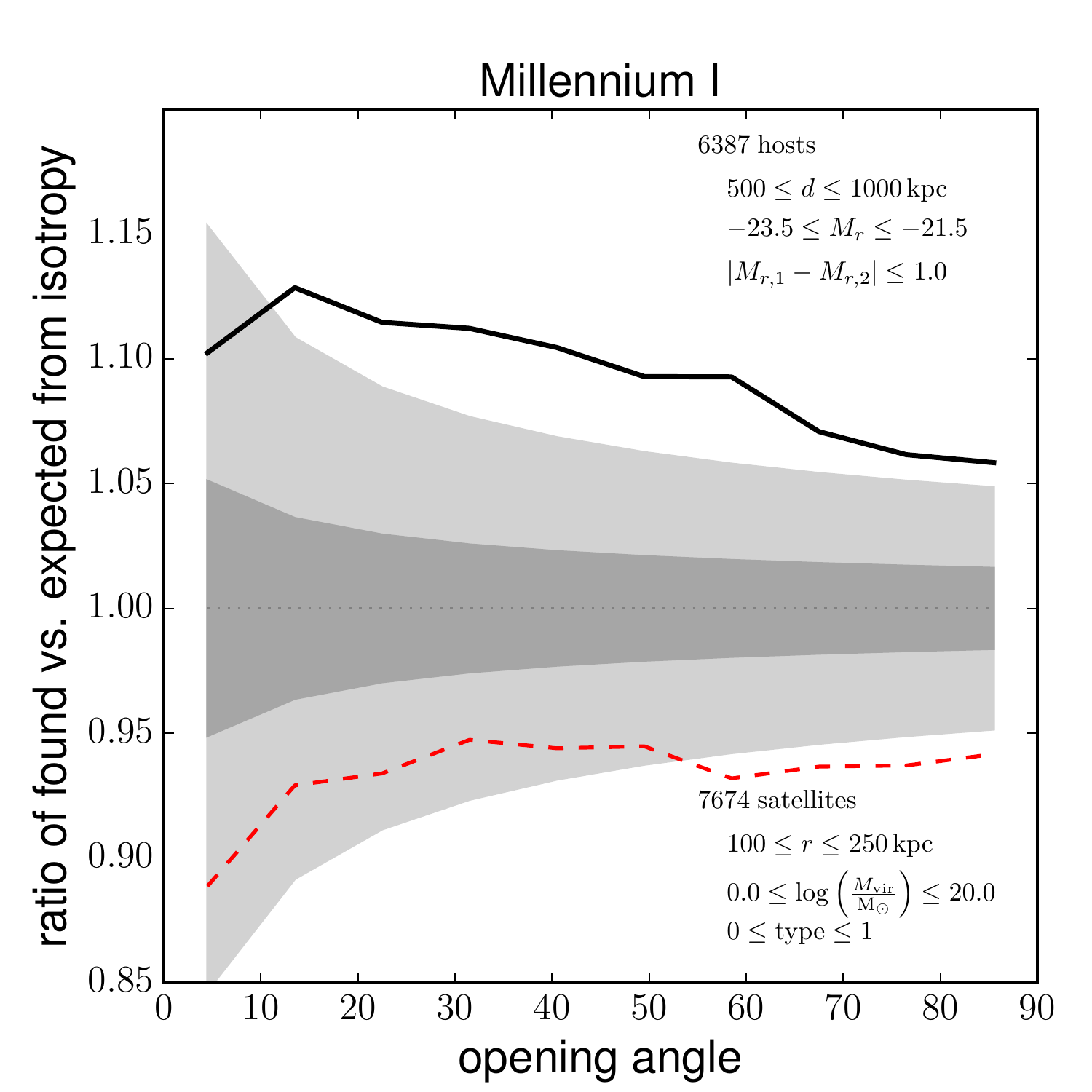}{0.25\textwidth}{(l)}
          }
\gridline{\fig{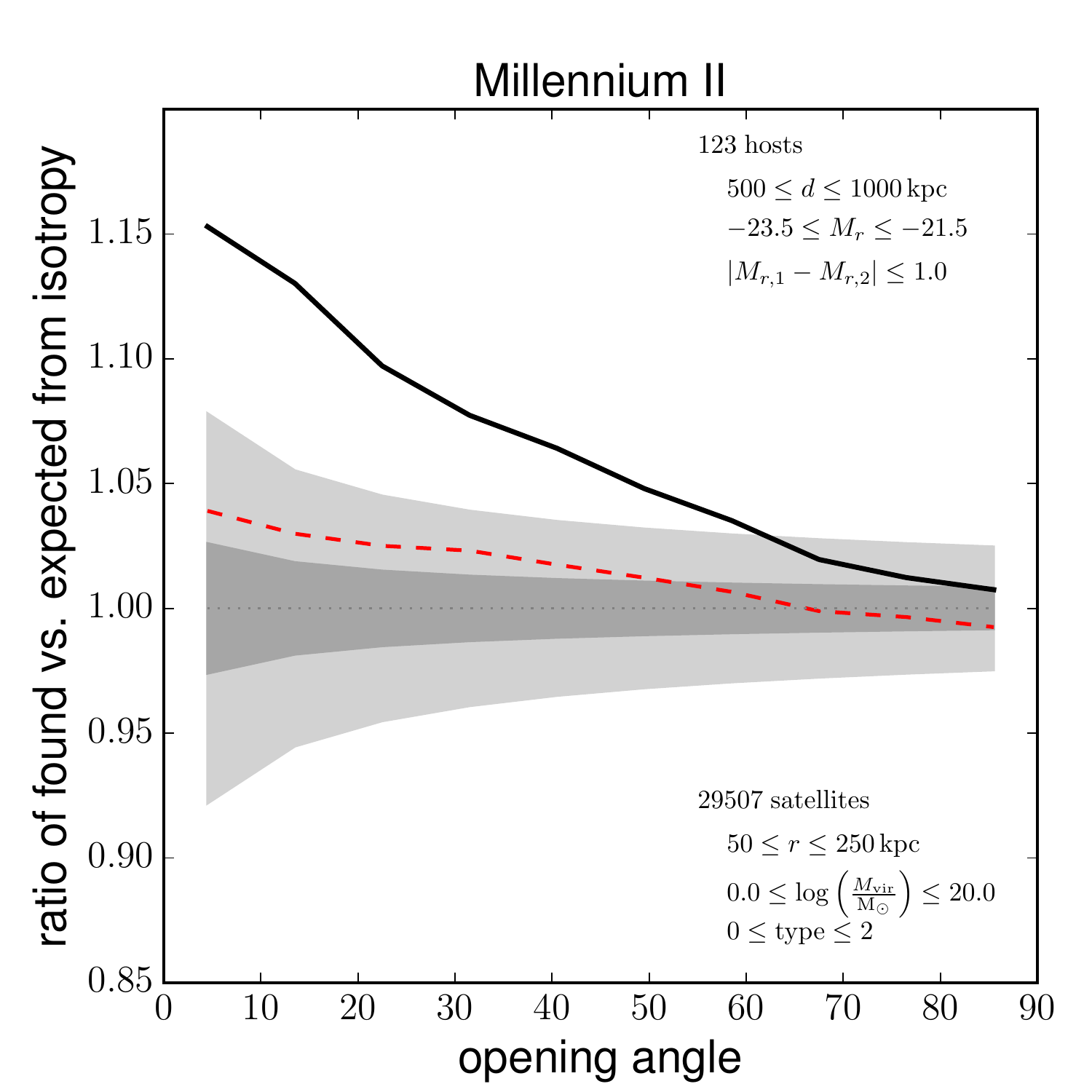}{0.25\textwidth}{(m)}
          \fig{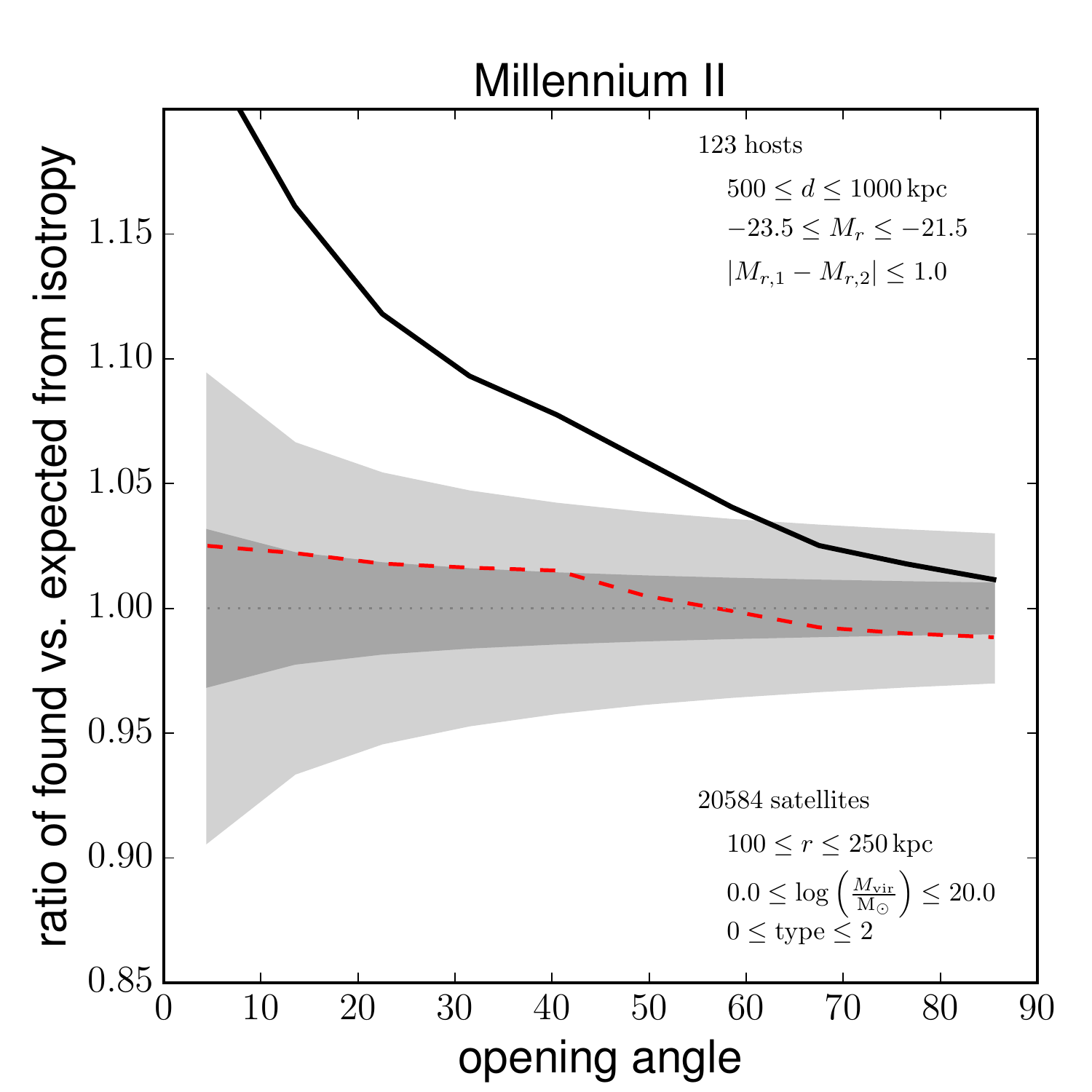}{0.25\textwidth}{(n)}
          \fig{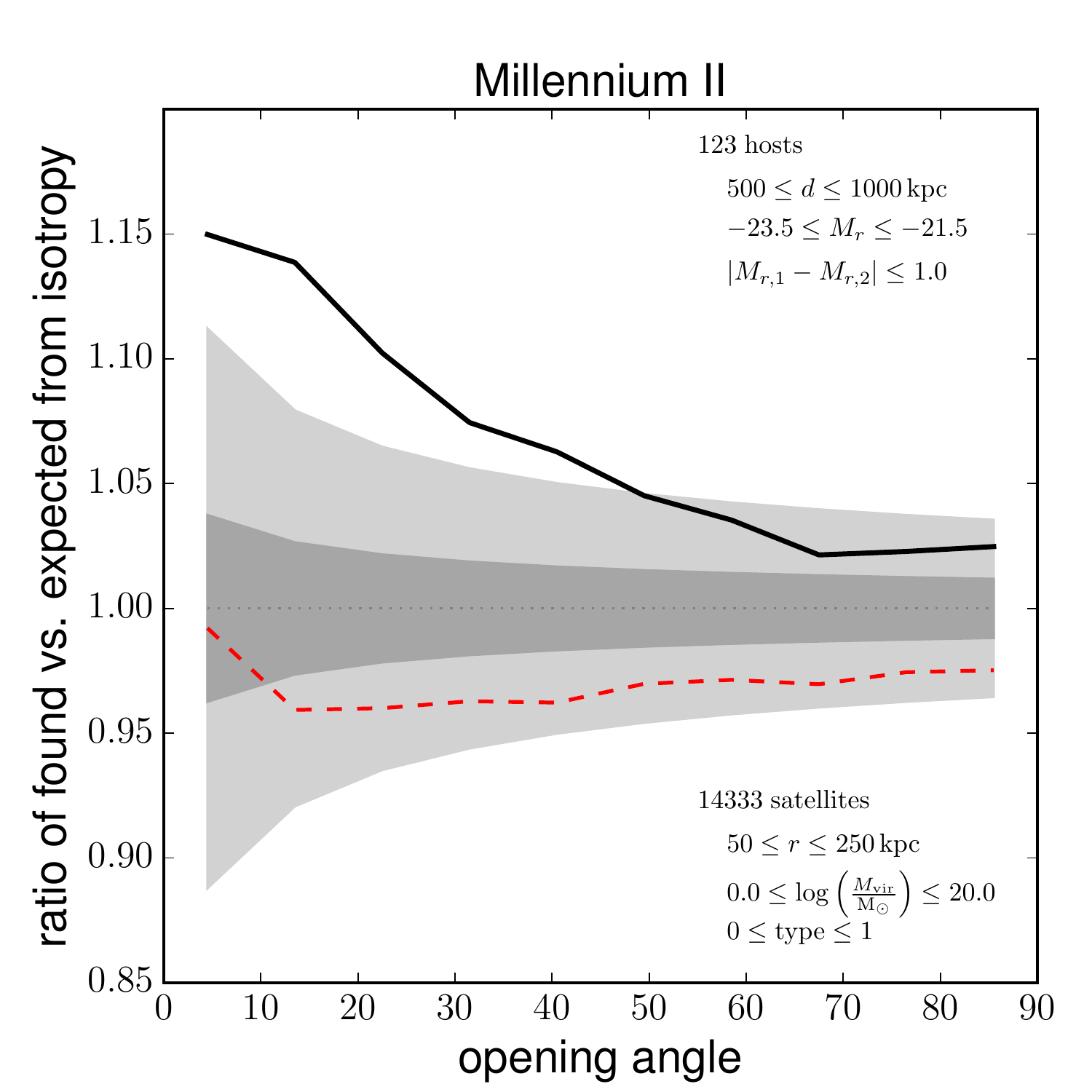}{0.25\textwidth}{(o)}
          \fig{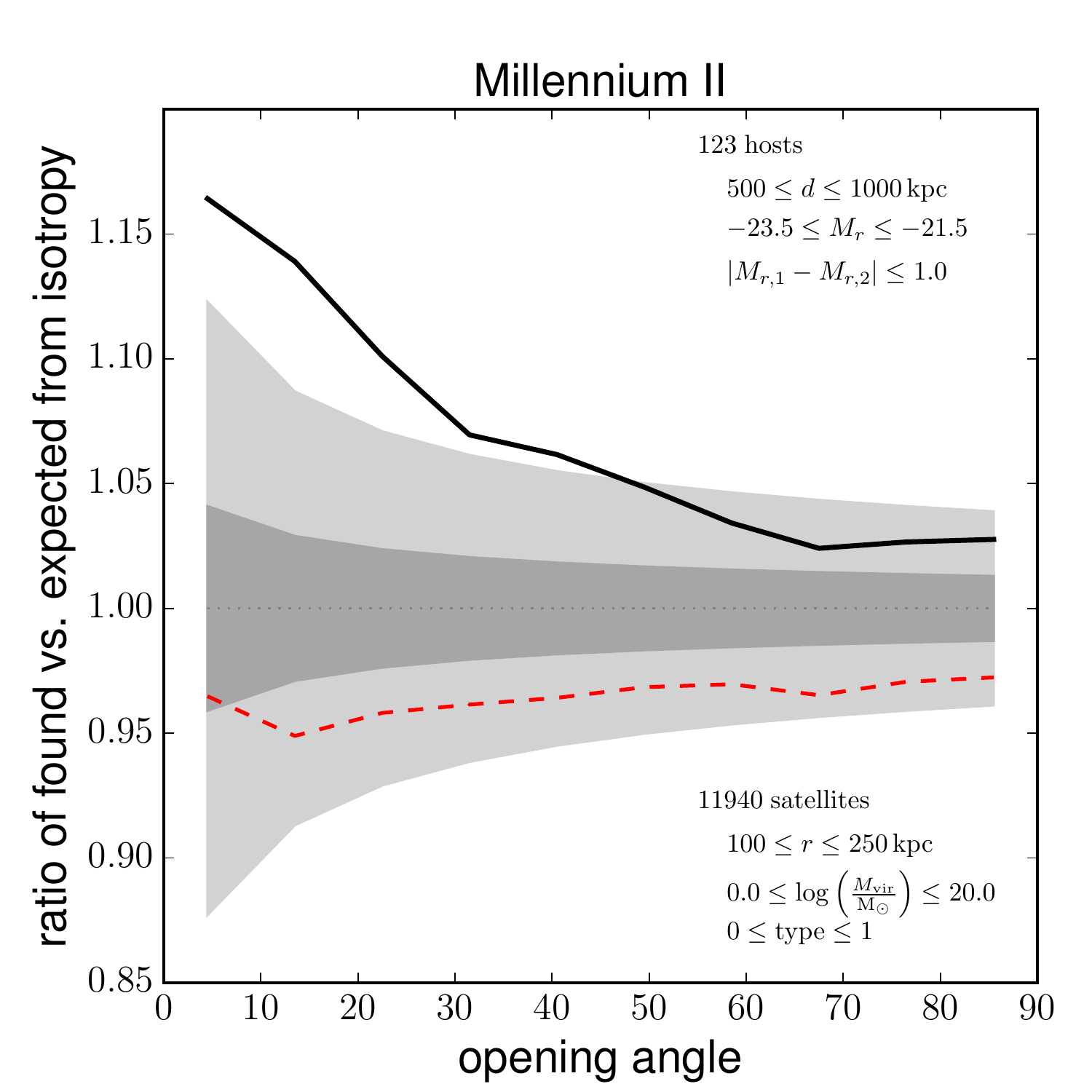}{0.25\textwidth}{(p)}
          }
\caption{
Same as Fig. \ref{fig:masssubsamples}, but for satellite sub-samples selected to have a minimum distance of 50\,kpc (first and third column) or 100\,kpc (second and fourth column) from their host.
\label{fig:radiussubsamples}}
\end{figure*}

One concern is that a comparison between the observed and the simulated systems does not consider satellites of similar luminosity, because only the most luminous satellites will be identified by SDSS. The previous sections already revealed some indications that the lopsided signal is preserved, and possibly enhanced, for more massive satellites, which on average can be expected to be more luminous. To address this question in more detail, we now split up the satellite galaxies in the Millennium-I and Millennium-II simulations into high- and low-mass samples. The samples are selected by the peak virial mass $M_\mathrm{vir}$\ of the satellites, and the masses seperating the two regimes are chosen to result in roughly similar sample sizes such that similar statistics are obtained. 

We choose to split by $M_\mathrm{vir}$\ and not by the luminosity of the satellites because mass is more directly related to the underlying dynamics, and independent of the semi-analytic galaxy formation model applied to the DMO simulations. However, we expect that in particular the most massive satellites follow a relation between stellar and dark halo mass according to which galaxies in more massive halos tend to be more luminous. We have checked and confirmed that the results are indeed virtually unchanged if we split the satellite samples according to their $r$-band magnitude instead of $M_\mathrm{vir}$.

The resulting cumulative angular distributions are shown in Fig. \ref{fig:masssubsamples}.
For Millennium-II, we separate the satellites at $M_\mathrm{vir} = 10^{9}\,\mathrm{M}_{\odot}$. Therefore, satellites such as the Small Magellanic Cloud (SMC) clearly fall in the more massive subset. This subset tends to show a stronger signal, with the exception of the 2D projected case without orphans. However, even in that case the simulated signal still substantially exceedes the observed one. For Millennium-I, the lowest peak virial masses in the simulation are $M_\mathrm{vir} \approx 10^{10}\,\mathrm{M}_{\odot}$, such that all satellites are in the SMC regime or above. Splitting the samples at $M_\mathrm{vir} = 10^{11}\,\mathrm{M}_{\odot}$\ again reveals a stronger excess of lopsided satellites in the higher-mass subset. Overall, the lopsided signal thus appears to increase for more massive satellites. 
While this is not exactly a cut to match observed luminosities, which for Millennium-II would not leave enough statistics, it is reassuring that the lopsided signal, if different, should be stronger for the more luminous objects.

Another concern is the physical separation between the satellites and their host. In particular in the inner regions of a dark matter halo, non-gravitational interactions and additional tidal effects caused by the host galaxy can affect the spatial distribution and survivability of satellite galaxies \citep{2017arXiv170103792G}. If the lopsided signal were dominated by satellites in this region, the simulation results would have to be considered unreliable. To test this, we analyse subsamples of the satellite distributions by excluding all satellites within 50 or 100\,kpc of their host. The results are plotted in Fig. \ref{fig:radiussubsamples} for both the 3D and the projected 2D distributions. In the latter case, satellites within a projected radius of 50 or 100\,kpc are excluded. The lopsided signal clearly remains strong if the innermost satellites are excluded. There is even some indication that the signal is stronger for more distant satellites, but the differences are too small to allow a firm conclusion. These findings indicate that the lopsided signal found in the dark-matter-only  simulations is not dominated by the innermost satellites. We therefore consider the signal to be reliable, and expect that it will continue to be present even in more detailed hydrodynamic cosmological simulations that model non-gravitational interactions and additional tidal disruption due to more realistical host galaxies.

\section{Overlap Signal}\label{sec:overlap}

\begin{figure}
\plotone{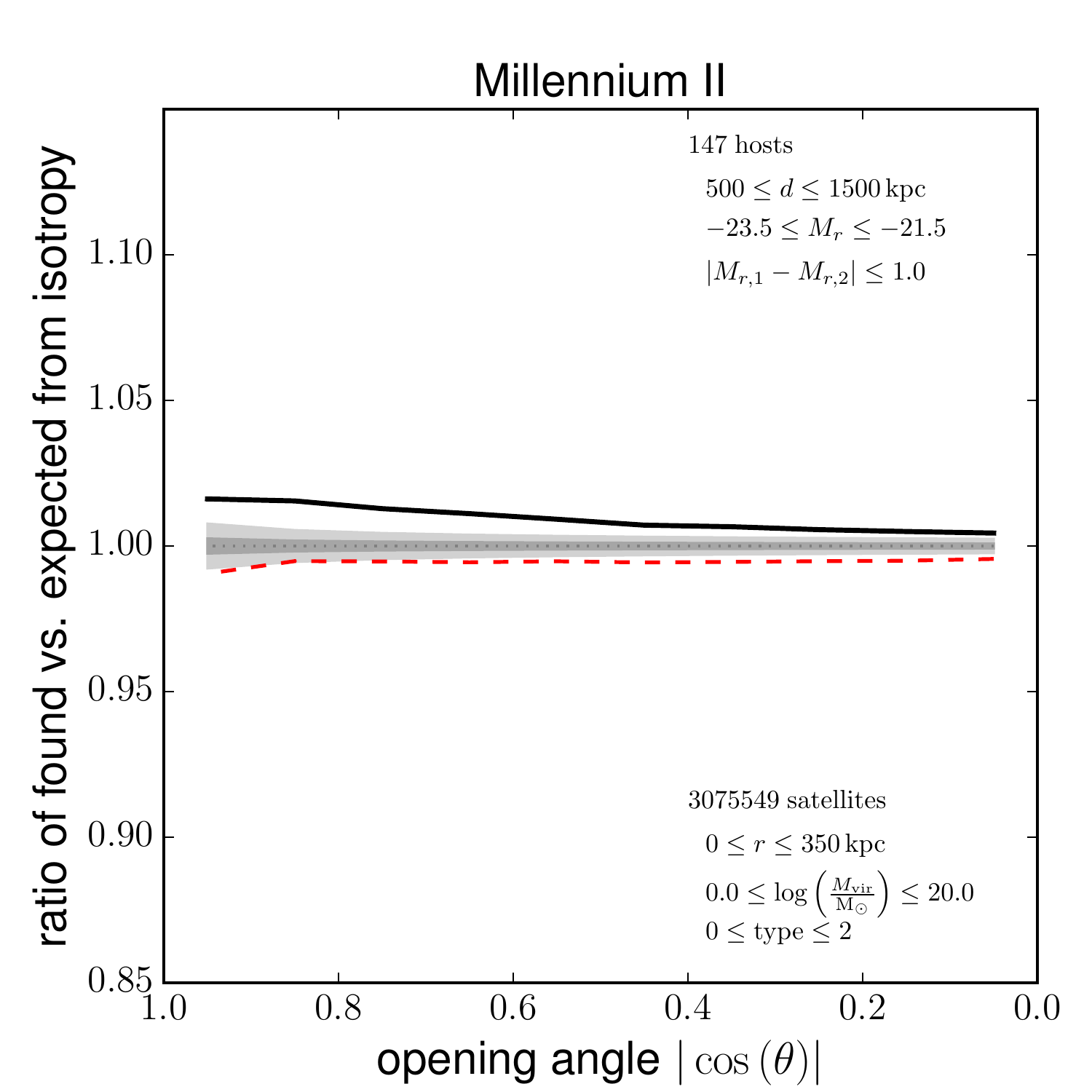}
\plotone{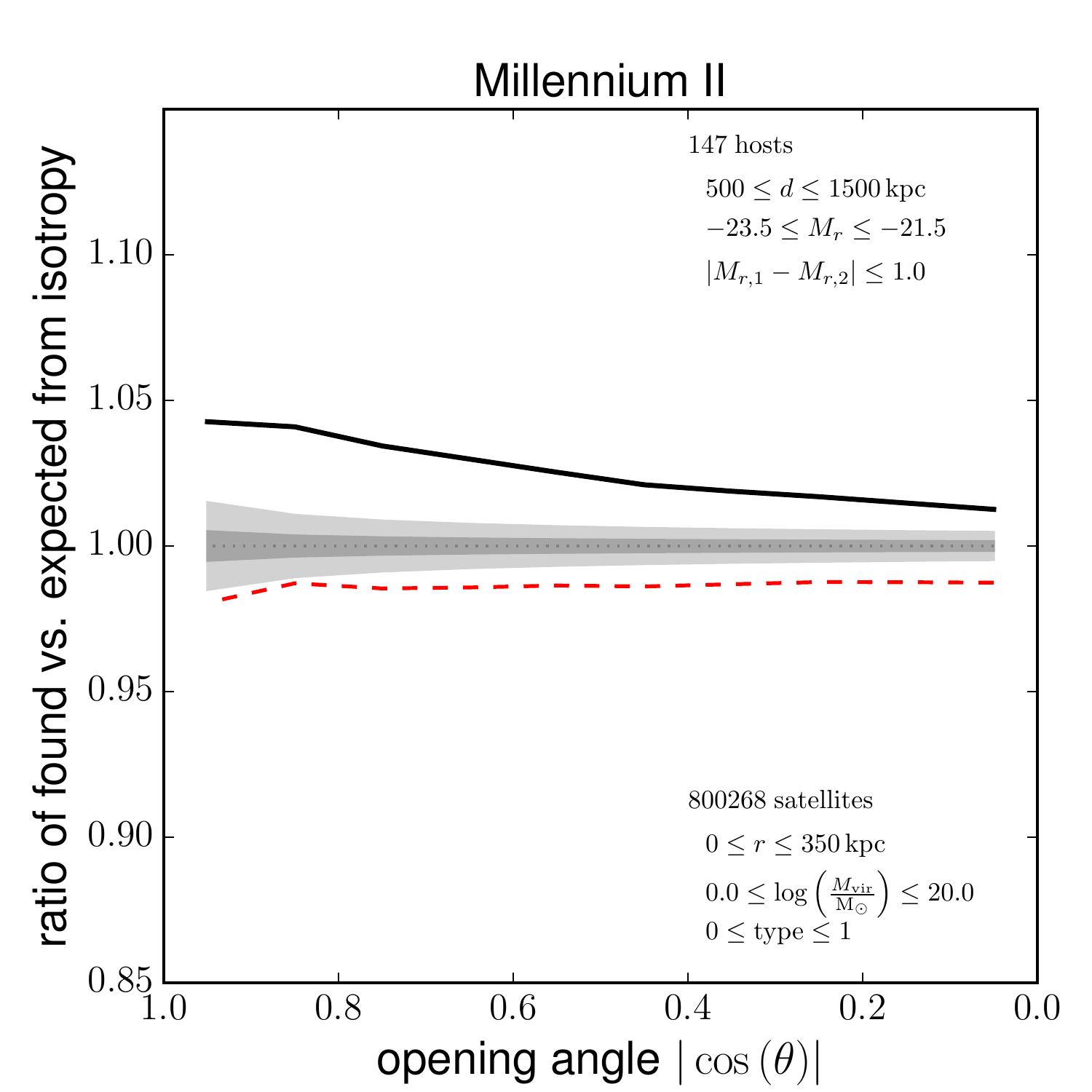}
\caption{
Signal due to an overlap in satellite systems around isolated hosts, selected to resemble the host galaxy pairs in magnitude and distance. As in Figures \ref{fig:masssubsamples} and \ref{fig:radiussubsamples}, black solid lines show the relative abundance of satellites on the side facing the partner primary, while red dashed lines indicate their relative abundance in the opposite direction. They are given relative to an isotropic distribution, whose 1 and $3\sigma$\ scatter is illustrated by the dark and light gray areas. The {\it upper} panel shows the signal including orphan galaxies, the {\it lower} panel the one when orphan galaxies are excluded. While part of the lopsidedness found in the simulations can be attributed to an overlap in the surrounding galaxy distributions, the observed excess is larger by a factor of $\gtrsim 6$\ (with orphan galaxies) to $\gtrsim 2$ (excluding orphan galaxies).
 \label{fig:fakepairs}}
\end{figure}

One possible explanation for the observed lopsidedness is a simple overlap of the galaxy distributions surrounding the hosts. If not only the satellite distribution but also the larger-scale galaxy distribution shows increasing number density with decreasing radial distance from the primaries, then placing two primaries in close proximity can result in an excess of galaxies towards the partner primary. To test whether this effect is sufficiently strong to explain the observed signal, we create artificial pairs of primaries by stacking isolated hosts and their surrounding galaxy distribution from the Millennium-II simulation. For this, we first identify all isolated host galaxies with  $-23.5 \leq M_r \leq -21.5$\ and no other galaxy with $M_r \leq -20.5$\ within 2\,Mpc. For each of our paired primaries, we chose those isolated hosts from this list that match the paired primaries the closest in $M_r$. These isolated galaxies with their surrounding galaxy population are then placed at the exact same distance as the original paired primaries. To ensure that the effect we observe is only due to the radial distribution of galaxies around the primaries, we randomize the positions of the surrounding galaxies relative to the primary by drawing them from an isotropic distribution while preserving their distance from the host. This is done 100 times for each galaxy to increase the statistics.

The resulting lopsided signal is plotted in Fig. \ref{fig:fakepairs}. As expected, the overlap in surrounding galaxies causes some lopsidedness in the distribution of galaxies around the artificial pairs. However, the excess is very small (less than 2 per cent) if orphan galaxies are included, and about twice as high if they are excluded. This difference is expected due to the considerably smaller average distance of orphan galaxies from their host (higher radial concentration). They dominate close to their respective host where the randomization of their positions results in an isotropic signal, but do not contribute significantly to the galaxies at larger distances, which cause the overlap signal in the partner primary. 
Overall, the overlap signal is smaller than the signal found for the simulated primary pairs. We thus conclude that this effect is unlikely to be the sole cause of the found lopsidedness.

\section{Discussion and Conclusion}\label{sec:conclusion}

Based on an analysis of the stacked distributions of possible satellite galaxies around host galaxy pairs in SDSS, \citetalias{2016ApJ...830..121L} found a significant lopsidedness of satellite galaxy systems towards the second primary. Motivated by this, we searched for a similar signal in $\Lambda$CDM simulations. We find that the satellite distributions in the Millennium-I and Millennium-II simulations are indeed lopsided towards their partner primary. This anisotropy is highly significant (up to $5.0 \sigma$\ in 3D, $5.3 \sigma$\ in 2D). The degree of the lopsidedness exceeds the observed signal by up to a factor of two (see Figure \ref{fig:obsvstheory}). We argue that this is consistent with the existence of uniformly distributed fore- and background contamination in the observed sample, which is based on photometrically selected satellite galaxy candidates. Tentative evidence indicates that the signal is stronger for more massive satellites. A simple overlap of the galaxy distributions surrounding the paired primaries is not sufficient to explain the signal.

The sub-halo distributions of Local-Group-like galaxy pairs in the ELVIS suite display a similar lopsided signal as their lower-resolution counterparts. The satellite systems within the hydrodynamic Illustris-1 simulation are consistent with these findings, but are also consistent with a uniform distribution. This unsatisfactory result is due to the low number of host galaxy pairs and resolved satellite galaxies. Hydrodynamic simulations providing a larger sample of host pairs and a larger number of satellites will be required to clarify the situation.

The lopsidedness of satellite systems around galaxy pairs is potentially related to dark matter filaments connecting the primaries. The weak gravitational lensing analysis of stacked pairs of luminous red galaxies by \citet{2017MNRAS.468.2605E} revealed a significant signal indicative of filaments connecting the pairs. While their systems are more massive and more extended than the ones studies here, and their analysis focussed on the distribution of mass in the region beyond the virial radii of the hosts, their results offer a potential explanation for the lopsidedness because satellites are preferentially accreted along such filaments \citep{2004ApJ...603....7K,2011MNRAS.411.1525L}. It might therefore be of interest to study the distribution of satellite galaxies around the luminous red galaxy pairs of \citet{2017MNRAS.468.2605E} to observationally investigate a potential correlation with the filament signal. 
One would expect that if satellite galaxies are accreted along filaments which are more narrow than the host halo (as expected for massive halos situated at the end of filaments), then the satellites will have very radial orbits. Such orbits will bring the satellites close to the host galaxy, where they can be more easily disrupted. This can potentially help to explain why a signal in the direction opposite to the companion host galaxy can be found and is present only when orphan satellites are considered since these are by definition disrupted sub-halos.

It is intriguing that the $\Lambda$CDM model can reproduce this particular observation of lopsided satellite systems, given its struggles with other small-scale problems \citep{2010A&A...523A..32K, 2011MNRAS.415L..40B, 2011ApJ...742...20W, 2014ApJ...784L...6I, 2014MNRAS.442.2362P, 2017ApJ...836..152L,2017ARA&A..55..343B}. The present work can thus only be a first step in trying to understanding the lopsided nature of satellite distributions around host galaxy pairs with a theoretical approach. What is the dynamical origin of this effect? Is it unique to $\Lambda$CDM, dependent on the type of dark matter, or simply a general property of pairs of similar-mass galaxies in any gravitational dynamics? Do hydrodynamic effects influence the spatial distribution of satellite galaxies? Why do orphan satellites show a second excess in the direction facing away from the other primary?
Answering these questions will require large-volume, high-resolution cosmological simulations, as well as more focussed dynamical studies of paired host galaxies and their satellite systems.

\acknowledgments
We thank the anonymus referee for their thoughtful and helpful comments. MSP acknowledges that support for this work was provided by NASA through Hubble Fellowship grant \#HST-HF2-51379.001-A. JSB was supported by NSF grant AST-1518291,  HST theory programs AR-13921, AR-13888, and AR-14282.001, as well as program number HST-GO-13343. All of our Hubble-related support is  awarded by the Space Telescope Science Institute, which is operated by the Association of Universities  for  Research  in  Astronomy,  Inc.,  for  NASA,  under  contract  NAS5-26555. 
The Millennium Simulation databases used in this paper and the web application providing online access to them were constructed as part of the activities of the German Astrophysical Virtual Observatory (GAVO).

\end{document}